\documentclass[sigconf]{acmart}

\usepackage{acmart-taps,mydefs}

\usepackage{colortbl}
\usepackage{subcaption}
\usepackage{fontawesome5}
\usepackage{enumitem}
\usepackage{listings}
\usepackage{fancyvrb}
\usepackage{multirow}
\usepackage{stfloats}
\usepackage{makecell}
\usepackage{booktabs,tabularx,array}
\usepackage{verbatimbox}

\usepackage{booktabs,tabularx,array}
\usepackage{verbatimbox}
\usepackage{longtable}
\usepackage{float}
\usepackage{enumitem}
\usepackage{graphicx}
\usepackage{array}
\usepackage{geometry}
\usepackage{listings}
\usepackage{fancyvrb}
\usepackage{xcolor}
\usepackage{wrapfig}
\usepackage{fvextra}
\usepackage{multirow}
\usepackage{stfloats} 
\usepackage{placeins}
\newcolumntype{Y}{>{\raggedright\arraybackslash}X}

\usepackage{tikz}

\DefineVerbatimEnvironment{MyVerbatim}{Verbatim}
  {frame=single,fontsize=\small,breaklines=true,breaksymbolleft={} }

\definecolor{magma5}{HTML}{484848} %
\definecolor{magma4}{HTML}{636363}
\definecolor{magma3}{HTML}{909090} %
\definecolor{magma2}{HTML}{B4B4B4}
\definecolor{magma1}{HTML}{D4D4D4} %
\definecolor{mygray}{HTML}{58595b}
\definecolor{darkred}{HTML}{be1e2d}

\newcommand{\dotraise}{0.20ex}        %
\newcommand{\dotscale}{1.25}          %

\newcommand{\revision}[1]{#1}

\newcommand{\stageRaise}{-1ex}       %
\newcommand{\stageBox}{1.45em}
\newcommand{\stageDotScale}{3}
\newcommand{\stageTextRaise}{1.25ex}    %
\newcommand{\stageText}{\scriptsize\sffamily\bfseries}

\AtBeginDocument{%
  }

\copyrightyear{2026}
\acmYear{2026}
\setcopyright{cc}
\setcctype{by}
\acmConference[CHI '26]{Proceedings of the 2026 CHI Conference on Human Factors in Computing Systems}{April 13--17, 2026}{Barcelona, Spain}
\acmBooktitle{Proceedings of the 2026 CHI Conference on Human Factors in Computing Systems (CHI '26), April 13--17, 2026, Barcelona, Spain}
\acmDOI{10.1145/3772318.3790330}
\acmISBN{979-8-4007-2278-3/2026/04}

\begin{document}

\title{\textsc{Texterial}: A Text-as-Material Interaction Paradigm for LLM-Mediated Writing }

\author{Jocelyn Shen}
\affiliation{%
  \institution{Microsoft Research}
  \city{Redmond}
  \state{WA}
  \country{USA}
}
\affiliation{%
  \institution{MIT Media Lab}
  \city{Cambridge}
  \state{MA}
  \country{USA}
}
\email{joceshen@mit.edu}

\author{Nicolai Marquardt}
\orcid{0000-0002-5473-2448}
\affiliation{\institution{Microsoft Research}
\city{Redmond}
\state{WA}
\country{USA}}
\email{nicmarquardt@microsoft.com}

\author{Hugo Romat}
\orcid{0000-0002-9194-8889}
\affiliation{\institution{Microsoft Research}
\city{Redmond}
\state{WA}
\country{USA}}
\email{romathugo@microsoft.com}

\author{Ken Hinckley}
\orcid{0000-0002-4733-4927}
\affiliation{\institution{Microsoft Research}
\city{Redmond}
\state{WA}
\country{USA}}
\email{kenneth.p.hinckley@gmail.com}

\author{Nathalie Riche}
\orcid{0000-0003-1759-7512}
\affiliation{\institution{Microsoft Research}
\city{Redmond}
\state{WA}
\country{USA}}
\email{nath@microsoft.com}

\author{Fanny Chevalier}
\affiliation{%
  \institution{Microsoft Research}
  \city{Redmond}
  \state{WA}
  \country{USA}
}
\affiliation{%
  \institution{University of Toronto}
  \city{Toronto}
  \state{ON}
  \country{Canada}
}
\email{fanny@dgp.toronto.edu}

\renewcommand{\shortauthors}{Shen et al.}

\begin{abstract}
What if text could be sculpted and refined like clay---or cultivated and pruned like a plant? \textsc{Texterial} reimagines \textbf{text as a material} that users can grow, sculpt, and transform. Current generative-AI models enable rich text operations, yet rigid, linear interfaces often mask such capabilities. We explore how the text-as-material metaphor can reveal AI-enabled operations, reshape the writing process, and foster compelling user experiences. A formative study shows that users readily reason with text-as-material, informing a conceptual {framework} that explains how material metaphors shift mental models and bridge gulfs of envisioning, execution, and evaluation in LLM-mediated writing. We present the design and evaluation of two technical probes: \textit{Text as Clay}, where users refine text through gestural sculpting, and \textit{Text as Plants}, where ideas grow serendipitously over time. This work expands the design space of writing tools by treating text as a living, malleable medium.
\end{abstract}


\begin{teaserfigure}
 \includegraphics[width=\textwidth]{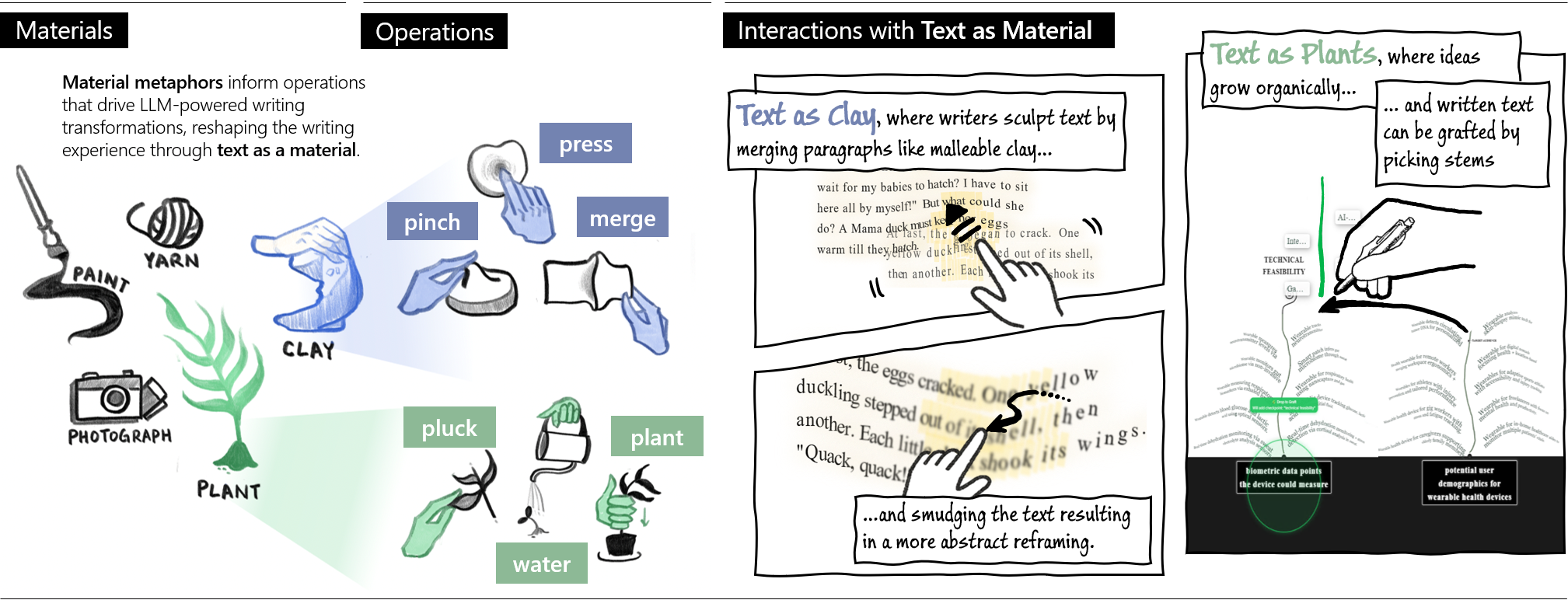}
 \caption{We explore how materials can inspire new interactions with text. Building on insights from a formative study, we propose \textsc{Texterial}, a conceptual {framework} and prototype implementations that treat ~\textit{text-as-material}. Our work uses materiality to foster a new way of interacting with generative AI.}
 \Description{Shows a framework for material-inspired text interaction. On the left, icons for materials (paint, yarn, clay, photograph, plant) are paired with operations (e.g., pinch, merge, pluck, plant, water). In the center, a plant metaphor is illustrated where dragging creates a branching fern of words. On the right, pressing on text highlights and emphasizes localized words. Together, the diagram illustrates how materials, operations, and interactions combine to treat text as a malleable medium that fosters a new way of interacting with generative AI.}
 \label{fig:teaser}
     \vspace{10pt}
\end{teaserfigure}

\maketitle

\section{Introduction}

Current large language models (LLMs) offer unprecedented capabilities to transform and generate text, from abstracting meaning \cite{huang_towards_2023} and remixing ideas \cite{gero_sparks_2022}, to suggesting alternatives \cite{lu_whatelse_2025} or drafting entire passages from a single seed phrase~\cite{wei_emergent_2022, brown_language_2020}. Yet, these capabilities are not always used to their full potential. Although natural language {instructions provide versatility, crafting effective prompts and steering the results is often difficult~\cite{zamfirescu2023johnny}. In particular, understanding what the model can do and how to instruct it within the constraints of linear chat interfaces, limits the ways in which people fluidly think \textit{with} the model and bottlenecks their wide-ranging capabilities~\cite{subramonyam2024bridging}.}

To better surface the expressive potential of LLMs, we propose a conceptual shift for designing {generative AI-powered} writing experiences: \textbf{reimagining text as a material} --- something that can be directly manipulated, blended, squashed, or shattered. {Our framework is inspired by metaphors to physical materials (clay, paint, or even plants), which map LLM operations to expressive touch interactions.} {For example, imagine a creative story writer refining their draft. Instead of prompting, “\textit{Make this paragraph more vivid, while keeping the introduction and conclusion the same, and merge it with the style of Shakespeare...}” the user might stretch the text like clay to expand the text, pinch it to make certain phrases more concrete, and smear one of Shakespeare's sonnets into their writing.} 
{Through this perspective, the material properties, and interactions afforded by the material, help externalize the writing process by leveraging intuitive sensorimotor skills and embodied reasoning (e.g., ripping, mixing, smoothing) in ways that prompts cannot} ~\cite{neale_chapter_1997, malafouris_how_2013, KirshIntelligentUseOfSpace1995, HutchinsCognitionWild1996}.

{To expand this idea of text-as-material, \textsc{Texterial} generalizes HCI design metaphors to LLM-mediated writing.}
{We envision these techniques \textit{enriching} and \textit{complementing} existing writing techniques rather than replacing them, opening up a design space of material-inspired interactions that can infuse the wider scope of writing activities at different stages---early ideation, editing, or iterative refinement.}
{To explore this design space,} we first conducted a formative study to understand how individuals might conceptualize text as a material, gaining insight into what metaphors resonate with people, and how they would approach writing under such a paradigm (Section~\ref{sec:formative-study}). %
Building on these insights, we distilled important dimensions and considerations for material-inspired LLM-driven writing tools into a conceptual framework, from the lens of the seven-stages-of-action model~\cite{norman2013design} in Section~\ref{sec:framework}.
Our conceptual {framework} articulates the layers on which LLMs operate (semantics, structure, style), operations on these layers (compose, abstract, ideate, condense, transform); and mapping execution of operations to material-like interactions. 
This {framework} formalizes {LLM functional capabilities} and metaphor-interface matches~\cite{neale_chapter_1997}, providing interaction designers with a structure and language to discuss how materials afford different AI-powered operations and inspire new writing interfaces that help bridge the gulfs of envisioning, execution, and evaluation~\cite{subramonyam2024bridging, norman2013design}.

As concrete demonstration of this conceptual {framework}, we developed two technical probes that {embody text as material for writing tasks with LLMs}:
\textit{Text as Clay}
and \textit{Text as Plants} (Section~\ref{sec:probes}).
\textit{Text as Clay} reimagines the text itself as clay-like material that can be pinched, stretched, pulled, merged, and smoothed through direct multi-touch gestures, to combine and refine text fragments.
\textit{Text as Plants} treats writing as an act of {gardening}: ideas are planted, watered, and grafted, and the system uses a language model to introduce variations, novel ideas, and leaves that branch and {evolve over time}. 
{Both of these technical probes are positioned to explore how material-inspired interactions can complement existing writing practices, with the potential to address distinct phases in the writing process.}
We used these probes to evaluate our approach in a qualitative focus-group study (n=10, 4 groups; Section \ref{sec:study}). Our findings indicate that framing text-as-material influences mental models.
{Gestural manipulations were found to be intuitive, expressive, and language-agnostic, which invited direct engagement with text transformations. The spatial and visual layouts of our prototypes supported non-linear exploration and non-committal ideation, workflows that are rarely afforded by linear LLM interfaces. Finally, participants found that our prototypes infused creative experimentation and playfulness in the process, and envisioned extensions of the metaphor to collaborative writing interactions.}

In sum, \textsc{Texterial} contributes the following:
\begin{itemize}
    \item \textbf{Empirical insights from a formative study} of how people conceptualize text-as-material, illustrating how material metaphors can naturally map to writing operations and phases of the writing process.
    \item \textbf{A conceptual {framework}} of the text-as-material approach that links material affordances to text manipulations with LLMs, providing {foundations for interaction designers of} novel, AI-driven writing tools.
    \item \textbf{Two diverse technical probes}---\textit{Text as Clay} and \textit{Text as Plants}---expanding the design space of writing tools by treating text as a malleable, interactive medium to embody underlying LLM operations.
    \item \textbf{Findings from a qualitative focus-group study} of these technical probes, offering evidence that material metaphors shape mental models for manipulating text, foster novel workflows, raise different expectations for different people, and invite creative and playful explorations.
\end{itemize}

Our research envisions a new way of writing with AI: one that invites users to craft and play with text as if it were a physical material {and imagines interactions that enable direct engagement with the writing process, rather than restricting users to linear, chat-based interfaces that pull the writer away from the act of writing itself}. 
{With the structured framework, technical probe explorations, and user study insights, our work is intended to inform and inspire designers and developers of writing interfaces, providing a set of enriching and complementary techniques that broaden the expressive possibilities of AI-mediated writing.}

\section{Background and Related Work}
Our work applies the conceptual lens of metaphors in HCI to writing interfaces. 
We review literature on materiality and design metaphors, conceptual frameworks that inform writing interfaces, and {the design and evaluation of} digital writing systems {previously introduced in the field}.

\subsection{Materiality and Design Metaphors}
HCI has long used metaphors to make digital systems more understandable by grounding them in familiar concepts \cite{neale_chapter_1997, carroll1988interface, blackwell_reification_2006}. For example, early graphical interfaces embraced metaphors literally through skeuomorphs---interface elements that mimic the aesthetics of a familiar physical medium to convey affordances~ \cite{kimball1982designing,jung_metaphors_2017, marcus_principles_1995, AdarBenevolentDeception2013}---introducing desktop, files, and trash can digital counterparts to physical desktops to help  users understand and appropriate new technologies~\cite{gross_skeu_nodate}. Although metaphors are a powerful scaffold for developing one's mental model and providing intuition and context \cite{Renom2023InteractionKnowledge, renom2022exploring, neale_chapter_1997}, interaction designers caution against letting metaphors hinder innovation by constraining themselves too closely to the physical world
\cite{gessler1998skeuomorphs}.

Artists and craftspeople often think through materials: the affordances and constraints of a medium like clay or wood actively shape the creative process, often suggesting new directions \cite{malafouris_at_2008, mccullough_abstracting_1998}. 
Design metaphors concerned with materials have inspired work in HCI, such as tangible user interfaces \cite{ishii_tangible_1997, piper_illuminating_2002} and direct manipulation techniques \cite{hutchins_direct_1985}. As highlighted in theories of material engagement, materials allow users to apply their intuitive spatial or physical reasoning to interaction \cite{malafouris_how_2013}, i.e. a form of \textit{embodied} interaction~\cite{dourish2001action}. Materials provide a slow and iterative creative process and rapport to the medium itself, allowing for direct manipulation of objects on a screen \cite{malafouris_at_2008, malafouris_how_2013}. 

Our work builds on this rich history of materiality and design metaphors in HCI to explore how we might interact with \textit{text} as a material \cite{schmid_empowering_2013}. This departs from the standard keyboard-screen model of writing, much as tangible user interfaces departed from WIMP interfaces \cite{angelini_tangible_2015}. By conceptualizing text as a material, we aim to surface the capabilities of underlying generative models in writing interfaces through natural gestures and manipulations of the material. %

\subsection{Conceptual Frameworks for Writing Interfaces}
Prior works have proposed conceptual frameworks that categorize and inspire new writing interfaces. Such frameworks typically delineate the stages of the writing process, following the model of planning, translating, and revising in some form \cite{hayes_new_2000, flower_cognitive_nodate, wallas_art_2014, csikszentmihalyi_creativity_2010}. Review articles in the field find a dearth of writing tools targeted specifically for the pre-writing phase (e.g., idea generation)
\cite{zhao_making_2025}. 

Recent works provide design spaces and frameworks that are helpful for describing \textit{AI-driven} writing tools, drawing inspiration from creative practices \cite{frich_mapping_2019-1,lee_coauthor_2022, lee_design_2024, kim_authors_2024}. These frameworks are often categorized by author attributes, the writing process, the paradigm or interaction system, the technology used to assist writing, or the overall ecosystem where writing occurs \cite{lee_design_2024}. For instance, 
Buschek~\cite{buschek_collage_2024} proposes approaching the design of modern LLM-driven writing tools through a "collage" process: interfaces fragment text into pieces, juxtaposed with different voices or content sources, and shift the author's role from composing text to arranging it. The \textit{Cells, Generators, and Lenses} structure the design space from an algorithmic, object-oriented perspective \cite{kim_cells_2023}. In this model, 
the key components of text generation are reified into interactive objects: cells (text units or prompts), generators (configurable model instances), and lenses (output displays or filters). By conceptualizing writing tools as interactive, composable, and persistent UI elements, the design framework focuses primarily on the iterative process of creation \cite{sawyer_iterative_2021}. Beyond the text domain, ``AI Instruments'' enable making a user’s intent into a tangible, direct-manipulation tool~\cite{riche_ai-instruments_2025}. These frameworks provide lenses to both \textit{generate and understand} novel experiences in the age of generative AI. 

Overall, conceptual frameworks in the field offer both \textit{analytical} tools to critique existing AI writing UIs, and \textit{constructive} tools to inspire new designs focused on embracing fragmentation and recombination in writing. {Our framework instead provides a structured approach for designing interaction techniques from the perspective of the human model of action~\cite{norman2013design}. It does so by articulating functional capabilities of LLMs and conceptualizing interaction design in terms of materiality and its affordances~\cite{norman1999affordance, hartson2003cognitive}.}

\subsection{Writing Interfaces in HCI}
Prior works have long explored developing novel writing interfaces that move beyond the static "typewriter" paradigm. Early hypertext systems, such as Bolter and Joyce’s interactive fiction editor 
{let writers interact with a diagrammatic view of their text~\cite{bolter_hypertext_1987}, paving the way for non-linear, "fluid text" documents that acted as canvases which could be dynamically rearranged to make room for additional information beyond the text itself \cite{zellweger_impact_2000}}. Interest in writing support tools have resurged with modern AI, with many early ideas being viable with today’s LLMs \cite{mahlow_writing_2023}. 

A first line of work emphasizes direct manipulation for the task of \textbf{text editing}. These interfaces often focus on addressing the challenge of {maintaining consistency in textual documents and} controllable text manipulation. For example, Textlets~\cite{han_textlets_2020} reify text selections into persistent entities, which can aid in search-and-replace operations. {Through focus-group studies, this work found that ``reification'' (turning commands into first class objects or instruments), ``polymorphism'' (applying instruments to different types of objects), and ``reuse'' (making user inputs and system outputs reusable) encourages writers to use functions they might otherwise avoid with traditional text editors. With LLMs, recent interfaces focus primarily on leveraging models to perform writing tasks with greater controllability, while reducing the burden of drafting complex prompts or task sequences to achieve the user's editing goals \cite{lee_coauthor_2022}. For example, DirectGPT~\cite{masson_directgpt_2024} maps prompts to direct manipulation actions and minimizes both time and length of prompts written to edit text, code (and also images), when compared to baseline ChatGPT. LMCanvas~\cite{kim_lmcanvas_2023} reifies prompts and model actions into composable interface elements, allowing writers to spatially reorganize ideas and invoke LLM text operations in place, rather than opening separate chat windows. Composable tools like Phraselette~\cite{calderwood_phraselette_2025} provide a visual palette to inspire more creative language, where authors can specify constraints applied to snippets of AI-generated text. By making small chunks of content malleable, these interfaces are designed to give authors interactive control over LLM edits without significant overhead of drafting highly specific prompts for steerability.
}
 
{Beyond interfaces that support text editing,  other works} use \textbf{visualization and information structuring} to give the user better "gists" of their current writing or {to explore idea spaces. 
For example, Lexichrome~\cite{kim_lexichrome_2020} investigates how relationships between colors and words can support writers quickly grasp editorial decisions. 
Researchers have also introduced systems that leverage LLMs to generate dynamic summaries that assist writers with \textit{global} text coherence. For example, Dang et al.\cite{dang_beyond_2022} proposed an editor that displays continuous AI-generated summaries for each paragraph alongside a written draft. They find through interview studies that AI summaries help authors reflect on the structure and alignment of their arguments as they write. Other works generate more structured gists through visual node-link maps that can help writers identify gaps in reasoning or support narrative construction
\cite{zhang_visar_2023}. For example, Masson et al.~\cite{masson_visual_2025} introduced "visual story-writing" and developed a text editor that automatically visualizes a graph of entity interactions, movement between locations, and a timeline of story events such that interactions with the visual representations translate back to written text. For idea generation and sensemaking, tools such as Luminate~\cite{suh_luminate_2024} and Sensecape~\cite{suh_sensecape_2023} let users explore a semantic “landscape” of AI-suggested ideas. In these works, the model generates and clusters concepts related to the user’s topic, which the writer can browse or interpolate to discover new angles on their writing. Similarly, WhatELSE~\cite{lu_whatelse_2025} uses AI to create a "possibility space" from user-defined narrative constraints, which allows writers to navigate variations in text and different levels of abstraction.
Such tools support writer's sensemaking when dealing with a large volume of suggestions~\cite{gero_supporting_2024}.}

{Most closely related to our work are} recent LLM-powered tools that \textbf{adapt design metaphors to support text editing or text generation}. For example, Textoshop~\cite{masson_textoshop_2025} approaches text editing through the analogy of {photo editing software such as Photoshop}, using smudge or tone brushes to smooth transitions or apply stylistic features. {The authors found that Textoshop improved users’ editing efficiency and satisfaction, preferring the interface over conventional document editors with ChatGPT}. 
Other works draw analogies to layering~\cite{siddiqui_scriptshift_2025, masson_textoshop_2025}, sketching \cite{chung_talebrush_2022} and painting \cite{chung_promptpaint_2023}, or even toy play \cite{chung_toyteller_2025} to steer text generation. {As one example, Script\&Shift~\cite{siddiqui_scriptshift_2025} presents a layered document interface where writers can separate \textit{content} layers from \textit{stylistic} layers, using modular units that encapsulate elements of the writing process to minimize disruption to the workflow. As another design metaphor example, TaleBrush~\cite{chung_talebrush_2022} enables writers to draw a story arc on a digital canvas, then uses a LLM to generate narrative text that follows the curve. This interaction technique allows authors to "doodle" or "sketch" stories rather than typing full sentences. Similarly, ToyTeller~\cite{chung_toyteller_2025} treats narrative construction as a form of toy play, where users manipulate toy characters and props to generate story text. Through qualitative analysis of think-aloud statements and interviews, these studies often find that such metaphors make the creative process more approachable, rather than leaving the user to wrestle with a blank page.}

Across these systems, HCI researchers have explored diverse metaphors to leverage the capabilities of generative models and shape the writing process. Our work builds on and extends this trajectory. {Moving beyond isolated metaphors, we propose a text-as-material perspective that synthesizes (1) the core functions and possibilities of AI-mediated writing, and (2) the way material affordances can shape how people understand and work with these tools.}

\section{Formative Study}
\label{sec:formative-study}

We first explored how people respond to the idea of treating text as  material. In a formative study,  participants imagined text as a handcraft medium and described the manipulations a digital tool could support in this paradigm. {Our goal was to assess whether} 
conceptualizing text as materials lends itself to text manipulations, inspires novel workflows, and what specific metaphors resonate with individuals{; rather than than exhaustively cataloging all possible ideas. We further explore our concept following a research-through-design approach~\cite{zimmerman2007research} using technical probes (\autoref{sec:probes}).}

\begin{figure}[htb]    
  \centering
  \vspace{10pt}
  \includegraphics[width=\linewidth]{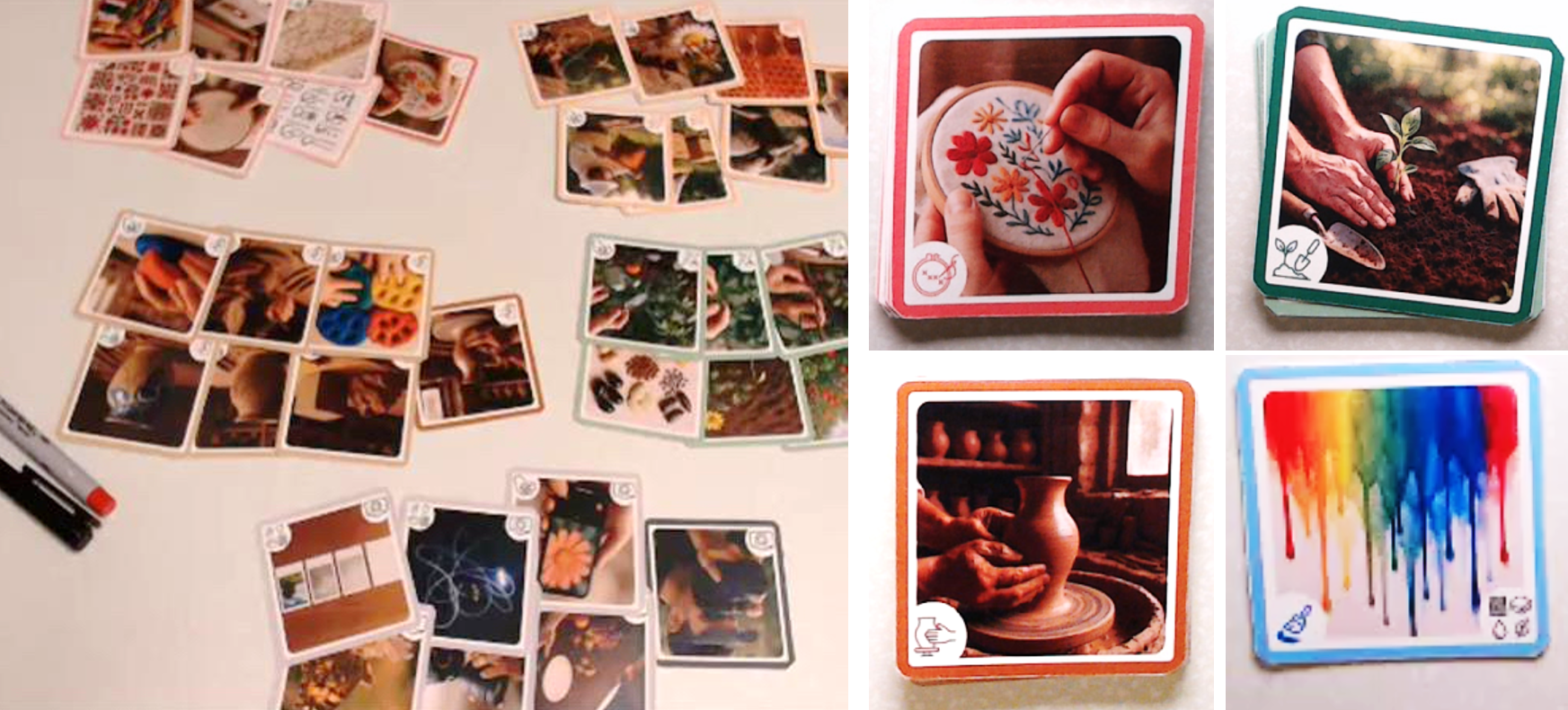} \caption{Elicitation materials from our formative study. The inspirational cards include entry-level crafts (gardening, pottery, photography) and advanced ones (embroidery, beekeeping), as well as inanimate (clay, threads), living  (plants, bees), and hybrid materials (photographed scenes).}
  \Description{Depicts a collection of square inspirational cards used in a formative study. The cards feature images of crafts and materials such as embroidery, gardening, pottery, photography, threads, plants, bees, and colorful paint. They represent a mix of entry-level and advanced crafts, inanimate materials, living elements, and hybrid photographed scenes.}
  \label{fig:elicitationmaterials}
\end{figure}

\subsection{Methodology \& Procedure}
The facilitator briefed participants in a walkthrough, illustrating the expected thinking using paint as an example material: \emph{``If text were paint, what would mixing paint mean? %
Would the same text manipulations apply if text were dry or wet paint?''}

Our methodology builds on ideation techniques that use metaphor as a generative tool~\cite{lockton2019new, tomitsch2018design}. To elicit ideas, we created inspirational cards~\cite{lockton2019new} featuring various common handcrafts and their associated materials (\autoref{fig:elicitationmaterials}). We suggested a loose explanatory structure~\cite{tomitsch2018design} inviting participants to \textit{`tell the metaphor's story'}--explain how text is manipulated through the metaphor; \textit{`elaborate the triggering concept'}--identify what aspect of the material prompted the conceptual transfer; and \textit{`elaborate assumptions'}--describe how this should work in practice. Participants were invited to use the cards freely and suggest their own handcraft-materials. We provided prototyping tools (paper, pencils, post-its) to help them express their ideas. The facilitator intervened only to clarify or probe further. {Details about our study methodology and materials used are included in our Supplementary Material.}

Each session lasted about one hour, and video and audio recordings were collected. Participants received \$50. The study was approved by the ethics board of [redacted]. We performed a thematic analysis of the videos and transcripts, using our guiding questions as deductive codes, while emerging themes were inductively captured. We focus our analysis on existence, following an interpretivist approach to reflexive thematic analysis~\cite{braunThematicAnalysisPractical2022}.

\begin{table*}[t]
\small
\caption{Formative study participants. The creativity score (min: 1, max: 50, higher is more creative) and experience with writing score (min: 6, max: 30, higher is more experienced) are based on Renom et al.~\cite{renom2022exploring}.}
\Description{Formative study participants. Creativity scores range from 1 to 50 (higher means more creative), and writing experience scores range from 6 to 30 (higher means more experienced).
P1: Model maker with 15 years of experience. Creativity score 38, writing experience score 12. No regular type of writing.
P2: Designer with over 20 years of experience. Creativity score 43, writing experience score 21. Writes creative briefs and presentations.
P3: Principal PM Manager with 1 year of experience. Creativity score 37, writing experience score 25. Writes emails, essays, scientific articles, and journaling.
P4: NLP Researcher with over 20 years of experience. Creativity score 45, writing experience score 29. Writes scientific articles, short stories, and poetry.}
\label{tab:formative-study}

\resizebox{0.8\linewidth}{!}{
\begin{tabular}{@{}lcccc@{}}
\hline
\textbf{ID} & \textbf{Occupation} & \textbf{Creativity} & \textbf{Experience} & \textbf{Type of Writing} \\
\hline
P1 & Model maker (15 yrs) & 38 & 12 & N/A \\
P2 & Designer (20+ yrs) & 43 & 21 & Creative briefs, presentations \\
P3 & Principal PM Manager (1 yr) & 37 & 25 & Emails, essays, scientific articles, journaling \\
P4 & NLP Researcher (20+ yrs) & 45 & 29 & Scientific articles, short stories, poetry \\
\hline
\end{tabular}
}

\end{table*}

\subsection{Participants}
We recruited four participants through convenience sampling at a tech company (age ranges: 30-39 (1) and 50-59 (3), 2 males and 2 females), targeting creative individuals and communication professionals, and verified alignment using
creativity and text-experience questionnaires from Renom et al.~\cite{renom2022exploring}---see \autoref{tab:formative-study}. %
{We recruited participants with varied backgrounds and writing experience to capture different perspectives on material-inspired interactions across different writing contexts. While our sample is small, our study confirmed that the concept of text-as-material has strong potential, which our work further explores through functional probes.}

\subsection{Results}

{\textbf{Writing without words---Treating text as a material enables a non-verbal approach to writing, which can offer several practical advantages.}} Participants contrasted this with traditional writing using a pen or keyboard, which demands deliberate word choice. The material metaphor instead allows emotions to be enacted physically and visually, by assembling and manipulating fragments of ideas. For example, P1 and P3 described sketching as a way to \textit{``let the emotion out,''} and P2 emphasized the satisfaction of physical expression: \textit{``That’s why I want to play with clay. I just slam it down and pick it up.''} Participants felt that such embodied, primal, organic experiences better matched the \textit{``non-linear''}, \textit{``messy and very unstructured creative part''} of the process (P2, P4) and introduces \textit{``humanness in the process.''} As P1 put it: \textit{``I'm not there just to convey information. I'm there to elicit an emotional response.''} P2 and P4 also remarked that skeuomorphic representations of text could allow them to \textit{``zoom out''} (P2) and visually capture properties such as redundancies (P4) in ways not possible with words.

{\textbf{Working with materials can offer an easy entry into the writing process.}}
 For many, writing can feel intimidating. Using alternative handcrafts as a proxy for expressing meaning could make the process feel less fraught.
Participants described how approaching text from the art of a familiar handcraft could help make it ``\textit{instant}'' (P4). As P1 detailed \textit{``It feels like I'm trying to capture that moment because I know it's going to be gone and I think a lot of... text is like that.''} For this, the metaphor needs to leverage a craft that the author has a reasonable understanding of and expectations about. P4, for whom writing is the most comfortable craft, said they would not benefit from the material approach, but ``\textit{I agree with this philosophy. It's just that for me [other crafts] are much more narrow.}'' They rejoin P2 in that such techniques can help people overcome the writer’s block (P2, P4) and fear of failure (P2, P1).

{\textbf{Different materials lend themselves to different writing tasks}}, such as idea generation, writing (the actual translation of putting words down), and editing, as identified by P4. Gardening was linked to ideation (P1, P4) where the notion of ``\textit{growing text}'' captured co-creation:  \emph{``there's another element that's providing suggestions''} (P1). P1 contrasted this %
with bonsai which they described as \textit{``arrogant''}---``\textit{you're absolutely forcing it to be something}''. This contrast highlights the different forms of authorial agency between material and craftsperson that can be afforded within the same metaphor. Inanimate materials, like clay (P2, P3, P4), metal (P2), paint (P4), jelly paint (P2), and fabric (P4) inspired narrative shaping and editing. Participants moved from rough assembly---\textit{``I roughly know what I want to write. So I'm gonna throw that hunk of clay down.} (P2)---to refinement, adjusting content and style---\textit{ ``%
I need to make it constructive, so I'm just going to sort of slowly take a look at what I have and I'm just going to soothe out, smooth out any rough edges I have in the content''} (P2). Clay was explicitly valued for its forgiveness: ``\textit{I love [the clay] because there's some imperfection here that's turned into beauty}'' (P2). In contrast,  metaphors like embroidery felt less accessible, ``\textit{requiring too much extreme hands on}'' (P1), and a ``\textit{disciplined process}'' (P2).

{\textbf{Working with materials might unlock LLM-mediated writing support.}} 
Participants \textit{enacted} material interactions that leverage the generative power of LLMs in ways that are otherwise difficult to articulate through prompts. For instance, P2 and P3 put pieces of raw clay (i.e. ideas) together in different spatial assemblies to shape a narrative, adjusting order, emphasis, or even framing by forming a semantic shape. Tearing clay removed text; pinching the edges refined it to make it sharper; smoothing polished and clarified ideas. All these manipulations correspond to shaping ideas and building a narrative. Glazing was linked to editing through precise refinements, though participants did not elaborate on specific approaches. For the garden, most described planting a seed (i.e. idea) for it to sprout and evolve (P1, P2, P4), while P1 imagined grafting as a way to combine ideas. P3 likened text development to a polaroid revealing its content---\textit{``as it's developing, the story and the text also morphs and takes more shape and gets clearer over time}''. This ambiguity, they said, could help ``\textit{open your mind to the possibilities}'' as ``\textit{a wonderful way to help people see and perceive the world differently.''}

{\textbf{Material malleability shapes creator control and mindset.} }
A common theme was the material malleability and how physical states influence a writer's mindset and processes. P4 noted that ``\textit{you can't really edit pottery once you've fired the clay. You can sort of when it's in the wet stage. But you're kind of stuck with whatever it is [when it's fired]. Whereas writing, it stays pliant, you know, to the last minute.}'' P1 and P2 described this as a helpful constraint \textit{``that forces me then to stop fiddling with it.''} P2 added that physical materials make it easier to start fresh, whereas text often invites endless revision and  attachment, which can hinder progress.

{\textbf{Material metaphors prompted LLM-mediated writing for co-creation and exploration, less so for precise execution.}} %
Participants (P2, P3) noted that text as material can make it difficult to express intent while maintaining control and predictability. Material metaphors were most concretely envisioned as a medium to invoke LLM co-creators with authorial agency ``\textit{suggesting fresh ideas}'' (P1), ``\textit{finding links between ideas}'' (P3), or ``\textit{smashing ideas together}'' (P2), rather than an input technique with high control and predictability. P1 described the unpredictability and imperfections of materials forces commitment, which could be liberating and generative. Evolving materials like growing plants and developing photographs were praised for their potential to inspire new perspectives (all), help get unstuck (P1, P2), or mitigate bias (P2). A common pattern emerged: When authors welcome input or are uncertain what they want, a material that actively contributes to creation in a non-neutral way---much like the bristles of a paintbrush producing a unique stroke---is positive, because such a process is forgiving and allows for inspiring surprises. During the narrative-shaping phase, %
participants preferred guiding materials organically from rough to refined forms. For fine edits, they imagined switching to more delicate tools, such as a fine brush. These transitions highlight where materiality comes in at different stages of the creation process.

\section{%
\textsc{Texterial}: Conceptual {Framework}}
\label{sec:framework}

Our formative study offers insight into the potential of text-as-material in LLM-mediated interaction, and the aspects of material thinking at play. In this section, we introduce a conceptual {framework} {that provides a structured way to approach the design of metaphor-based interaction with LLMs, grounded in interaction and materiality theories.}

\subsection{{Interacting with LLMs:} Cognition and Action}
\label{sec:cog-action}

\begin{figure}[htb]   
  \centering
  \includegraphics[width=0.8\linewidth]{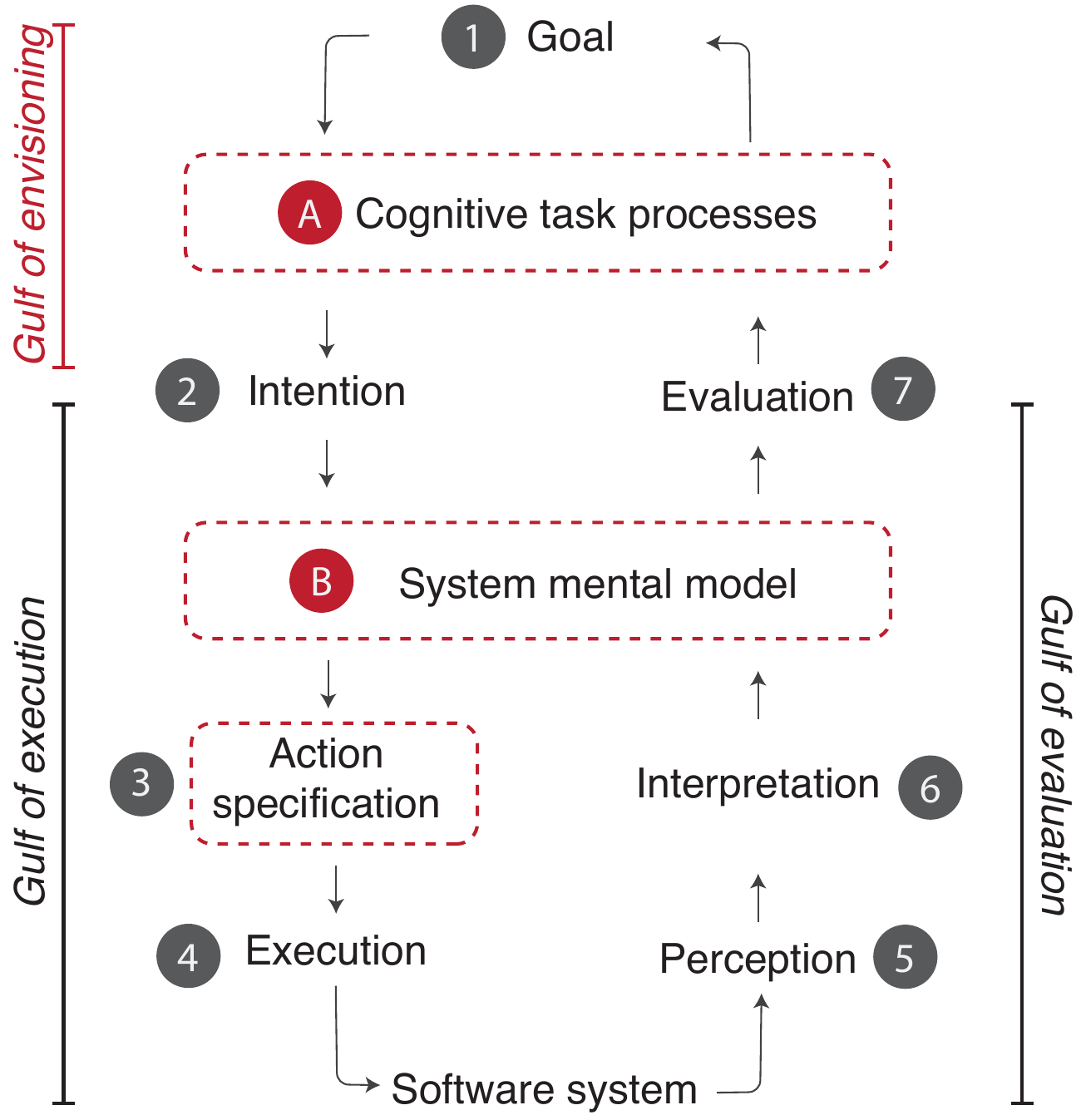} \caption{{Norman's stages-of-action~\cite{norman2013design} model expanded, with areas (in red) where interactions with LLMs are particularly difficult as identified by Subramonyam et al.~\cite{subramonyam2024bridging}.}}
  \Description{The figure presents an expanded version of Norman’s stages-of-action model, illustrating where interactions with large language models (LLMs) are especially difficult. The diagram is arranged vertically, beginning at the top with “Goal,” followed by “Intention,” “Action specification,” “Execution,” “Perception,” “Interpretation,” and finally “Evaluation,” which loops back toward the goal. These seven stages are connected with arrows to show the iterative process of acting on and assessing a system.

Two large vertical brackets frame the process. On the left side, the upper portion is labeled “Gulf of envisioning,” and the lower portion is labeled “Gulf of execution.” On the right side, a bracket spans the perceptual and interpretive steps and is labeled “Gulf of evaluation.” At the bottom of the diagram, the “Software system” is shown as the target of execution and the source of perceptual feedback.

The figure highlights two regions with dashed red outlines to indicate areas where interactions with LLMs are particularly challenging, as identified by Subramonyam et al. The first highlighted region, labeled “A: Cognitive task processes,” spans the space between a user’s goal and intention formation. The second region, labeled “B: System mental model,” covers the area where a user forms an intention, specifies an action, and attempts to interpret the system’s behavior. These regions represent steps that depend heavily on how users conceptualize both their own cognitive tasks and the system’s internal workings—areas that become more complex when dealing with probabilistic, opaque AI models.}
  \label{fig:stages-action}
  \vspace{-1em}
\end{figure}

Norman's stages-of-action model~\cite{norman2013design} describes the cognitive steps individuals go through when interacting with a system {(\autoref{fig:stages-action})}. %
The model is useful for identifying aspects where usability challenges may occur. Central to it are the \emph{gulf of execution} and \emph{gulf of evaluation} which refer to the disconnect between a user’s mental model on how to execute and evaluate a goal with a system, and how the system actually operates. Subramonyam et al.~\cite{subramonyam2024bridging} extend this model (see red additions in ~\autoref{fig:stages-action}) by {noting key factors that widen these gulfs when working with LLM-based systems. First, users often lack the understanding about how LLMs work (Fig~\ref{fig:stages-action}-\protect\stagered{B}); leading to uncertainties about what the LLM can do (gulf of execution) and what to expect from it (gulf of evaluation). Second, LLMs use unconstrained language for instruction, complicating action specification \protect\stage{3} (gulf of execution). These issues also affect the cognitive processes involved in forming intentions (Fig~\ref{fig:stages-action}-\protect\stage{1}$\to$
\protect\stagered{A}$\to$\protect\stage{2}).} Subramonyam et al. term this added difficulty the \emph{gulf of envisioning}. 

Bridging {these gulfs in interface design is typically achieved through well-designed \emph{perceived affordances} that enable users to act upon the system's \emph{functional affordances}~\cite{norman2013design, hartson2003cognitive}. Perceived affordances include \textit{cognitive} affordances --- design feature that helps users in knowing
something, \textit{physical} affordances -- design feature that helps users in doing a physical action in the interface, and \textit{sensory} affordances --- design feature that helps users sense something. Functional affordances refer to the functional capabilities of the system.}

Building {on these theories, our framework aims to support the design of interactions whose perceived affordances help users build a mental model of LLM functional capabilities and the means of instructing them. We first articulate what the LLM can help writers with (\S\ref{sec:llm-capabilities}), then reflect on the contrast between prompt-based and material-based interaction paradigms (\S\ref{sec:paradigms}) in surfacing and interfacing with these functionalities. Finally, we discuss how to match functional capabilities with perceived affordances of materials specifically (\S\ref{sec:match-material}).}

\subsection{{LLM Functional Capabilities}}
\label{sec:llm-capabilities}

\begin{figure}[htb]   
  \centering
  \includegraphics[width=0.8\linewidth]{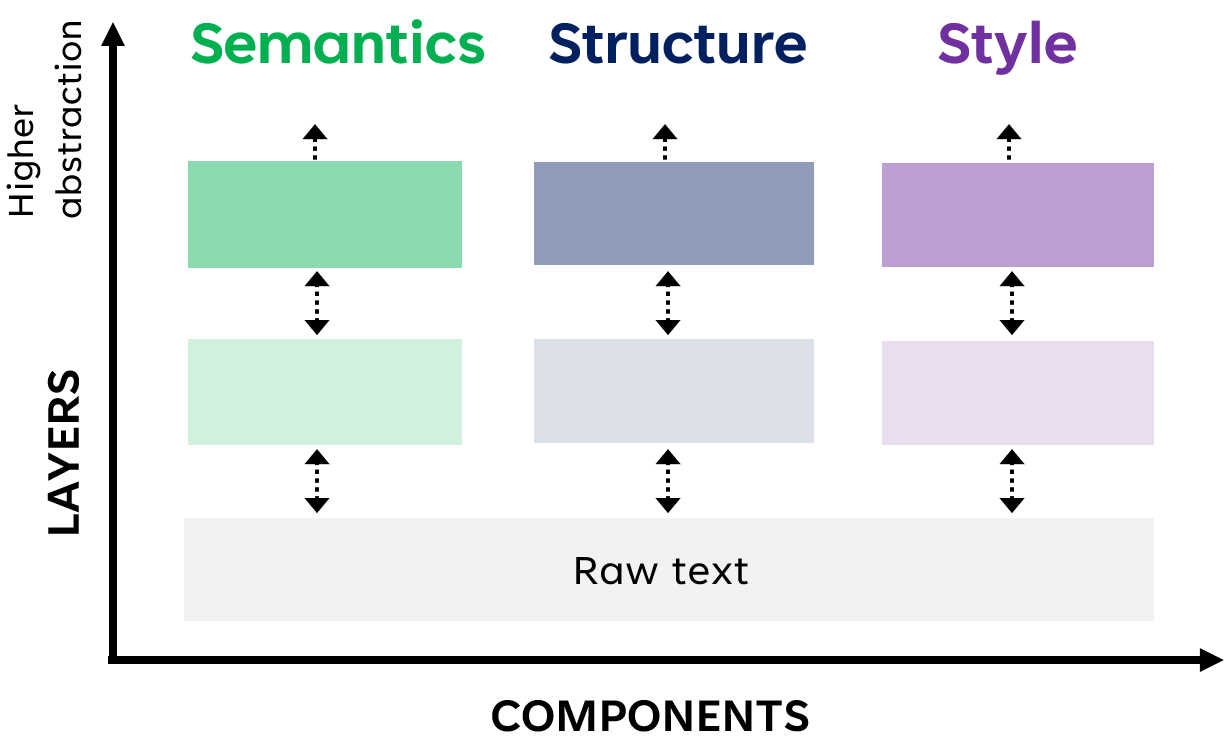} \caption{\textsc{Texterial} conceptual {framework}: {LLMs allow users to operate on the semantics, structure and style components of text at different levels of abstraction}.}

  \Description{The diagram shows three components—Semantics, Structure, and Style—arranged horizontally. Each component is represented as a stack of layered blocks above raw text, with arrows indicating movement between layers. The vertical axis represents levels of abstraction, from raw text at the bottom to higher abstraction at the top.}
  \label{fig:framework-layers}
  \vspace{1em}
\end{figure}

\begin{figure*}[t!]
  \includegraphics[width=\textwidth]{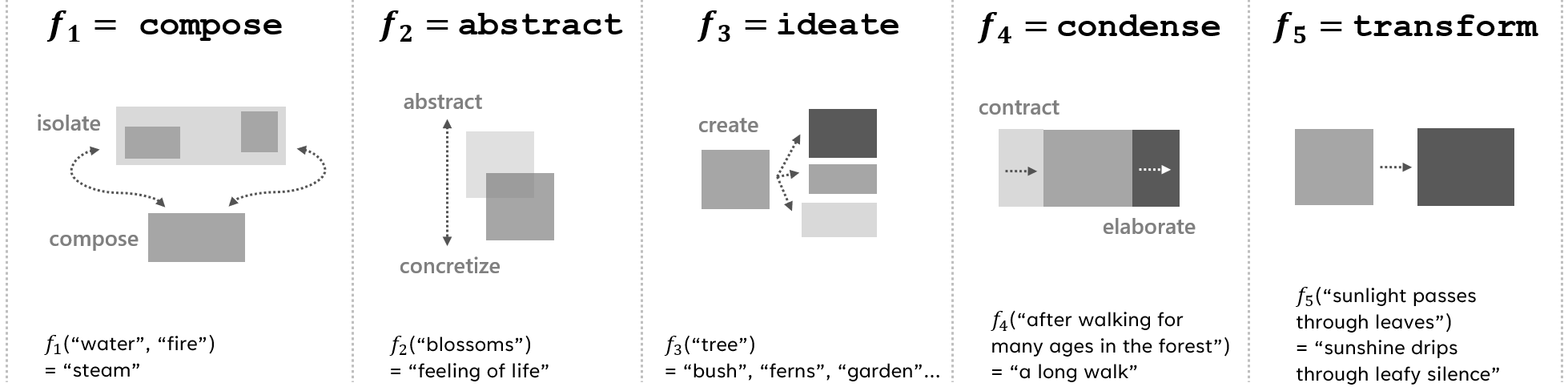}
  \caption{\textsc{Texterial} conceptual {framework}: Text operations users/models perform. {\textsc{composing} combines textual components together; \textsc{abstracting} moves to a meta-layer of any of the components; \textsc{ideating} takes a single seed and generates new concepts; \textsc{condensing} makes a component's content shorter; and finally \textsc{transforming} preserves the main qualities of a component, but expresses the content in a different way. Many of these operations have inverse functions: for example, instead of \textsc{composing}, users might \textsc{isolate} a single component into multiple distinct components.}}
  \Description{Framework of five text operations performed by users or models.

f₁ = compose: Combining or isolating elements. Example: f₁(“water”, “fire”) = “steam”.

f₂ = abstract: Moving between abstract and concrete. Example: f₂(“blossoms”) = “feeling of life”.

f₃ = ideate: Creating related alternatives. Example: f₃(“tree”) = “bush”, “ferns”, “garden”….

f₄ = condense: Contracting or elaborating text. Example: f₄(“after walking for many ages in the forest”) = “a long walk”.

f₅ = transform: Changing form or expression. Example: f₅(“sunlight passes through leaves”) = “sunshine drips through leafy silence”.

Each operation is illustrated with simple block diagrams showing the direction of composition, abstraction, creation, contraction, elaboration, or transformation.}
  \label{fig:framework}
\end{figure*}

Functional {capabilities of AI-mediated text manipulation can be organized into} the high-level components an LLM can assist writers with{, distilled from literature} (\autoref{fig:framework-layers}): the \textit{semantics} (i.e. inherent meaning) of text \cite{devlin_bert_2019, mikolov_efficient_2013}, the \textit{structure} including coherence and flow \cite{gros_attention_nodate}, and the \textit{style}---that is, how language signals characteristics of the writer via tone and literary style \cite{biber_variation_1988, shen_style_2017}. {LLMs also enable instructions at multiple levels of abstraction (y-axis).} In AI-mediated text authoring, this capability gives users significant flexibility: they can control the degree of authorial influence while leveraging the models' potential for emergent ideas~\cite{lu_whatelse_2025}. For instance, when an author prompts a model to ``write a story about an animal'', they keep most of the specifics of the semantics, structure, and style abstract and underspecified, with the intent that the model resolves uncertainties by choosing which animal, what happens to it, and how to tell the story. {More concretely, } we identify ~\textit{composing}, ~\textit{abstracting}, ~\textit{ideating}, ~\textit{condensing}, and ~\textit{transforming} as powerful generic "verbs" (operations) that leverage LLMs to perform expressive modifications of the text semantics, structure, or style at different levels of abstraction. See an illustration in ~\autoref{fig:framework} on the semantics component.

These {constituents form the \textbf{first building block of our conceptual framework: identifying and structuring LLM functional capabilities}. This allows that AI-mediated writing systems intentionally design for perceived affordances so that: 1) users identify these functionalities exist (cognitive affordance), supporting task processes involved in forming a meaningful intention (Fig.~\ref{fig:stages-action}-\stagered{A}$\to$\stage{2}) and building a mental model of the system (Fig.~\ref{fig:stages-action}-\stagered{B}); and 2) users can act on them (Fig.~\ref{fig:stages-action}-\stage{3},\stage{4}). In the next section, we contrast two interaction paradigms through this lens.}

\subsection{{Contrasting Interaction Paradigms: How We Interface with LLMs Functional Capabilities}}
\label{sec:paradigms}

\begin{table*}[t!]
\vspace*{10pt}
\Description{Analysis of how text-as-material has the potential to modify the interaction loop when working with LLMs for text manipulations, compared to a traditional chat interface. Three gulfs are discussed: envisioning, execution, and evaluation.
Gulf of Envisioning – Question: Do I know what it can do and how to communicate my intent?
Chat interface: Rated high gulf. A blank prompt can be intimidating; users may not know what to ask or how to begin, requiring prior intent or inspiration.
Text-as-material: Rated very low gulf. The interface invites play and experimentation; users can begin without a clear goal or assumptions, supporting emergent thinking and discovery.
Gulf of Execution – Question: Can I do what I want to do?
Chat interface: Rated moderate gulf. Users must translate intent into language, and success depends on prompt clarity. Requires mental effort to craft effective instructions.
Text as material: Rated low gulf. Users manipulate text directly with gestures; actions are intuitive and discoverable, giving users a stronger sense of control.
Gulf of Evaluation – Question: Did the system do what I expected?
Chat interface: Rated moderate gulf. Feedback is purely textual and may be verbose or ambiguous, making it harder to judge correctness or relevance.
Text as material: Rated low gulf. Provides immediate visual and tactile feedback, with changes visible and interpretable in real time.}
\caption{Analysis of how text-as-material has the potential to modify the interaction loop when working with LLMs for text manipulations, compared to a traditional chat interface.}
\label{tab:gulfs}
\small
\centering
\resizebox{\linewidth}{!}{%

\begin{tabular}{p{4.5cm}p{6cm}p{6cm}}
\hline
\textbf{Gulf} & \textbf{Chat Interface} & \textbf{Text-as-Material} \\
\hline
\hline
\textbf{Gulf of Envisioning~\cite{subramonyam2024bridging}} & \outlinedcirc{black} \outlinedcirc{black}
\outlinedcirc{black}
\filledcirc{magma4} 
\outlinedcirc{black}
High gulf  &   \filledcirc{magma4} 
\outlinedcirc{black}
\outlinedcirc{black} \outlinedcirc{black} \outlinedcirc{black} Very low gulf \\
\textit{Do I know what it can do and how to communicate my intent?} 
& Blank prompt can be intimidating; users may not know what to ask or how to begin. Requires prior intent or inspiration.
& The interface invites play and experimentation; users can begin without a clear goal or assumptions. Supports emergent thinking and discovery. \\
\hline
\textbf{Gulf of Execution~\cite{norman2013design}}
& \outlinedcirc{black} \outlinedcirc{black} \filledcirc{magma4} \outlinedcirc{black} \outlinedcirc{black} Moderate gulf
& \outlinedcirc{black}  \filledcirc{magma4} \outlinedcirc{black} \outlinedcirc{black} \outlinedcirc{black} Low gulf \\
\textit{Can I do what I want to do?}
&  Users must translate intent into language; success depends on prompt clarity. Requires mental effort to formulate effective instructions.
& Users manipulate text directly with gestures; actions are intuitive and discoverable. Users feel in control through embodied interaction. \\
\hline

\textbf{Gulf of Evaluation~\cite{norman2013design}} & \outlinedcirc{black} \outlinedcirc{black} \filledcirc{magma4} \outlinedcirc{black} \outlinedcirc{black} Moderate gulf  & \outlinedcirc{black}  \filledcirc{magma4} \outlinedcirc{black} \outlinedcirc{black} \outlinedcirc{black} Low gulf \\
\textit{Did the system do what I expected?} 
& Feedback is textual and may be verbose or ambiguous. Users may struggle to judge correctness or relevance.
& Immediate visual and tactile feedback; changes are visible and interpretable. Users can assess results in real time. \\
\hline
\end{tabular}}
\end{table*}

\begin{figure*}[t!]
    \vspace{12pt}
  \includegraphics[width=\textwidth]{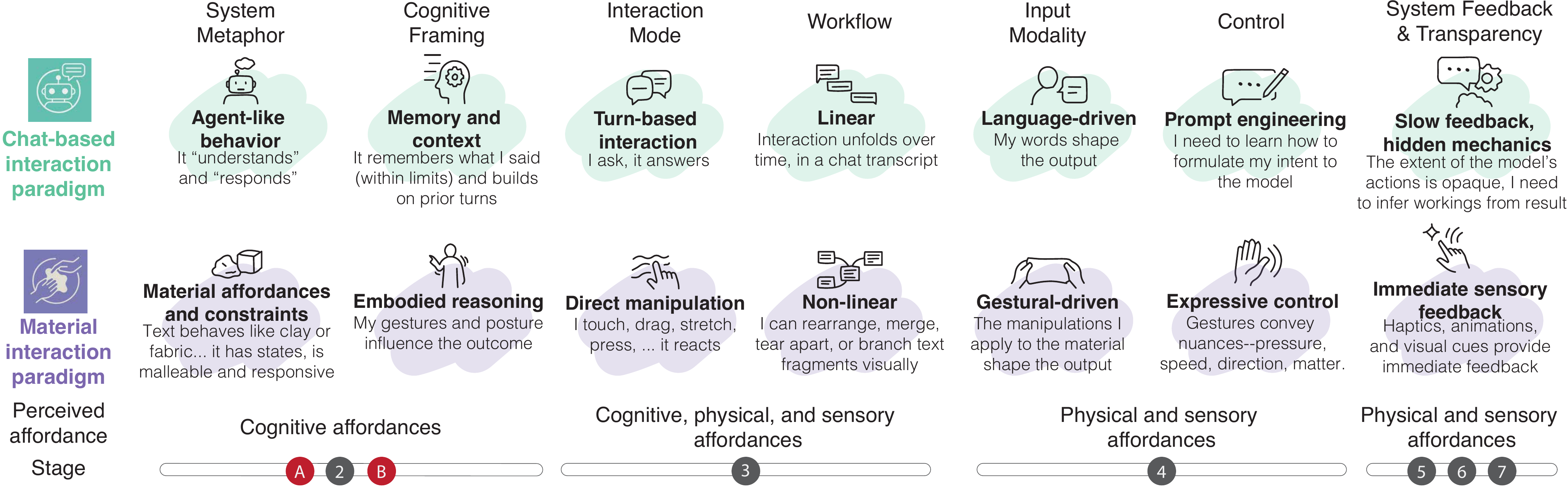}
  \caption{{Examples of some of the} key conceptual differences between AI-mediated text manipulation when using a prompt-based (green) vs. material-based (purple) interaction paradigm, and the perceived affordances~\cite{norman2013design, hartson2003cognitive} and the stages of the human action model~\cite{norman2013design} they primarily tie to.}
  \Description{The figure compares conceptual and mental shifts between prompt-based and material-based interactions with LLMs, mapping each to different stages of the human action model. The prompt-based paradigm (shown in green) frames the system through an agent-like metaphor: the model “understands,” responds, and uses memory and context from prior turns. Interaction is turn-based—I ask, it answers—and unfolds linearly as a transcript. Input is language-driven, requiring users to shape output through words, and control depends on prompt engineering, where users learn how to phrase their intent. Feedback is slow and often opaque, with hidden model mechanics that must be inferred.

The material-based paradigm (shown in purple) treats text as a manipulable material with affordances and constraints, behaving like clay or fabric. Cognitive framing shifts toward embodied reasoning, where gestures, posture, and physical movement influence outcomes. Interaction takes the form of direct manipulation—touching, dragging, stretching, or pressing—and the workflow is non-linear, allowing users to rearrange, merge, tear apart, or branch text fragments visually. Input is gestural, and control becomes expressive, with gestures conveying nuance such as pressure, direction, or speed. Feedback is immediate and sensory, supported by visual cues, animations, or haptics.

Along the bottom of the figure, a color-coded legend links each paradigm to stages of Norman’s action model, including intention, action specification, execution, perception, and evaluation, showing which stages each type of affordance (cognitive, physical, and sensory) primarily supports.}
  \label{fig:characteristics}
\end{figure*}

We examine the differences that exist between prompting LLMs, and treating text-as-material for AI-mediated text manipulation. We posit that contrast between the core concepts of different paradigms---a conversational partner that
responds to prompts with generated text; or a tactile, spatial, and gestural way to shape
text like a material---results in a fundamental shift in the users' mental model {because of their distinct perceived affordances}. Prompts invite to reason in terms of conversation, questions, and replies, whereas materials invites to reason in terms of shaping, layering, and
composing~\cite{richard_sennett_craftsman_nodate} (Section~\ref{sec:formative-study}). These differences directly impact the gulfs of envisioning, execution and evaluation. We provide a summary of this contrast in \autoref{tab:gulfs}, along with {illustrative examples} of the conceptual mechanisms influencing perceived affordances in \autoref{fig:characteristics}, {which we derive from the} literature in materiality~\cite{malafouris_at_2008, malafouris_how_2013,mccullough_abstracting_1998}, cognitive psychology~\cite{alibali_gesture_2024,dourish2001action} and HCI~\cite{hutchins_direct_1985}.

\subsection{Mapping LLM Functional Capabilities to Material Affordances}
\label{sec:match-material}

{Building on the above observations, we now introduce \textbf{the second component of our framework: mapping LLM functional capabilities to the perceived affordances of materials}.} This is where interaction design through metaphor comes into play. It involves defining a set of material-based inputs (e.g., gestures) and determining the corresponding LLM operations each input triggers.

{From stone and paint, to fabric, plants, or water, many materials can inspire interaction metaphors that embody text as a material. Choosing among these options and designing the input vocabulary requires consideration of:} 

\begin{itemize}
\item [a.] \textbf{The material’s affordances and constraints in the physical world} (with some flexibility). {The physical properties of materials influence the cognitive affordances users derive from them (Fig.~\ref{fig:stages-action}-\stagered{A}$\to$\stage{2}$\to$\stagered{B}). For instance, malleable materials such as fabric or wet paint suggest that if text were conceived as this material, it would too be malleable; whereas if text were conceived as ceramics or stone, it would instead fracture into parts.} The choice of the primary material for the interaction metaphor should enable users to rely on intuitive assumptions.

\item[b.] \textbf{The strength of the conceptual match between material verb and the intended LLM operation}. Here, metaphorical language can guide this process~\cite{gibbs2006imagining} (e.g., we ``smooth the tone,'' ``weave an argument,'' ``polish the phrasing,'' or ``tie loose ends''). {Materials that are already associated with such language can create stronger cognitive affordances (Fig.~\ref{fig:stages-action}-\stagered{A},\stage{2},\stagered{B},\stage{3}), e.g. we can smooth clay, and weave threads.}

 \item[c.] \textbf{The expressivity and usability of the input itself}. {The fidelity with with which physical and sensorial affordances are achieved in the digital metaphor, i.e. how the material is reified in the digital system, and whether it supports similar manipulations as in the real world, matters (Fig.~\ref{fig:stages-action}-\stage{3}$\to$\stage{4}).} Factors like hand approach, touch pressure, and execution speed may enhance expressivity by achieving higher fidelity with the physical counterpart, but could also reduce discoverability or ease-of-use.

 \item[d.] \textbf{The look-and-feel design matters}. Striking the right balance between being too literal and not enough, which requires experimentation~\cite{urbano2022skeuomorphism}. Animations, haptics, and visual cues can effectively convey perceived affordances (Fig.~\ref{fig:stages-action}-\stage{3},\stage{4}) and provide clear, immediate feedback (Fig.~\ref{fig:stages-action}-\stage{5},\stage{6},\stage{7}), and should be carefully designed.
 \end{itemize}

This \textsc{Texterial} {framework} aims to provide a structure and design vocabulary that are both descriptive---how text interfaces leverage model capabilities, and generative---the framework can be used to inspire novel text-based interactions with LLMs. It is important to note that we do not claim to have developed a comprehensive or definitive framework, but rather, we provide a concrete preliminary {foundation} for a metaphor-based approach to interacting with generative AI. For concrete implementation and instantiations of our framework, please see the Supplementary Material.

\section{\textsc{Texterial}: Technical Probes of Text as Material}
\label{sec:probes}

\begin{figure*}[t!]
    \centering
    \includegraphics[width=\linewidth]{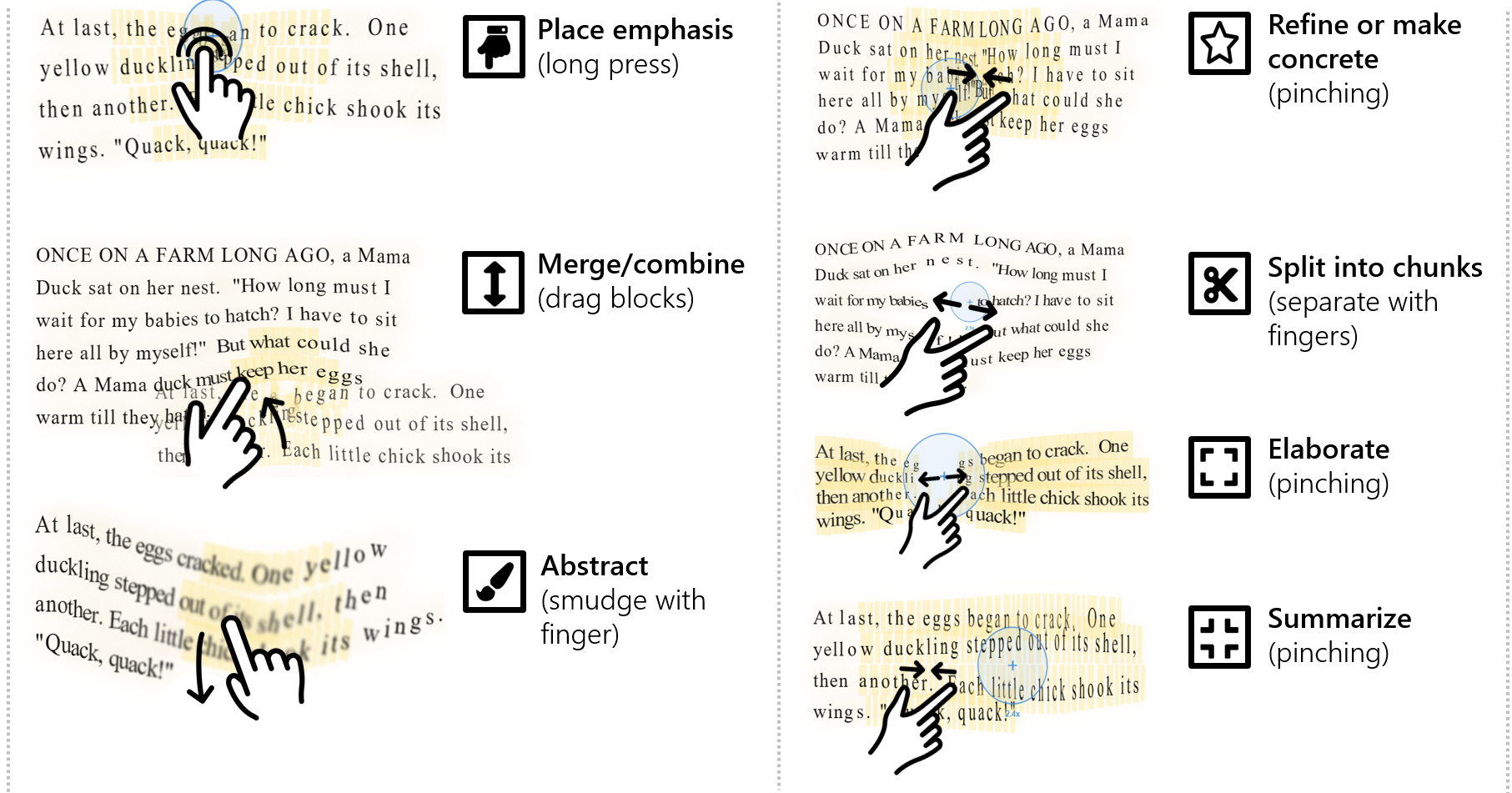}
    \caption{In our \textit{Text as Clay} interface, users tangibly sculpt existing drafts through expressive touch gestures. Localized, typographic animations emphasize the clay-like effect and highlight different edits the model made to the text. Users can interact with voice to avoid detracting from the tactile experience.}
    \Description{Shows the “Text as Clay” interface, where users sculpt drafts through touch gestures.
Gestures and their functions include:
Place Emphasis: Long press highlights text.
Merge/Combine: Drag blocks together.
Abstract: Smudge with a finger.
Refine: Pinch inward to tighten or clarify.
Split into Chunks: Separate text with fingers.
Elaborate: Stretch fingers outward to expand content.
Summarize: Squash fingers inward to compress text.
Typography animates locally to emphasize the clay-like quality of edits, while voice interaction is supported to complement tactile control.}
    \label{fig:clay}
\end{figure*}

\begin{figure*}[t!]
  \includegraphics[width=\linewidth]{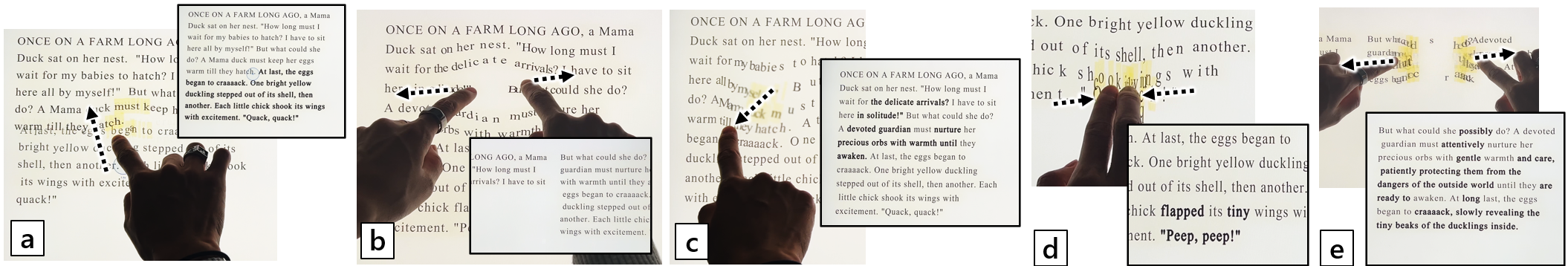}
  \caption{\textit{Text as Clay }Storyboard. The user is editing their story to make it more engaging: (a) They merge a paragraph into the existing block to seamlessly combine the two, and (b) separate another paragraph into two pieces by ripping. (c) They can use their finger to smudge text, making it more abstract, and (d) pinching makes text more concrete. Finally, (e) stretching out elaborates on the text.}
  \Description{Storyboard of the “Text as Clay” interface. (a) User drags text blocks to merge them into one. (b) User rips text apart to split it into two chunks. (c) User smudges text with a finger to abstract or blur meaning. (d) User pinches text inward to refine and tighten expression. (e) User stretches text outward to elaborate, making it longer and more detailed. The sequence demonstrates how tactile gestures—dragging, ripping, smudging, pinching, and stretching—sculpt text in a clay-like manner.}
  \label{fig:claystoryboard}
\end{figure*}

We demonstrate the use of the \textsc{Texterial} conceptual {framework} with two technical probes: (1) \textit{Text as Clay} and (2) \textit{Text as Plants}. 
We focus on these two prototypes since the metaphors resonated with participants of the formative study, and the material choices span contrasting affordances which lend themselves to different phases of the writing process. Our prototypes are used to demonstrate feasibility and to evaluate users' experiences with our approach in our focus group study (Section \ref{sec:study}). See the video figure for a demonstration of the probes, and {detailed figures for each interaction technique are included in the Supplementary Material.}

\subsection{Text as Clay}

\textit{Text as Clay} reimagines the text itself as a deformable, malleable material that can be pinched, stretched, pulled, merged, and smoothed through direct touch manipulations (Figures~\ref{fig:clay} and ~\ref{fig:claystoryboard}).

The text-as-clay techniques {support revision, with the metaphor reflecting an incremental, material-oriented process. Writers start }with rough isolated fragments, then iteratively refining through localized control. Similar to working with physical clay, text can be composed, pressed, pulled, smoothed, or torn apart. For visual representation and system feedback, we drew inspiration from typography and text-based animations, using only text and its responsiveness to convey clay matter and feel (see~\autoref{fig:clay}). Typographic bounce animations and bolding indicate inline the changes the model applied in response to material transformations.

Within our implemented \textit{Text as Clay} prototype, we provide the following touch-gesture operations supporting an expressive vocabulary of text sculpting through clay (Figure~\ref{fig:clay}) and the corresponding LLM operations (\autoref{fig:framework}):
{
\aptLtoX[graphic=no,type=html]{\begin{description}}{\setlength{\itemsep}{0.5\baselineskip}
\begin{description}[
  labelindent=0pt,
  leftmargin=0pt,
  labelwidth=2.8em,
  labelsep=-22.0001pt
]}  \item{} \begin{imageonly}\faHandPointDown\end{imageonly}\ \textbf{Squeeze (Place emphasis).}
  A firm press on a localized region emphasizes the span that was pressed. The
  touch area specifies the extent of text to emphasize (although we do not support
  it, pressure could indicate the extent of the emphasis). (\textsc{elaborate})

  \item{} \begin{imageonly}\faArrowsAltV\end{imageonly}\ \textbf{Merge / Move Blocks. }
  Dragging blocks combine text fragments. Drag direction sets the order for the
  combination, while overlap controls integration: a small overlap lightly
  combining fragments (\textsc{compose}) and a large overlap fully integrates them
  (\textsc{transform}). See \autoref{fig:gesture-example}.

  \item{} \begin{imageonly}\faPaintBrush\end{imageonly}\ \textbf{Smudge (Abstract). }
  A smearing gesture softens and abstracts language in the touched area, producing
  more impressionistic phrasing and looser semantics for higher-level reframing.
  (\textsc{abstract})

  \item{} \begin{imageonly}\faStar\end{imageonly}\ \textbf{Pinch (Refine / Make Concrete). }
  Pinching tightens passages, replacing vague wording with more precise phrasing
  and details. The stronger the pinch, the more specific and concrete the text
  becomes. (\textsc{concretize})

  \item{} \begin{imageonly}\faCut\end{imageonly}\ \textbf{Rip (Split into Chunks). }
  Tearing a fragment splits it into smaller pieces at the rip location, enabling
  finer-grained, localized edits and text re-structuring. (\textsc{isolate})

  \item{} \begin{imageonly}\faExpand\end{imageonly}\ \textbf{Stretch (Elaborate). }
  Pulling outwards with two fingers expands a span to elaborate ideas, introduce
  examples, or add connective tissue. Touch area indicates the scope of the
  fragment concerned. (\textsc{elaborate})

  \item{} \begin{imageonly}\faCompress\end{imageonly}\ \textbf{Squash (Summarize). }
  Pressing inwards with two fingers condenses a span into a tighter summary while
  preserving the main thrust of the passage. Touch area indicates the scope of the
  fragment concerned. (\textsc{condense})
  
  \item{} \begin{imageonly}\faMicrophone\end{imageonly}\ \textbf{Voice Input. } 
  New clay content blocks are added by speaking aloud, creating fresh clay-blocks that can then be integrated into the sculpted text.  
\end{description}}

\begin{figure*}[tb!]
  \centering
  \includegraphics[width=0.95\textwidth]{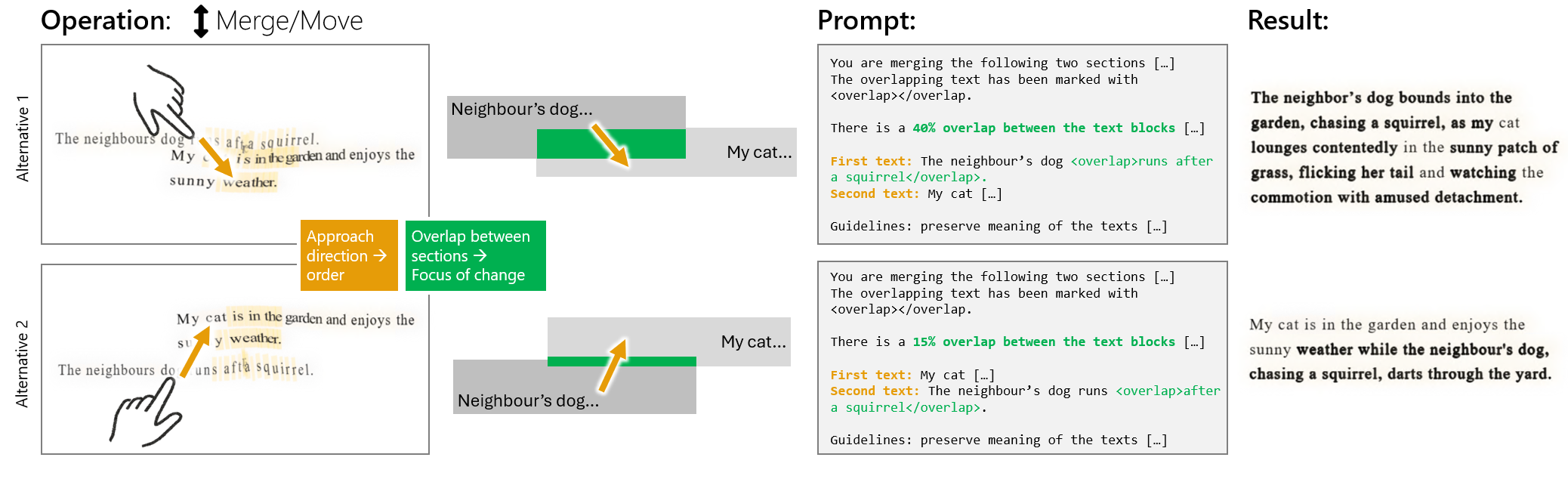} \caption{Example interaction flow when merging two "clay" text blocks ($f_1 = $ \textsc{compose}). The direction of approach and overlap between blocks is captured in the language model's prompt to guide generation.}
  \Description{Diagram showing how two text blocks (“My cat is in the garden…” and “The neighbour’s dog runs…”) can be merged using a clay-like interaction metaphor. The user drags one block toward the other, with the approach direction determining order and the overlap amount shaping the focus of change. Two alternatives are illustrated, each generating a different merged prompt. Prompts explicitly encode overlap percentage and order, producing final outputs where the two sentences are blended into a cohesive passage.}
  \label{fig:gesture-example}
\end{figure*}

To illustrate these techniques, imagine the following scenario (Figure~\ref{fig:claystoryboard}): 

\begin{quote} 
\textit{A writer is editing their new children's book story to make it more engaging. They merge the second paragraph into the existing block to seamlessly combine the two (\ref{fig:claystoryboard}a). They’re happy with the transition, but feel like the beginning of the story could be more poetic and abstract. To focus on just the beginning, they rip the paragraph into two pieces again (\ref{fig:claystoryboard}b), and use their finger to 'smudge' the text, which rewrites the text in a more abstract way (\ref{fig:claystoryboard}c). The user then feels that more detailed descriptions are needed in the beginning of the story, so they pinch it to add further touches (\ref{fig:claystoryboard}d). But now the model added too many details, so they smudge the paragraph a bit to make it less concrete. Finally, the end of the story needs expansion, and they expand this section by stretching it out (\ref{fig:claystoryboard}e).}
\end{quote}

This short scenario shows how the sculpting metaphor supports expressive and incremental iteration, positioning the writer as an artisan who continually reshapes their material until it takes on its desired form.

\subsection{Text as Plants}
\begin{figure*}[t!]
  \centering
  \aptLtoX[graphic,type=html]%
    {\xbox{aptbox}{\XMLaddatt{style}{max-width:70\%;display:block;margin:0 auto;}\includegraphics{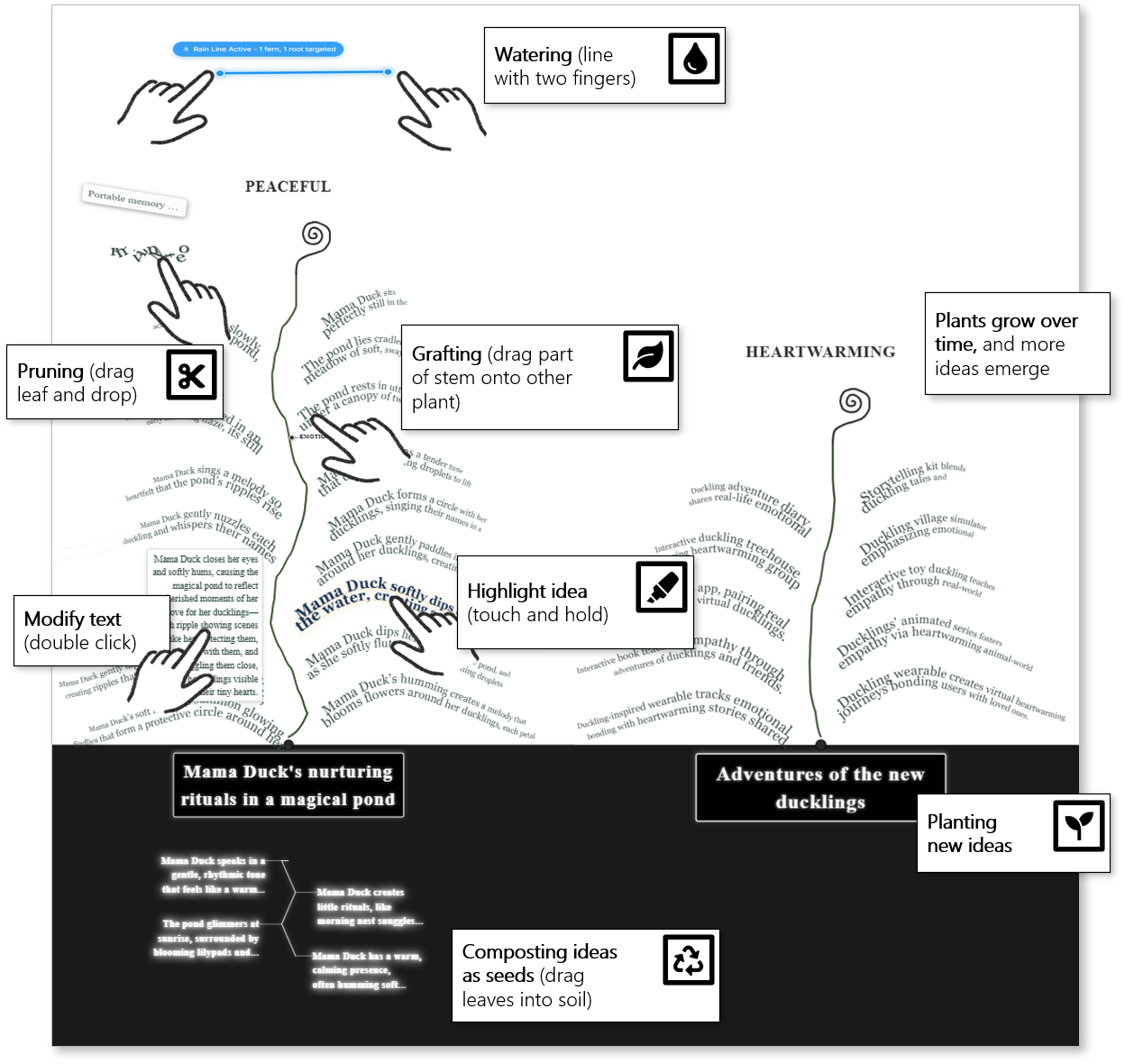}}}%
    {\includegraphics[width=0.9\linewidth]{figures/garden-overview.png}}
  \caption{Through our \textit{Text as Plants} interface, users continually shape their "garden" of ideas, watering, pruning, preserving, composting, and allowing ideas to grow serendipitously over time. Users can interact via voice and pen-touch gestures. Sound effects, such as plucking, and fluttering animations emphasize the materiality of the garden.}
  \Description{Shows the “Text as Garden” interface, where ideas are visualized as plants that users can nurture and reshape.

Users interact through gardening metaphors:

Pruning: Dragging a leaf and dropping it to remove.

Watering: Drawing a line with two fingers to help ideas grow.

Grafting: Dragging part of a stem onto another plant to connect ideas.

Composting: Dragging leaves into soil to turn old ideas into seeds.

Preserving: Touch and hold to keep an idea.

Modify text: Double-clicking to edit words.

Growth over time: Plants generate new ideas as they expand.

The garden metaphor is reinforced with animations (like leaves fluttering) and sound effects (like plucking), emphasizing text as a living, material medium.}
  \label{fig:garden}
\end{figure*}

\begin{figure*}[tb!]
  \includegraphics[width=\linewidth]{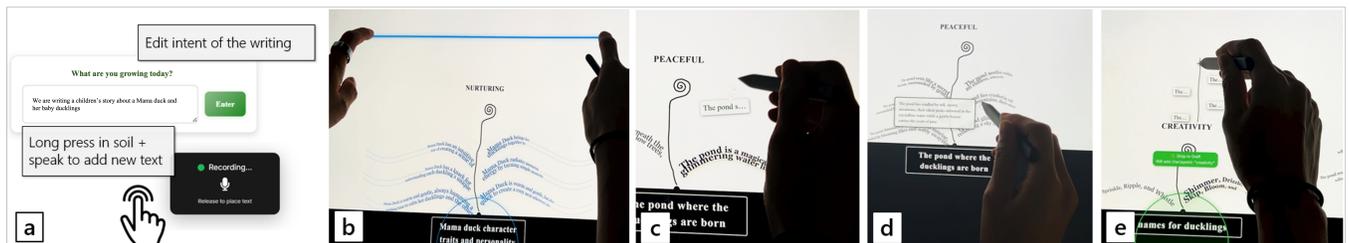}
  \caption{\textit{Text as Plants} Storyboard: (a) To write a children's story, a writer is planting new ferns of ideas for the setting, main character traits, and side characters. (b) They water plants to generate further ideas, and (c) pluck a few of the ideas they don’t like (d) They can read details of each generated idea, and directly edit the text. (e) Finally, they combine two ideation plant branches by grafting a part of one plant onto another.}
  \Description{Storyboard of the “Text as Plants” interface. (a) Users long-press on the soil and speak to add new text; intent of the writing can also be edited. (b) Watering gesture with two fingers helps ideas grow, here used to expand “Mama duck character traits and personality.” (c) A stem emerges labeled “PEACEFUL,” with text ideas branching around it. (d) Users rearrange and position text fragments to refine meaning. (e) New ideas are added, such as “names for ducklings,” reinforcing the garden metaphor for nurturing text.}
  \label{fig:gardenstoryboard}
\end{figure*}

\textit{Text as Plants} supports a slow, reflective engagement with text, where ideas can grow and evolve over time and be selectively nurtured or discarded (Figures \ref{fig:garden} and \ref{fig:gardenstoryboard}).
The plant metaphor specifically captures  temporality and emergence: Plants undergo organic and sometimes unpredictable growth, while still requiring care from the tender. With this metaphor, the writer tends to their ideas by watering, pruning, and shaping the landscape.

The \textit{Text as Plants} prototype conceptualizes ideation as a landscape where \textbf{concepts (as seeds) grow like plants, ideas (as leaves) can be combined through various modes, and any plant can be pruned or grafted}. 
Each plant encodes diverse ideas: The fern-like structures embody variation, and each frond represents a possible branch of the idea. Fiddleheads encode user-specified \textit{dimensions} of growth (e.g. the further the plant grows along the "creativity" dimension the more novel the ideas). 
We again drew inspiration from typography, calligrams, and typographic poetry art to design the look-and-feel such that text remains central. The interaction vocabulary is as follows (\autoref{fig:garden}):

{\aptLtoX[graphic=no,type=html]{\begin{description}}{\setlength{\itemsep}{0.5\baselineskip}
\begin{description}[
  labelindent=0pt,
  leftmargin=0pt,
  labelwidth=2.8em,
  labelsep=-22.0001pt
]}
  \item{} \begin{imageonly}\faSeedling\end{imageonly}\ \textbf{Plant. }
  Planting introduces new ideas. A long press in the soil combined with verbal
  input extracts a fresh seed and begins a new growth. (\textsc{create})

  \item{} \begin{imageonly}\faCut\end{imageonly}\ \textbf{Prune. }
  Pruning away a leaf removes the idea the user no longer wishes to pursue.
  (\textsc{isolate} and discard)

  \item{} \begin{imageonly}\faHighlighter\end{imageonly}\ \textbf{Highlight. }
  Tending to a leaf highlights the idea the user finds promising, making it pop
  out in the garden.

  \item{} \begin{imageonly}\faLeaf\end{imageonly}\ \textbf{Graft. }
  Plucking a leaf to graft it to another plant produces hybrids: subsequent
  generated ideas follow the combined history of the plants
  (\textsc{compose}, \textsc{create}).

  \item{} \begin{imageonly}\faTint\end{imageonly}\ \textbf{Water. }
  Watering accelerates growth: sketching a rain line nourishes the plants below,
  stimulating faster emergence of ideas (\textsc{create}).

  \item{} \begin{imageonly}\faRecycle\end{imageonly}\ \textbf{Compost. }
  Composting combines ideas from scratch: leaves or plant segments are returned
  to the soil, where their roots interweave and combine concepts automatically
  (\textsc{compose}, \textsc{create}).

\end{description}
}

In a short scenario (Figure~\ref{fig:gardenstoryboard}), the writer is ideating on a follow-up story.

\begin{quote}
\textit{The writer plants a new fern for each of setting, main character traits, and side characters concepts (\ref{fig:gardenstoryboard}a). They first explore the main character, generating ideas on this plant by watering it  (\ref{fig:gardenstoryboard}b). Meanwhile the “setting” plant has slowly grown some ideas about the location and setting of the story. They identify one idea that stands out, plucks the leaf and plant it for it to grow further as a new plant. The user then prunes a few of the new ideas they don't like, and will come back to it later (\ref{fig:gardenstoryboard}c). They see that the plant has grown (\ref{fig:gardenstoryboard}d), and wish to combine a compelling setting idea to make characters more memorable. To do this, they graft the setting ideas onto the side character plant (\ref{fig:gardenstoryboard}e).}
\end{quote}

Through \textit{Text as Plants}, users can let ideas evolve serendipitously over time, and revisit their garden to prune, compost, or graft---creating a slow and reflective ideation process with text as a material.

\subsection{Implementation}
The technical probes use TypeScript, React.js, and Vite. A Flask backend with Azure OpenAI integration runs the LLM-powered text generation and manipulation. Voice input is handled through the Web Speech API with continuous recognition, and both touch and pen interactions are supported. Undo/redo and interaction logging are done in React state snapshots.\footnote{All the prompts are provided in the supplemental material.}

\paragraph{\textit{Text as Clay}.}  Animations use character-level edits applied through CSS transforms and custom animation utilities. Collision detection uses bounding box calculations with configurable intensity thresholds, while pinch operations use distance-based influence calculations from touch center points. %
To apply LLM prompts to localized areas, we use tagging (<>) around touched text, and instruct the model to apply transformations primarily within the tagged areas, with more tags being added depending on the intensity of the operation being applied (e.g. $<$stretch$>$ with low intensity vs. $<<<<$stretch$>>>>$ with high intensity, as shown in Figure \ref{fig:gesture-example}).

\paragraph{Text as Plants.} 
We use SVG-based rendering with Fabric.js for curved text paths, with dynamic leaf positioning that varies based on the fern seed. Each fern maintains dimension checkpoints that track how the fiddlehead text (growth direction) has evolved over time, stored as an array of leaf pairs. Sound effects are implemented with the HTML5 Audio API, and animations are implemented using CSS and physics-based logic.

\section{Focus Group Study}
\label{sec:study}
To gain insight on the \textsc{Texterial} approach and interfaces, we ran four focus groups involving hands-on experience with our two technical probes. {Our goal was to gain a better understanding of our framework's ability to bridge gulfs of execution, envisioning and evaluation by probing 1) how material metaphors shape users' mental models for writing with generative AI, and 2) investigating affordances and limitations of material-inspired interactions with text.

We opted for a hybrid methodology, blending the focus group 
dynamics with elements of semi-structure interviewing. Recognizing and articulating mental models is often difficult for people as these are deeply internalized and most people may not even be aware of them~\cite{norman2013design}. By facilitating collective discussion, we enabled participants to build on each other’s perspectives, making it more likely for mental model shifts to surface and be articulated. The semi-structured elements allowed for neutral and systematic probing participants about perceived affordances (what can be done, how can it be done) and expected interactive response (what will happen) from the experimenter, while still supporting organic group interaction. }

\subsection{Procedure and Study Design}
Each session lasted approximately 60 minutes and was conducted in person. Focus groups comprised 2-3 participants with diverse experience in writing. One experimenter led the discussion while another provided guidance with the technical probes when needed. 
After getting participants' consent, and briefing of the study with an example (paint metaphor), we structured the focus group into three phases:
\begin{enumerate}
    \item \textbf{\textit{Text as Clay} discovery and experimentation} (20min). We prompted participants to guess what gestures they could perform if they considered text-as-clay, and what effect such gestures would have on text. After each guess, we encouraged them to try and experiment with the interface, asking about their expectations and  disagreements between participants. We also encouraged participants to provide feedback on the look-and-feel and interaction design.
    \item \textbf{\textit{Text as Plants} experimentation} (10min). Similarly, one experimenter provided a brief overview of possible interactions and participants experimented with text-as-plants, thinking out loud.  We encouraged participants to provide feedback on the look-and-feel and interaction design. 
    \item \textbf{Discussion on material metaphor value and mental model shift compared to chat interfaces} (20min). One experimenter prompted participants to reflect on \textit{text-as-material} overall, inviting them to verbalize  their mental model for this interaction paradigms, and how it differs from their current usage of LLMs in writing. The experimenter also probed the value and limitations of these types of interfaces and how such interaction techniques might fit into the writing process.   
\end{enumerate}

We collected audio and video data of participants' interactions with the prototypes.  Similar to our formative study, we used seed themes as deductive codes, and inductively captured emerging themes following an interpretivist approach to reflexive thematic analysis~\cite{braunThematicAnalysisPractical2022}. All participants received \$50 in compensation, and the study was approved by our ethics review board.

\subsection{Participants}
We recruited 10 participants through convenience sampling at a large technology company. Participants represented diverse roles and writing experience: researchers writing scientific papers, administrative staff preparing emails and documentation, product managers writing software specifications and presentations, and designers creating  briefs and communications for broad audiences. All were familiar with generative AI for ideation and/or writing.  Two participants (P9, P10) were follow-ups from our formative study (formerly P2, P3).

\subsection{Results}

We organize our findings around the four main themes as follows.

\subsubsection{Metaphors and Mental Models}\mbox{}\par

\textbf{Participants’ remarks suggest that different metaphors shape their mental model, acting as cognitive scaffold to guide thinking and approaches to writing.} They referred to ``sculpting'' (text-as-clay), ``growing'' (text-as-plants) text, and ``overlaying'' text (text-as-paint), using metaphor-based verbs that demonstrated material reasoning applied to text manipulations. 
P5 verbalized the conceptual mapping: \textit{``You can easily associate your tasks [...] with something that you're doing in clay sculpturing.''} %
All participants commented that different metaphors invited complementary writing phases---P6 explained that \textit{``the two metaphors here kind of suggest different writing styles [...] I'd use [text-as-plants] before going to the clay and very often I might even skip it depending on what I'm writing.''}.

\textbf{Metaphors inspired participants to rethink assumptions and engage in creative reasoning about text generation models.}
Metaphors inspired participants to come up with new operations and associations with generative models. For example, P2 explained how they would paint for ideation: \textit{``maybe I put some red, black, red, black, white right in the middle [...] and then my story is like, oh, what is happening here? You know, and it creates this rollercoaster of a novel.''}; and P3 expanded to other modalities:  \textit{``You could take another step to, say, flowers could be [...] like a link to a picture [text illustration]''}. The ease of direct manipulation also enticed them to experiment with building or refining their assumptions. For instance, %
P5 probed P4:\textit{``Maybe just place it in anywhere and see what it does?''} P1 explained how a metaphor encouraged them to think differently about how to approach text composition: ~\textit{``you put me in front of a typical [chat] interface and I don't know what to type. I don't know how to have it prompt me to think about other things, but you put me in front of something like this. It prompts me to think [...] encourages my brain to do different things.''}

\textbf{Metaphors may create different expectations for different people.}
We observed that a few participants had different expectations for the same gesture. For example, P2 expected the smudge gesture to rewrite the text in ~\textit{``simpler language, like jumping across readability level''}, while P3 expected it to use a different wording.  
P8 noted that interactions for the clay metaphor may be hard to discover, in part because of the 3D nature of the metaphor versus the 2D interaction on screen: \textit{``I’m not sure [mimicking shaping a bowl with clay with both hands] is doable with the clay [interface].''} In addition, P8 pointed out a disconnect between the physical metaphor and desired text editing behavior, \textit{``In clay [...] you're never gonna perfectly get the two pieces back. But with text [...] you can revert.''} These discrepancies between metaphor and operations may prove confusing at first. However, participants did not need to be prompted again on usage, suggesting that people may adapt to a specific gesture-operation mapping once discovered and understood.

\subsubsection{Direct Manipulation via Gestures}\mbox{}\par

\textbf{Gestures were seen as more direct and efficient than traditional input methods.}
All participants valued how gestures let them act on text without translating intentions into commands. P9 noted \textit{``[with material] I can see what I'm doing and I can tell again as I'm doing it that I'm getting the immediate feedback of like pushing this in is creating this dent or whatever.''} albeit noticing slight LLM-response lag. Gestures reduced the need for tedious selections as \textit{``It just really cuts down [...] highlighting, copying, pasting.''} (P4) while enabling \textit{``a lot more precision with just a couple gestures''}, referring to expressiveness of interaction \textit{``Depending on how much you pinch or how close you move to something [...] it changes that pattern of the highlighted text.''} %

\textbf{In all groups, participants contrasted familiar mobile gestures with metaphor-based gestures.}
In the first group, P2 wanted familiar mobile gestures such as double tap and pinch-to-zoom, but P1 argued \textit{``Are those the best gestures to use? Maybe there's something more intuitive… that fits our bodies and minds.''} They added, \textit{``Would this [clay metaphor] become your next nature rather than what you've been accustomed to with the phone?''} This exchange highlights a tension between a standard gesture vocabulary and one grounded in metaphorical framing. In the second group, P4 suggested \textit{``to keep it simple''} through multimodal touch-and-speech interaction: \textit{``I wanted to be able to say, oh take this area right here and then do this.''} P5 responded that metaphors allowed mapping different actions to a single gesture. %

\subsubsection{Spatial Layout and Visual Design}\mbox{}\par

\textbf{All participants valued the spatial layout and visual overview as a way to organize, explore, and reshape ideas.} Participants described how the interface supported early-stage ideation and structural experimentation. 
For clay, P7 explained they were doing something like it on paper already \textit{``One of the things I do before I start writing is write out a bunch of points on physical sheet of paper. I am kind of reshaping the text.''} P6 emphasized the ability to \textit{``Talk about specific parts of the text, separate ideas, and rearrange.''} P8 appreciated the overview a canvas provided: \textit{``I like that with interfaces like that you get some more global view maybe of everything, whereas you have to like scroll. You have your chat interactions and they are pretty long.''} 
For the text-as-plants, P4 repeatedly commented on organizing ideas: \textit{``It provides more of an organized timeline.''} P2, initially skeptical \textit{``I don't see why I would use this app','} later reflected, \textit{``The more I look at it, the more I like it [garden] [...] you can type in what you want [...] it grows [...] I don't necessarily want to read a whole bunch, like with ChatGPT''} appreciating how ideas were grouped and matured visually over time.

\textbf{The visual and artistic feel of the interface shaped expectations and interpretations.}
P4 remarked, \textit{``Everything here for me so far has been more of an artistic kind of fluid process than a defined methodical process.''}  %
This aesthetic framing led some to expect visual changes like font enlargement rather than semantic transformations.  
P7 wished for more abstracted look-and-feel design: \textit{``It doesn’t have to match the real world.''} whereas P3 argued for the opposite: \textit{``You could use real pieces of quick projection mapping.''} During discussions participants speculated that more literal interfaces (skeuomorphic cues) would aid discovery and support gesture recall, while P7 noted \textit{``I don't think I need realistic metaphors for things, I work very well with abstractions''}.

\subsubsection{Impacts on the Writing Process}\mbox{}\par

\textbf{Participants described how manipulating materials in space, not words, redefined their writing workflows, by enabling non-linear exploration, ambient ideation, fluid manipulation of ideas, and experimentation.}
Rather than composing text linearly, participants used spatial and metaphorical interactions to explore structure and meaning. P7 explained, \textit{``Writing actually does happen two-dimensionally for me''}, and P3 felt that \textit{``Having a wide open canvas [...] could be a really powerful thing.''}; it allows the user to \textit{``Talk about specific parts of the text, separate ideas [...] and rearrange,''} (P6) before combining them into a coherent whole. %
Several participants also described gestural interaction freed them from the constraints of language. P1 shared, \textit{``English is not my first language [...] Having something more intuitive, not through language could be very interesting.''} P9 noted how the interface invited manipulation and exploration: \textit{``I like the experimentation. I like that you can get in there and do it.''}

\textbf{Supporting slow, evolving engagement with ideas and introduced a new phase in the writing workflow.}
P5 felt that text-as-plants supported reflection, as it~\textit{``adds a visual aspect to how my thoughts are branching out into newer ideas or something. So yes, that's much more useful [than current use of LLMs].''} %
The plants material, as a living, non-neutral contributor of ideas in the background was described by P6 as being \textit{``a different experience and there's nothing like it.''} In particular, \textit{``The idea that something would slowly grow, over time. Somehow, maybe from, yeah, constantly listening to input or like whatever you're doing to tend to it, then that could be a completely different way to generate ideas or to interact with things.''} (P6)

\textbf{Metaphorical interfaces invite collaboration}. In the last focus group, participants discussed at length the lack of collaborative support in current use of LLMs today. When experimenting simultaneously with the probe, they discussed the value of supporting a collaborative canvas to shape ideas together. P10 stated \textit{``I feel like this where there's a collaborative thing where we can all kind of manipulate it. And in a way that isn't just like, go up and do your own prompting''} while P9 described her current process was do asynchronous brainstorming in their group, each using LLMs individually.

\textbf{Participants also highlighted aspects of the writing process amiss compared to chat interfaces} such as the ability to gain awareness of steps taken by LLMs: P10~\textit{``[chat interfaces] show you their reasoning [...] and you can kind of figure out if it's going off track because it's like now I'm checking these sources or these sources.''}. Another aspect discussed by P10 was the ability to use LLMs to critique their ideas and writing. P9, on the other hand, noted the ability to conduct with a prompt: \textit{``you can be very, very specific and surgical about swapping out small things.''}

\section{Discussion}

\subsection{Affordances of Materials are Context-Dependent and Multimodal}
As demonstrated in the formative study, participants expressed \textbf{diverse preferences for material metaphors depending on the task at hand}. For expert writers, writing itself may already feel like working with material, making additional metaphors redundant or even distracting. This subjectivity means that material metaphors may not generalize uniformly across all users, and future systems should carefully weigh the affordances and limitations of certain material properties, echoing similar cautions as in skeuomorphic design \cite{gross_skeu_nodate}. This was again corroborated in our focus group study, where participants found the metaphor of clay helpful in refining or sculpting text (supporting verbs in our framework, such as \textsc{Composing} and \textsc{Condensing}), while the metaphor of the garden lent itself well to growth and emergence (with focus on operations like \textsc{Ideation}).

A sentiment generally shared was that \textbf{gestures and visual feedback, as enabled by material thinking, allow users to interact with models in a more direct and hands-on manner, similar to the way craftsmen interact with physical materials} \cite{hutchins_direct_1985}.  Leveraging direct manipulation and spatial interaction, users could sculpt, layer, or arrange media elements in ways that reflect their intent without needing to translate it into language \cite{10.1145/985692.985731}. For example, participants appreciated gestural expression when they could not find the right words, whether due to complexity or language barriers. 
Furthermore, spatial layout and visual overviews provided a different lens to engage with writing. 

This suggests a future direction where the material metaphor could act as a powerful bridge for integrating non-textual media (e.g. visuals, audio, video) into creating  multi-media artifacts \cite{birchfield_embodiment_2008, chung_promptpaint_2023}.  Just as participants expressed the benefit of bypassing language to shape text, it may help when working with visuals and sounds that are often difficult to articulate with words alone. Concepts like mood, rhythm, or atmosphere can be challenging to describe. Metaphorical and gestural interactions could be beneficial tools for externalized thinking and creation \cite{malafouris_at_2008}.

\subsection{Design Tensions of Metaphors as Mental Models}

Throughout our conceptualization of text as material, and in both our studies, we find that different material metaphors afford different \textit{mental models} when approaching interaction with generative AI. A recurring tension in participants’ reflections was {beyond the context-dependent nature of different materials, interactions should strike an intentional } \textbf{balance between literal and abstract design choices} \cite{gessler1998skeuomorphs}. Some participants argued for more skeuomorphic cues to aid discovery, while others favored more abstract designs, making the use of familiar mobile gestures such as double-tap. This tension reflects a broader design challenge: how much fidelity to the metaphor is useful versus constraining, which is a long-standing question in HCI metaphors \cite{hutchins1987metaphors, jung_metaphors_2017}. Literal designs may facilitate discovery, understanding and recall, especially for users unfamiliar with abstract interaction paradigms. However, they risk introducing false equivalences between physical properties and digital operations. For instance, one participant noted that when merging pieces of clay and ripping them apart, you could not really get the same pieces again, whereas in digital interfaces it would be possible and probably desired to reverse interaction. These findings highlight that users do not only need metaphors to be discoverable, but also require a clear mapping between the components of text they wish to manipulate (semantics, structure, style) and the operations they expect to perform (compose, abstract, ideate, condense, transform) \cite{neale_chapter_1997}.

Participants also \textbf{encountered semantic mismatches between gestures and their expected outcomes}. The same gesture (e.g., smudge) was interpreted differently across users (P2 expected simplification, while P3 anticipated rewording) suggesting that metaphor-driven gestures may not always yield shared mental models. Our framework surfaces these ambiguities, showing how material meta\-phors both inspire interaction and risk overloading verbs, reinforcing the need for clarity in operation mapping. While learning through trial-and-error can partly address this, these discrepancies underscore the importance of designing for clear affordances and feedback that enable users to recover from initial incorrect expectations.

Ultimately, metaphorical interfaces must strike a balance between evocative framing and functional clarity. It is also interesting to consider \textit{progressive} metaphor fidelity, where initial interactions are grounded in familiar metaphors for ease of entry, but gradually abstracted to support expert use as core concepts are assimilated and gestures memorized.

\subsection{Shaping the Writing Process by Adopting a Craftsmanship Mindset}

Through observation and participants' comments in both studies, we found evidence that material metaphors of text \textbf{felt more creative and exploratory} than writing text with chat-based LLMs interfaces. Discussions from participants comparing metaphor-inspired interfaces to their prior use of AI indicate that this is because these experiences are closer to \textit{craftsmanship} rather than collaboration with a model \cite{richard_sennett_craftsman_nodate}. We gathered evidence that multiple factors contributed to this crafting mindset: directness of the metaphor-based gestures, feedback of AI-driven manipulations, the rich visual look and feel provided by text animations and deformations, and the use of a two-dimensional workspace to engage with the AI.
The crafting mindset we observed aligns with how writers could use our interfaces to operate across levels of abstraction in our conceptual {framework}---sometimes reshaping semantics (changing ideas themselves), other times reworking style (tone and phrasing) in the same way an artist might shift from composition to stylistic touches.

Participants mentioned \textbf{fluidity, speed and the ability to shape text through embodied interaction} as core components that made these experiences feel more creative. The probes invited different angles of experimentation, and we observed participants shaping, arranging, and revisiting ideas alone and together throughout the session. Their interactions highlighted how the interfaces could support ``non-committal ideation,'' where craftsmen quickly sketch and discard ideas, without becoming so attached to them. While some participants desired the ability to revert actions, others appreciated that materials can be \textit{predictably unpredictable} (e.g., a watercolor painter might not know exactly where the paint is going to spread, but know that it will spread more the more water is added to the paint). This loosening of control could lead to new discoveries during the writing process.

When contrasting this experience with their current use of generative models, it became clear that \textbf{participants approached chat-based interactions with a different mindset}. They described engaging with the models as they would with a collaborator, to get critique on their writing ideas, as well as issuing specific instructions for rewriting \cite{buschek_collage_2024}. Their discussions surfaced the additional cognitive load of handling discussions with the model and needing to clarify their intent through prompting, highlighting an important distinction between users focusing on a task vs. talking to a \textit{model} that is doing the task (diegetic vs. non-diegetic text) \cite{zamfirescu2023johnny}.
Implications for the design of future writing experiences include the need to accommodate and shift between different mindsets in the creation process. Understanding how interfaces can help writers transition between immersive, craft-oriented creation using these material metaphors, and more structured, dialogic critique with LLMs is critical to support holistic writing workflows.

\subsection{Limitations and Opportunities} 

{We acknowledge limitations in the small sample size of our formative study, and for our summative study, we did not evaluate specific techniques proposed against traditional editors or AI-assisted tools. Follow-up evaluations could study relative performance, efficiency, usability tradeoffs. Such controlled comparisons of each technique can offer valuable insights on their strengths and weaknesses within the broader landscape of writing tools.

In this paper, we chose to begin addressing more foundational questions about the value of \textit{families of techniques} enabled by material metaphors, approaching this through functional technology probes. In our summative study, we set out to understand how such metaphors can bridge the gulfs of execution, envisioning and evaluation when writing with generative AI. 
Our focus groups revealed that material metaphors appear to not only shape mental models and approaches to generative AI writing but also foster creative, collaborative, and non-linear workflows that markedly differ from current chat-based interfaces. We recognize the limitations of a short-term study to assess the extent of mental model shifts and grasp the extent of their impact on writing workflows. Future work should focus on longitudinal studies to examine whether and how these conceptual shifts we observed persist, evolve, and ultimately transform the way users approach generative AI writing tools. Such studies could make use of measures of creativity that analyze the artifacts users produced, beyond the user's self-perception of creative experience.}

Our technical probes---\textit{Text as Plants }and \textit{Text as Clay}---represent only two instantiations of a much larger design space. They serve as provocations rather than exhaustive solutions. There is potential in other metaphors (e.g., text as paint, photograph, or liquid) to augment the writing experience, and we could study how multiple metaphors might interoperate.  {This is particularly important given the abstract nature of these metaphors may be challenging for some users, especially in tasks requiring precise editing. Future iterations should explore adjusting the complexity of metaphors based on the task at hand, providing users with a smoother transition between different stages of the writing process or combining across metaphors tailored to the user.}
Ultimately, extending \textsc{Texterial} opens opportunities to examine richer, multi-modal metaphors (e.g., combining visual sketching, tactile feedback, or embodied interaction) that blur boundaries between physical and digital materiality. Such explorations could deepen our understanding of material engagement in HCI and broaden how generative models support the writing process.

\section{Conclusion}

\textsc{Texterial} investigates how treating text as material can change the experience of human–AI writing interactions. Our formative study shows that people readily reason with metaphors such as clay and plants to describe how they might shape, grow, or refine ideas. Building on these insights, we introduced a conceptual {framework} grounded in interaction theory, which articulates how material metaphors can reshape users’ mental models and help bridge the gulfs of envisioning, execution, and evaluation in LLM-mediated writing. The {framework} serves as a design vocabulary that is both descriptive (making existing capabilities legible) and generative (inspiring new interaction concepts).
As instantiations of our conceptual {framework}, we designed and implemented two probes demonstrating how distinct materials cue different modes of writing. Participants viewed \textit{Text as Clay} as supporting localized, iterative refinement with a sense of  embodied control, whereas \textit{Text as Plants} supported slower ideation, emergence, and selective nurturing.

In sum, \textsc{Texterial} expands the design space of AI-mediated writing by making model capabilities graspable through material metaphors---inviting users to craft, experiment, and play with text as if it were a physical material.
We hope this perspective inspires future systems that augment the expressive possibilities of working with language, while preserving the deeply human and material nature of text.

\begin{acks}
We thank the participants of our user study for taking the time to provide feedback on this work, and the reviewers of this submission for concrete suggestions to improve this paper. We also thank Chu Li, Rima Cao, Ryan Yen, Kwon Ko, Jude Rayan, and Judith Amores for their feedback throughout this project.
\end{acks}

\bibliographystyle{ACM-Reference-Format}
\bibliography{bibliography}

@inproceedings{lockton2019new,
	title        = {New Metaphors: A Workshop Method for Generating Ideas and Reframing Problems in Design and Beyond},
	author       = {Lockton, Dan and Singh, Devika and Sabnis, Saloni and Chou, Michelle and Foley, Sarah and Pantoja, Alejandro},
	year         = 2019,
	booktitle    = {Proceedings of the 2019 Conference on Creativity and Cognition},
	location     = {San Diego, CA, USA},
	publisher    = {Association for Computing Machinery},
	address      = {New York, NY, USA},
	series       = {C\&C '19},
	pages        = {319--332},
	doi          = {10.1145/3325480.3326570},
	isbn         = 9781450359177,
	url          = {https://doi.org/10.1145/3325480.3326570},
	numpages     = 14,
}

@book{tomitsch2018design,
	title        = {Design. think. make. break. repeat. A handbook of methods},
	author       = {Tomitsch, Martin and Wrigley, Cara and Borthwick, Madeleine and Ahmadpour, Naseem and Frawley, Jessica and Kocaballi, A Baki and N{\'u}nez-Pacheco, Claudia and Straker, Karla},
	year         = 2018,
	publisher    = {BIS publishers},
	address      = {Amsterdam, The Netherlands},
}

@article{gibbs2006imagining,
	title        = {Imagining metaphorical actions: Embodied simulations make the impossible plausible},
	author       = {Gibbs Jr, Raymond W and Gould, Jessica J and Andric, Michael},
	year         = 2006,
	journal      = {Imagination, Cognition and Personality},
	publisher    = {SAGE Publications Sage CA: Los Angeles, CA},
	volume       = 25,
	number       = 3,
	pages        = {221--238},
}

@inproceedings{zamfirescu2023johnny,
	title        = {Why Johnny Can’t Prompt: How Non-AI Experts Try (and Fail) to Design LLM Prompts},
	author       = {Zamfirescu-Pereira, J.D. and Wong, Richmond Y. and Hartmann, Bjoern and Yang, Qian},
	year         = 2023,
	booktitle    = {Proceedings of the 2023 CHI Conference on Human Factors in Computing Systems},
	location     = {Hamburg, Germany},
	publisher    = {Association for Computing Machinery},
	address      = {New York, NY, USA},
	series       = {CHI '23},
	doi          = {10.1145/3544548.3581388},
	isbn         = 9781450394215,
	url          = {https://doi.org/10.1145/3544548.3581388},
	articleno    = 437,
	numpages     = 21,
}

@article{birchfield_embodiment_2008,
	title        = {Embodiment, {Multimodality}, and {Composition}: {Convergent} {Themes} across {HCI} and {Education} for {Mixed}-{Reality} {Learning} {Environments}},
	shorttitle   = {Embodiment, {Multimodality}, and {Composition}},
	author       = {Birchfield, David and Thornburg, Harvey and Megowan-Romanowicz, M. Colleen and Hatton, Sarah and Mechtley, Brandon and Dolgov, Igor and Burleson, Winslow},
	year         = 2008,
	journal      = {Advances in Human-Computer Interaction},
	volume       = 2008,
	number       = 1,
	pages        = 874563,
	doi          = {10.1155/2008/874563},
	issn         = {1687-5907},
	url          = {https://onlinelibrary.wiley.com/doi/abs/10.1155/2008/874563},
	urldate      = {2025-09-11},
	copyright    = {Copyright © 2008 David Birchfield et al.},
	note         = {\_eprint: https://onlinelibrary.wiley.com/doi/pdf/10.1155/2008/874563},
	language     = {en},
}

@inproceedings{10.1145/985692.985731,
	title        = {I/O brush: drawing with everyday objects as ink},
	author       = {Ryokai, Kimiko and Marti, Stefan and Ishii, Hiroshi},
	year         = 2004,
	booktitle    = {Proceedings of the SIGCHI Conference on Human Factors in Computing Systems},
	location     = {Vienna, Austria},
	publisher    = {Association for Computing Machinery},
	address      = {New York, NY, USA},
	series       = {CHI '04},
	pages        = {303–310},
	doi          = {10.1145/985692.985731},
	isbn         = 1581137028,
	url          = {https://doi.org/10.1145/985692.985731},
	numpages     = 8,
}

@incollection{alibali_gesture_2024,
	title        = {Gesture in reasoning: {An} embodied perspective},
	shorttitle   = {Gesture in reasoning},
	author       = {Alibali, Martha W. and Boncoddo, Rebecca and Hostetter, Autumn B.},
	year         = 2024,
	booktitle    = {The {Routledge} {Handbook} of {Embodied} {Cognition}},
	publisher    = {Routledge},
	note         = {Num Pages: 13},
	edition      = 2,
address = {Abingdon, Oxon, UK and New York, NY, USA},
}

@inproceedings{subramonyam2024bridging,
	title        = {Bridging the Gulf of Envisioning: Cognitive Challenges in Prompt Based Interactions with LLMs},
	author       = {Subramonyam, Hari and Pea, Roy and Pondoc, Christopher and Agrawala, Maneesh and Seifert, Colleen},
	year         = 2024,
	booktitle    = {Proceedings of the 2024 CHI Conference on Human Factors in Computing Systems},
	location     = {Honolulu, HI, USA},
	publisher    = {Association for Computing Machinery},
	address      = {New York, NY, USA},
	series       = {CHI '24},
	doi          = {10.1145/3613904.3642754},
	isbn         = 9798400703300,
	url          = {https://doi.org/10.1145/3613904.3642754},
	articleno    = 1039,
	numpages     = 19,
}

@inproceedings{devlin_bert_2019,
	title        = {{BERT}: Pre-training of Deep Bidirectional Transformers for Language Understanding},
	author       = {Devlin, Jacob  and Chang, Ming-Wei  and Lee, Kenton  and Toutanova, Kristina},
	year         = 2019,
	month        = jun,
	booktitle    = {Proceedings of the 2019 Conference of the North {A}merican Chapter of the Association for Computational Linguistics: Human Language Technologies, Volume 1 (Long and Short Papers)},
	publisher    = {Association for Computational Linguistics},
	address      = {Minneapolis, Minnesota},
	pages        = {4171--4186},
	doi          = {10.18653/v1/N19-1423},
	url          = {https://aclanthology.org/N19-1423/},
	editor       = {Burstein, Jill  and Doran, Christy  and Solorio, Thamar},
}

@article{kimball1982designing,
	title        = {Designing the Star user interface},
	author       = {Kimball, Ralph and Harslem, B Verplank E},
	year         = 1982,
	journal      = {Byte},
	volume       = 7,
	number       = 1982,
	pages        = {242--282},
}

@book{dourish2001action,
	title        = {Where the action is: the foundations of embodied interaction},
	author       = {Dourish, Paul},
	year         = 2001,
	publisher    = {MIT press},
	address      = {Cambridge, MA},
}

@book{malafouris_how_2013,
	title        = {How {Things} {Shape} the {Mind}: {A} {Theory} of {Material} {Engagement}},
	shorttitle   = {How {Things} {Shape} the {Mind}},
	author       = {Malafouris, Lambros},
	year         = 2013,
	month        = jul,
	publisher    = {The MIT Press},
	address      = {Cambridge, MA},
	doi          = {10.7551/mitpress/9476.001.0001},
	isbn         = {978-0-262-31566-1},
	url          = {https://direct.mit.edu/books/monograph/2994/How-Things-Shape-the-MindA-Theory-of-Material},
	urldate      = {2025-08-13},
	language     = {en},
}

@misc{richard_sennett_craftsman_nodate,
	title        = {The {Craftsman}},
	author       = {{Richard Sennett}},
	year         = 2008,
	journal      = {Yale University Press},
	url          = {https://yalebooks.yale.edu/book/9780300151190/the-craftsman/},
	urldate      = {2025-08-18},
	language     = {en-US},
}

@book{mccullough_abstracting_1998,
	title        = {Abstracting {Craft}: {The} {Practiced} {Digital} {Hand}},
	shorttitle   = {Abstracting {Craft}},
	author       = {McCullough, Malcolm},
	year         = 1998,
	month        = jul,
	publisher    = {MIT Press},
	address      = {Cambridge, MA, USA},
	isbn         = {978-0-262-63189-1},
	language     = {en},
}

@incollection{malafouris_at_2008,
	title        = {At the {Potter}’s {Wheel}: {An} {Argument} for {Material} {Agency}},
	shorttitle   = {At the {Potter}’s {Wheel}},
	author       = {Malafouris, Lambros},
	year         = 2008,
	booktitle    = {Material {Agency}: {Towards} a {Non}-{Anthropocentric} {Approach}},
	publisher    = {Springer US},
	address      = {Boston, MA},
	pages        = {19--36},
	doi          = {10.1007/978-0-387-74711-8_2},
	isbn         = {978-0-387-74711-8},
	url          = {https://doi.org/10.1007/978-0-387-74711-8_2},
	urldate      = {2025-08-18},
	language     = {en},
	editor       = {Knappett, Carl and Malafouris, Lambros},
}

@inproceedings{shen_style_2017,
	title        = {Style transfer from non-parallel text by cross-alignment},
	author       = {Shen, Tianxiao and Lei, Tao and Barzilay, Regina and Jaakkola, Tommi},
	year         = 2017,
	booktitle    = {Proceedings of the 31st International Conference on Neural Information Processing Systems},
	location     = {Long Beach, California, USA},
	publisher    = {Curran Associates Inc.},
	address      = {Red Hook, NY, USA},
	series       = {NIPS'17},
	pages        = {6833–6844},
	isbn         = 9781510860964,
	numpages     = 12,
}

@misc{mikolov_efficient_2013,
	title        = {Efficient {Estimation} of {Word} {Representations} in {Vector} {Space}},
	author       = {Mikolov, Tomas and Chen, Kai and Corrado, Greg and Dean, Jeffrey},
	year         = 2013,
	month        = sep,
	publisher    = {arXiv},
	doi          = {10.48550/arXiv.1301.3781},
	url          = {http://arxiv.org/abs/1301.3781},
	urldate      = {2025-08-18},
	note         = {arXiv:1301.3781 [cs]},
}

@article{gros_attention_nodate,
	title        = {Attention, intentions, and the structure of discourse},
	author       = {Grosz, Barbara J. and Sidner, Candace L.},
	year         = 1986,
	month        = jul,
	journal      = {Comput. Linguist.},
	publisher    = {MIT Press},
	address      = {Cambridge, MA, USA},
	volume       = 12,
	number       = 3,
	pages        = {175–204},
	issn         = {0891-2017},
	issue_date   = {July-September 1986},
	numpages     = 30,
}

@book{biber_variation_1988,
	title        = {Variation across {Speech} and {Writing}},
	author       = {Biber, Douglas},
	year         = 1988,
	publisher    = {Cambridge University Press},
	address      = {Cambridge},
	doi          = {10.1017/CBO9780511621024},
	isbn         = {978-0-521-42556-8},
	url          = {https://www.cambridge.org/core/books/variation-across-speech-and-writing/A546CF5ED8F8E62F1432CB2F369CF356},
	urldate      = {2025-08-18},
}

@inproceedings{suh_luminate_2024,
	title        = {Luminate: {Structured} {Generation} and {Exploration} of {Design} {Space} with {Large} {Language} {Models} for {Human}-{AI} {Co}-{Creation}},
	shorttitle   = {Luminate},
	author       = {Suh, Sangho and Chen, Meng and Min, Bryan and Li, Toby Jia-Jun},
	year         = 2024,
	month        = may,
	booktitle    = {Proceedings of the {CHI} {Conference} on {Human} {Factors} in {Computing} {Systems}},
	publisher    = {Association for Computing Machinery},
	address      = {New York, NY, USA},
	series       = {{CHI} '24},
	pages        = {1--26},
	doi          = {10.1145/3613904.3642400},
	isbn         = {979-8-4007-0330-0},
	url          = {https://dl.acm.org/doi/10.1145/3613904.3642400},
	urldate      = {2024-06-04},
}

@misc{siddiqui_scriptshift_2025,
	title        = {Script\&{Shift}: {A} {Layered} {Interface} {Paradigm} for {Integrating} {Content} {Development} and {Rhetorical} {Strategy} with {LLM} {Writing} {Assistants}},
	shorttitle   = {Script\&{Shift}},
	author       = {Siddiqui, Momin and Pea, Roy and Subramonyam, Hari},
	year         = 2025,
	month        = feb,
	publisher    = {arXiv},
	doi          = {10.48550/arXiv.2502.10638},
	url          = {http://arxiv.org/abs/2502.10638},
	urldate      = {2025-05-19},
	note         = {arXiv:2502.10638 [cs]},
	language     = {en},
}

@misc{lu_whatelse_2025,
	title        = {{WhatELSE}: {Shaping} {Narrative} {Spaces} at {Configurable} {Level} of {Abstraction} for {AI}-bridged {Interactive} {Storytelling}},
	shorttitle   = {{WhatELSE}},
	author       = {Lu, Zhuoran and Zhou, Qian and Wang, Yi},
	year         = 2025,
	month        = feb,
	publisher    = {arXiv},
	doi          = {10.48550/arXiv.2502.18641},
	url          = {http://arxiv.org/abs/2502.18641},
	urldate      = {2025-05-19},
	note         = {arXiv:2502.18641 [cs]},
	language     = {en},
}

@inproceedings{chung_talebrush_2022,
	title        = {{TaleBrush}: {Sketching} {Stories} with {Generative} {Pretrained} {Language} {Models}},
	shorttitle   = {{TaleBrush}},
	author       = {Chung, John Joon Young and Kim, Wooseok and Yoo, Kang Min and Lee, Hwaran and Adar, Eytan and Chang, Minsuk},
	year         = 2022,
	month        = apr,
	booktitle    = {Proceedings of the 2022 {CHI} {Conference} on {Human} {Factors} in {Computing} {Systems}},
	publisher    = {Association for Computing Machinery},
	address      = {New York, NY, USA},
	series       = {{CHI} '22},
	pages        = {1--19},
	doi          = {10.1145/3491102.3501819},
	isbn         = {978-1-4503-9157-3},
	url          = {https://dl.acm.org/doi/10.1145/3491102.3501819},
	urldate      = {2025-05-18},
}

@inproceedings{masson_textoshop_2025,
	title        = {Textoshop: {Interactions} {Inspired} by {Drawing} {Software} to {Facilitate} {Text} {Editing}},
	shorttitle   = {Textoshop},
	author       = {Masson, Damien and Kim, Young-Ho and Chevalier, Fanny},
	year         = 2025,
	month        = apr,
	booktitle    = {Proceedings of the 2025 {CHI} {Conference} on {Human} {Factors} in {Computing} {Systems}},
	publisher    = {Association for Computing Machinery},
	address      = {New York, NY, USA},
	series       = {{CHI} '25},
	pages        = {1--14},
	doi          = {10.1145/3706598.3713862},
	isbn         = {979-8-4007-1394-1},
	url          = {https://dl.acm.org/doi/10.1145/3706598.3713862},
	urldate      = {2025-05-18},
}

@inproceedings{chung_toyteller_2025,
	title        = {Toyteller: {AI}-powered {Visual} {Storytelling} {Through} {Toy}-{Playing} with {Character} {Symbols}},
	shorttitle   = {Toyteller},
	author       = {Chung, John Joon Young and Roemmele, Melissa and Kreminski, Max},
	year         = 2025,
	month        = apr,
	booktitle    = {Proceedings of the 2025 {CHI} {Conference} on {Human} {Factors} in {Computing} {Systems}},
	publisher    = {Association for Computing Machinery},
	address      = {New York, NY, USA},
	series       = {{CHI} '25},
	pages        = {1--23},
	doi          = {10.1145/3706598.3713435},
	isbn         = {979-8-4007-1394-1},
	url          = {https://dl.acm.org/doi/10.1145/3706598.3713435},
	urldate      = {2025-05-18},
}

@inproceedings{buschek_collage_2024,
	title        = {Collage is the {New} {Writing}: {Exploring} the {Fragmentation} of {Text} and {User} {Interfaces} in {AI} {Tools}},
	shorttitle   = {Collage is the {New} {Writing}},
	author       = {Buschek, Daniel},
	year         = 2024,
	month        = jul,
	booktitle    = {Proceedings of the 2024 {ACM} {Designing} {Interactive} {Systems} {Conference}},
	publisher    = {Association for Computing Machinery},
	address      = {New York, NY, USA},
	series       = {{DIS} '24},
	pages        = {2719--2737},
	doi          = {10.1145/3643834.3660681},
	isbn         = {979-8-4007-0583-0},
	url          = {https://dl.acm.org/doi/10.1145/3643834.3660681},
	urldate      = {2025-05-20},
}

@inproceedings{chung_promptpaint_2023,
	title        = {PromptPaint: Steering Text-to-Image Generation Through Paint Medium-like Interactions},
	author       = {Chung, John Joon Young and Adar, Eytan},
	year         = 2023,
	booktitle    = {Proceedings of the 36th Annual ACM Symposium on User Interface Software and Technology},
	location     = {San Francisco, CA, USA},
	publisher    = {Association for Computing Machinery},
	address      = {New York, NY, USA},
	series       = {UIST '23},
	doi          = {10.1145/3586183.3606777},
	isbn         = 9798400701320,
	url          = {https://doi.org/10.1145/3586183.3606777},
	articleno    = 6,
	numpages     = 17,
}

@inproceedings{han_textlets_2020,
	title        = {Textlets: {Supporting} {Constraints} and {Consistency} in {Text} {Documents}},
	shorttitle   = {Textlets},
	author       = {Han, Han L. and Renom, Miguel A. and Mackay, Wendy E. and Beaudouin-Lafon, Michel},
	year         = 2020,
	month        = apr,
	booktitle    = {Proceedings of the 2020 {CHI} {Conference} on {Human} {Factors} in {Computing} {Systems}},
	publisher    = {ACM},
	address      = {Honolulu HI USA},
	pages        = {1--13},
	doi          = {10.1145/3313831.3376804},
	url          = {https://dl.acm.org/doi/10.1145/3313831.3376804},
	urldate      = {2025-05-22},
}

@inproceedings{huang_towards_2023,
	title        = {Towards {Reasoning} in {Large} {Language} {Models}: {A} {Survey}},
	shorttitle   = {Towards {Reasoning} in {Large} {Language} {Models}},
	author       = {Huang, Jie and Chang, Kevin Chen-Chuan},
	year         = 2023,
	month        = jul,
	booktitle    = {Findings of the {Association} for {Computational} {Linguistics}: {ACL} 2023},
	publisher    = {Association for Computational Linguistics},
	address      = {Toronto, Canada},
	pages        = {1049--1065},
	doi          = {10.18653/v1/2023.findings-acl.67},
	url          = {https://aclanthology.org/2023.findings-acl.67/},
	urldate      = {2025-05-22},
	editor       = {Rogers, Anna and Boyd-Graber, Jordan and Okazaki, Naoaki},
}

@inproceedings{kim_authors_2024,
	title        = {Authors' Values and Attitudes Towards AI-bridged Scalable Personalization of Creative Language Arts},
	author       = {Kim, Taewook and Han, Hyomin and Adar, Eytan and Kay, Matthew and Chung, John Joon Young},
	year         = 2024,
	booktitle    = {Proceedings of the 2024 CHI Conference on Human Factors in Computing Systems},
	location     = {Honolulu, HI, USA},
	publisher    = {Association for Computing Machinery},
	address      = {New York, NY, USA},
	series       = {CHI '24},
	doi          = {10.1145/3613904.3642529},
	isbn         = 9798400703300,
	url          = {https://doi.org/10.1145/3613904.3642529},
	articleno    = 31,
	numpages     = 16,
}

@inproceedings{riche_ai-instruments_2025,
	title        = {{AI}-{Instruments}: {Embodying} {Prompts} as {Instruments} to {Abstract} \& {Reflect} {Graphical} {Interface} {Commands} as {General}-{Purpose} {Tools}},
	shorttitle   = {{AI}-{Instruments}},
	author       = {Riche, Nathalie and Offenwanger, Anna and Gmeiner, Frederic and Brown, David and Romat, Hugo and Pahud, Michel and Marquardt, Nicolai and Inkpen, Kori and Hinckley, Ken},
	year         = 2025,
	month        = apr,
	booktitle    = {Proceedings of the 2025 {CHI} {Conference} on {Human} {Factors} in {Computing} {Systems}},
	publisher    = {Association for Computing Machinery},
	address      = {New York, NY, USA},
	series       = {{CHI} '25},
	pages        = {1--18},
	doi          = {10.1145/3706598.3714259},
	isbn         = {979-8-4007-1394-1},
	url          = {https://dl.acm.org/doi/10.1145/3706598.3714259},
	urldate      = {2025-05-27},
}

@inproceedings{kim_cells_2023,
	title        = {Cells, {Generators}, and {Lenses}: {Design} {Framework} for {Object}-{Oriented} {Interaction} with {Large} {Language} {Models}},
	shorttitle   = {Cells, {Generators}, and {Lenses}},
	author       = {Kim, Tae Soo and Lee, Yoonjoo and Chang, Minsuk and Kim, Juho},
	year         = 2023,
	month        = oct,
	booktitle    = {Proceedings of the 36th {Annual} {ACM} {Symposium} on {User} {Interface} {Software} and {Technology}},
	publisher    = {ACM},
	address      = {San Francisco CA USA},
	pages        = {1--18},
	doi          = {10.1145/3586183.3606833},
	isbn         = {979-8-4007-0132-0},
	url          = {https://dl.acm.org/doi/10.1145/3586183.3606833},
	urldate      = {2025-05-28},
	language     = {en},
}

@inproceedings{lee_design_2024,
	title        = {A {Design} {Space} for {Intelligent} and {Interactive} {Writing} {Assistants}},
	author       = {Lee, Mina and Gero, Katy Ilonka and Chung, John Joon Young and Shum, Simon Buckingham and Raheja, Vipul and Shen, Hua and Venugopalan, Subhashini and Wambsganss, Thiemo and Zhou, David and Alghamdi, Emad A. and August, Tal and Bhat, Avinash and Choksi, Madiha Zahrah and Dutta, Senjuti and Guo, Jin L.C. and Hoque, Md Naimul and Kim, Yewon and Knight, Simon and Neshaei, Seyed Parsa and Shibani, Antonette and Shrivastava, Disha and Shroff, Lila and Sergeyuk, Agnia and Stark, Jessi and Sterman, Sarah and Wang, Sitong and Bosselut, Antoine and Buschek, Daniel and Chang, Joseph Chee and Chen, Sherol and Kreminski, Max and Park, Joonsuk and Pea, Roy and Rho, Eugenia Ha Rim and Shen, Zejiang and Siangliulue, Pao},
	year         = 2024,
	month        = may,
	booktitle    = {Proceedings of the 2024 {CHI} {Conference} on {Human} {Factors} in {Computing} {Systems}},
	publisher    = {Association for Computing Machinery},
	address      = {New York, NY, USA},
	series       = {{CHI} '24},
	pages        = {1--35},
	doi          = {10.1145/3613904.3642697},
	isbn         = {979-8-4007-0330-0},
	url          = {https://dl.acm.org/doi/10.1145/3613904.3642697},
	urldate      = {2025-05-27},
}

@article{flower_cognitive_nodate,
	title        = {A Cognitive Process Theory of Writing},
	author       = {Flower, Linda and Hayes, John R.},
	year         = 1981,
	journal      = {College Composition and Communication},
	publisher    = {National Council of Teachers of English},
	volume       = 32,
	number       = 4,
	pages        = {365--387},
	doi          = {10.2307/356600},
	url          = {https://doi.org/10.2307/356600},
	language     = {en},
}

@inproceedings{brown_language_2020,
	title        = {Language models are few-shot learners},
	author       = {Brown, Tom B. and Mann, Benjamin and Ryder, Nick and Subbiah, Melanie and Kaplan, Jared and Dhariwal, Prafulla and Neelakantan, Arvind and Shyam, Pranav and Sastry, Girish and Askell, Amanda and Agarwal, Sandhini and Herbert-Voss, Ariel and Krueger, Gretchen and Henighan, Tom and Child, Rewon and Ramesh, Aditya and Ziegler, Daniel M. and Wu, Jeffrey and Winter, Clemens and Hesse, Christopher and Chen, Mark and Sigler, Eric and Litwin, Mateusz and Gray, Scott and Chess, Benjamin and Clark, Jack and Berner, Christopher and McCandlish, Sam and Radford, Alec and Sutskever, Ilya and Amodei, Dario},
	year         = 2020,
	month        = dec,
	booktitle    = {Proceedings of the 34th {International} {Conference} on {Neural} {Information} {Processing} {Systems}},
	publisher    = {Curran Associates Inc.},
	address      = {Red Hook, NY, USA},
	series       = {{NIPS} '20},
	pages        = {1877--1901},
	isbn         = {978-1-7138-2954-6},
	urldate      = {2025-06-04},
}

@incollection{neale_chapter_1997,
	title        = {Chapter 20 - {The} {Role} of {Metaphors} in {User} {Interface} {Design}},
	author       = {Neale, Dennis C. and Carroll, John M.},
	year         = 1997,
	month        = jan,
	booktitle    = {Handbook of {Human}-{Computer} {Interaction} ({Second} {Edition})},
	publisher    = {North-Holland},
	address      = {Amsterdam},
	pages        = {441--462},
	doi          = {10.1016/B978-044481862-1.50086-8},
	isbn         = {978-0-444-81862-1},
	url          = {https://www.sciencedirect.com/science/article/pii/B9780444818621500868},
	urldate      = {2025-06-04},
	editor       = {Helander, Marting G. and Landauer, Thomas K. and Prabhu, Prasad V.},
}

@inproceedings{masson_directgpt_2024,
	title        = {{DirectGPT}: {A} {Direct} {Manipulation} {Interface} to {Interact} with {Large} {Language} {Models}},
	shorttitle   = {{DirectGPT}},
	author       = {Masson, Damien and Malacria, Sylvain and Casiez, Géry and Vogel, Daniel},
	year         = 2024,
	month        = may,
	booktitle    = {Proceedings of the 2024 {CHI} {Conference} on {Human} {Factors} in {Computing} {Systems}},
	publisher    = {Association for Computing Machinery},
	address      = {New York, NY, USA},
	series       = {{CHI} '24},
	pages        = {1--16},
	doi          = {10.1145/3613904.3642462},
	isbn         = {979-8-4007-0330-0},
	url          = {https://doi.org/10.1145/3613904.3642462},
	urldate      = {2025-06-04},
}

@inproceedings{kim_lmcanvas_2023,
	title        = {{LMCanvas}: {Object-Oriented} Interaction to Personalize Large Language Model-Powered Writing Environments},
	author       = {Kim, Tae Soo and Sarkar, Arghya and Lee, Yoonjoo and Chang, Minsuk and Kim, Juho},
	year         = 2023,
	month        = apr,
	booktitle    = {CHI 2023 Workshop on Generative AI and HCI},
	location     = {Virtual},
	publisher    = {ACM},
	address      = {New York, NY, USA},
	pages        = {1--4},
	numpages     = 4,
	language     = {en},
}

@inproceedings{zhang_visar_2023,
	title        = {{VISAR}: {A} {Human}-{AI} {Argumentative} {Writing} {Assistant} with {Visual} {Programming} and {Rapid} {Draft} {Prototyping}},
	shorttitle   = {{VISAR}},
	author       = {Zhang, Zheng and Gao, Jie and Dhaliwal, Ranjodh Singh and Li, Toby Jia-Jun},
	year         = 2023,
	month        = oct,
	booktitle    = {Proceedings of the 36th {Annual} {ACM} {Symposium} on {User} {Interface} {Software} and {Technology}},
	publisher    = {Association for Computing Machinery},
	address      = {New York, NY, USA},
	series       = {{UIST} '23},
	pages        = {1--30},
	doi          = {10.1145/3586183.3606800},
	isbn         = {979-8-4007-0132-0},
	url          = {https://dl.acm.org/doi/10.1145/3586183.3606800},
	urldate      = {2025-06-04},
}

@inproceedings{zellweger_impact_2000,
	title        = {The impact of fluid documents on reading and browsing: an observational study},
	shorttitle   = {The impact of fluid documents on reading and browsing},
	author       = {Zellweger, Polle T. and Regli, Susan Harkness and Mackinlay, Jock D. and Chang, Bay-Wei},
	year         = 2000,
	month        = apr,
	booktitle    = {Proceedings of the {SIGCHI} conference on {Human} {Factors} in {Computing} {Systems}},
	publisher    = {Association for Computing Machinery},
	address      = {New York, NY, USA},
	series       = {{CHI} '00},
	pages        = {249--256},
	doi          = {10.1145/332040.332440},
	isbn         = {978-1-58113-216-8},
	url          = {https://dl.acm.org/doi/10.1145/332040.332440},
	urldate      = {2025-06-04},
}

@misc{mahlow_writing_2023,
	title        = {Writing {Tools}: {Looking} {Back} to {Look} {Ahead}},
	shorttitle   = {Writing {Tools}},
	author       = {Mahlow, Cerstin},
	year         = 2023,
	month        = mar,
	publisher    = {arXiv},
	doi          = {10.48550/arXiv.2303.17894},
	url          = {http://arxiv.org/abs/2303.17894},
	urldate      = {2025-06-04},
	note         = {arXiv:2303.17894 [cs]},
}

@misc{calderwood_phraselette_2025,
	title        = {Phraselette: {A} {Poet}'s {Procedural} {Palette}},
	shorttitle   = {Phraselette},
	author       = {Calderwood, Alex and Chung, John Joon Young and Sun, Yuqian and Roemmele, Melissa and Kreminski, Max},
	year         = 2025,
	month        = mar,
	publisher    = {arXiv},
	doi          = {10.48550/arXiv.2503.06335},
	url          = {http://arxiv.org/abs/2503.06335},
	urldate      = {2025-06-08},
	note         = {arXiv:2503.06335 [cs]},
	language     = {en},
}

@inproceedings{gero_supporting_2024,
	title        = {Supporting {Sensemaking} of {Large} {Language} {Model} {Outputs} at {Scale}},
	author       = {Gero, Katy Ilonka and Swoopes, Chelse and Gu, Ziwei and Kummerfeld, Jonathan K. and Glassman, Elena L.},
	year         = 2024,
	month        = may,
	booktitle    = {Proceedings of the 2024 {CHI} {Conference} on {Human} {Factors} in {Computing} {Systems}},
	publisher    = {Association for Computing Machinery},
	address      = {New York, NY, USA},
	series       = {{CHI} '24},
	pages        = {1--21},
	doi          = {10.1145/3613904.3642139},
	isbn         = {979-8-4007-0330-0},
	url          = {https://dl.acm.org/doi/10.1145/3613904.3642139},
	urldate      = {2025-06-04},
}

@inproceedings{dang_beyond_2022,
	title        = {Beyond {Text} {Generation}: {Supporting} {Writers} with {Continuous} {Automatic} {Text} {Summaries}},
	shorttitle   = {Beyond {Text} {Generation}},
	author       = {Dang, Hai and Benharrak, Karim and Lehmann, Florian and Buschek, Daniel},
	year         = 2022,
	month        = oct,
	booktitle    = {Proceedings of the 35th {Annual} {ACM} {Symposium} on {User} {Interface} {Software} and {Technology}},
	publisher    = {Association for Computing Machinery},
	address      = {New York, NY, USA},
	series       = {{UIST} '22},
	pages        = {1--13},
	doi          = {10.1145/3526113.3545672},
	isbn         = {978-1-4503-9320-1},
	url          = {https://dl.acm.org/doi/10.1145/3526113.3545672},
	urldate      = {2025-06-04},
}

@inproceedings{bolter_hypertext_1987,
	title        = {Hypertext and creative writing},
	author       = {Bolter, Jay David and Joyce, Michael},
	year         = 1987,
	month        = nov,
	booktitle    = {Proceedings of the {ACM} conference on {Hypertext}},
	publisher    = {Association for Computing Machinery},
	address      = {New York, NY, USA},
	series       = {{HYPERTEXT} '87},
	pages        = {41--50},
	doi          = {10.1145/317426.317431},
	isbn         = {978-0-89791-340-9},
	url          = {https://dl.acm.org/doi/10.1145/317426.317431},
	urldate      = {2025-06-04},
}

@book{norman2013design,
	title        = {The design of everyday things},
	author       = {Norman Donald, A},
	year         = 2013,
	publisher    = {MIT Press},
	address      = {Cambridge, MA},
}

@inproceedings{kim_lexichrome_2020,
	title        = {Lexichrome: {Text} {Construction} and {Lexical} {Discovery} with {Word}-{Color} {Associations} {Using} {Interactive} {Visualization}},
	shorttitle   = {Lexichrome},
	author       = {Kim, Chris and Hinrichs, Uta and Mohammad, Saif M. and Collins, Christopher},
	year         = 2020,
	month        = jul,
	booktitle    = {Proceedings of the 2020 {ACM} {Designing} {Interactive} {Systems} {Conference}},
	publisher    = {Association for Computing Machinery},
	address      = {New York, NY, USA},
	series       = {{DIS} '20},
	pages        = {477--488},
	doi          = {10.1145/3357236.3395503},
	isbn         = {978-1-4503-6974-9},
	url          = {https://dl.acm.org/doi/10.1145/3357236.3395503},
	urldate      = {2025-06-04},
}

@article{hutchins_direct_1985,
	title        = {Direct {Manipulation} {Interfaces}},
	author       = {Hutchins, Edwin L. and , James D., Hollan and and Norman, Donald A.},
	year         = 1985,
	month        = dec,
	journal      = {Human–Computer Interaction},
	volume       = 1,
	number       = 4,
	pages        = {311--338},
	doi          = {10.1207/s15327051hci0104_2},
	issn         = {0737-0024},
	url          = {https://doi.org/10.1207/s15327051hci0104_2},
	urldate      = {2025-06-04},
	note         = {Publisher: Taylor \& Francis \_eprint: https://doi.org/10.1207/s15327051hci0104\_2},
}

@techreport{hutchins1987metaphors,
	title        = {Metaphors for Interface Design},
	author       = {Hutchins, Edwin},
	year         = 1987,
	address      = {La Jolla, CA, USA},
	number       = {ICS Tech Report 8702},
	pages        = 34,
	url          = {https://eric.ed.gov/?id=ED287460},
	institution  = {Institute for Cognitive Science, University of California, San Diego},
}

@incollection{carroll1988interface,
	title        = {Interface metaphors and user interface design},
	author       = {Carroll, John M and Mack, Robert L and Kellogg, Wendy A},
	year         = 1988,
	booktitle    = {Handbook of human-computer interaction},
	publisher    = {Elsevier},
	address      = {Amsterdam},
	pages        = {67--85},
}

@misc{wei_emergent_2022,
	title        = {Emergent {Abilities} of {Large} {Language} {Models}},
	author       = {Wei, Jason and Tay, Yi and Bommasani, Rishi and Raffel, Colin and Zoph, Barret and Borgeaud, Sebastian and Yogatama, Dani and Bosma, Maarten and Zhou, Denny and Metzler, Donald and Chi, Ed H. and Hashimoto, Tatsunori and Vinyals, Oriol and Liang, Percy and Dean, Jeff and Fedus, William},
	year         = 2022,
	month        = oct,
	publisher    = {arXiv},
	doi          = {10.48550/arXiv.2206.07682},
	url          = {http://arxiv.org/abs/2206.07682},
	urldate      = {2025-06-04},
	note         = {arXiv:2206.07682 [cs]},
}

@book{hayes_new_2000,
	title        = {A new framework for understanding cognition and affect in writing},
	author       = {Hayes, John R.},
	year         = 2000,
	publisher    = {International Reading Association},
	address      = {Newark, DE, US},
	series       = {Perspectives on writing:  {Research}, theory, and practice},
	doi          = {10.1598/0872072681},
	isbn         = {978-0-87207-268-8},
	note         = {Pages: 44},
}

@inproceedings{gero_sparks_2022,
	title        = {Sparks: {Inspiration} for {Science} {Writing} using {Language} {Models}},
	shorttitle   = {Sparks},
	author       = {Gero, Katy Ilonka and Liu, Vivian and Chilton, Lydia},
	year         = 2022,
	month        = jun,
	booktitle    = {Proceedings of the 2022 {ACM} {Designing} {Interactive} {Systems} {Conference}},
	publisher    = {Association for Computing Machinery},
	address      = {New York, NY, USA},
	series       = {{DIS} '22},
	pages        = {1002--1019},
	doi          = {10.1145/3532106.3533533},
	isbn         = {978-1-4503-9358-4},
	url          = {https://doi.org/10.1145/3532106.3533533},
	urldate      = {2025-06-04},
}

@inproceedings{Renom2023InteractionKnowledge,
	title        = {Interaction Knowledge: Understanding the ‘Mechanics’ of Digital Tools},
	author       = {Renom, Miguel A. and Caramiaux, Baptiste and Beaudouin-Lafon, Michel},
	year         = 2023,
	booktitle    = {Proceedings of the 2023 CHI Conference on Human Factors in Computing Systems},
	location     = {Hamburg, Germany},
	publisher    = {Association for Computing Machinery},
	address      = {New York, NY, USA},
	series       = {CHI '23},
	doi          = {10.1145/3544548.3581246},
	isbn         = 9781450394215,
	url          = {https://doi.org/10.1145/3544548.3581246},
	articleno    = 403,
	numpages     = 14,
}

@inproceedings{angelini_tangible_2015,
	title        = {Tangible {Meets} {Gestural}: {Comparing} and {Blending} {Post}-{WIMP} {Interaction} {Paradigms}},
	shorttitle   = {Tangible {Meets} {Gestural}},
	author       = {Angelini, Leonardo and Lalanne, Denis and van den Hoven, Elise and Mazalek, Ali and Abou Khaled, Omar and Mugellini, Elena},
	year         = 2015,
	month        = jan,
	booktitle    = {Proceedings of the {Ninth} {International} {Conference} on {Tangible}, {Embedded}, and {Embodied} {Interaction}},
	publisher    = {Association for Computing Machinery},
	address      = {New York, NY, USA},
	series       = {{TEI} '15},
	pages        = {473--476},
	doi          = {10.1145/2677199.2683583},
	isbn         = {978-1-4503-3305-4},
	url          = {https://dl.acm.org/doi/10.1145/2677199.2683583},
	urldate      = {2025-08-20},
}

@inproceedings{renom2022exploring,
	title        = {Exploring Technical Reasoning in Digital Tool Use},
	author       = {Renom, Miguel A. and Caramiaux, Baptiste and Beaudouin-Lafon, Michel},
	year         = 2022,
	booktitle    = {Proceedings of the 2022 CHI Conference on Human Factors in Computing Systems},
	location     = {New Orleans, LA, USA},
	publisher    = {Association for Computing Machinery},
	address      = {New York, NY, USA},
	series       = {CHI '22},
	doi          = {10.1145/3491102.3501877},
	isbn         = 9781450391573,
	url          = {https://doi.org/10.1145/3491102.3501877},
	articleno    = 579,
	numpages     = 17,
}

@inproceedings{schmid_empowering_2013,
	title        = {Empowering materiality: inspiring the design of tangible interactions},
	shorttitle   = {Empowering materiality},
	author       = {Schmid, Magdalena and Rümelin, Sonja and Richter, Hendrik},
	year         = 2013,
	month        = feb,
	booktitle    = {Proceedings of the 7th {International} {Conference} on {Tangible}, {Embedded} and {Embodied} {Interaction}},
	publisher    = {Association for Computing Machinery},
	address      = {New York, NY, USA},
	series       = {{TEI} '13},
	pages        = {91--98},
	doi          = {10.1145/2460625.2460639},
	isbn         = {978-1-4503-1898-3},
	url          = {https://dl.acm.org/doi/10.1145/2460625.2460639},
	urldate      = {2025-08-20},
}

@inproceedings{zhao_making_2025,
	title        = {Making the {Write} {Connections}: {Linking} {Writing} {Support} {Tools} with {Writer} {Needs}},
	shorttitle   = {Making the {Write} {Connections}},
	author       = {Zhao, Zixin and Masson, Damien and Kim, Young-Ho and Penn, Gerald and Chevalier, Fanny},
	year         = 2025,
	month        = apr,
	booktitle    = {Proceedings of the 2025 {CHI} {Conference} on {Human} {Factors} in {Computing} {Systems}},
	publisher    = {Association for Computing Machinery},
	address      = {New York, NY, USA},
	series       = {{CHI} '25},
	pages        = {1--21},
	doi          = {10.1145/3706598.3713161},
	isbn         = {979-8-4007-1394-1},
	url          = {https://dl.acm.org/doi/10.1145/3706598.3713161},
	urldate      = {2025-08-20},
}

@article{blackwell_reification_2006,
	title        = {The reification of metaphor as a design tool},
	author       = {Blackwell, Alan F.},
	year         = 2006,
	month        = dec,
	journal      = {ACM Trans. Comput.-Hum. Interact.},
	volume       = 13,
	number       = 4,
	pages        = {490--530},
	doi          = {10.1145/1188816.1188820},
	issn         = {1073-0516},
	url          = {https://dl.acm.org/doi/10.1145/1188816.1188820},
	urldate      = {2025-08-20},
}

@inproceedings{frich_mapping_2019-1,
	title        = {Mapping the {Landscape} of {Creativity} {Support} {Tools} in {HCI}},
	author       = {Frich, Jonas and MacDonald Vermeulen, Lindsay and Remy, Christian and Biskjaer, Michael Mose and Dalsgaard, Peter},
	year         = 2019,
	month        = may,
	booktitle    = {Proceedings of the 2019 {CHI} {Conference} on {Human} {Factors} in {Computing} {Systems}},
	publisher    = {Association for Computing Machinery},
	address      = {New York, NY, USA},
	series       = {{CHI} '19},
	pages        = {1--18},
	doi          = {10.1145/3290605.3300619},
	isbn         = {978-1-4503-5970-2},
	url          = {https://dl.acm.org/doi/10.1145/3290605.3300619},
	urldate      = {2025-08-20},
}

@book{wallas_art_2014,
	title        = {The {Art} of {Thought}},
	author       = {Wallas, Graham},
	year         = 2014,
	publisher    = {Solis Press},
	address      = {London},
	isbn         = {978-1-910146-05-7},
	language     = {English},
}

@book{csikszentmihalyi_creativity_2010,
	title        = {Creativity: {Flow} and the {Psychology} of {Discovery} and {Invention}},
	shorttitle   = {Creativity},
	author       = {Csikszentmihalyi, Mihaly},
	year         = 2010,
	publisher    = {Harper Perennial},
	address      = {New York},
	isbn         = {978-0-06-092820-9},
	language     = {English},
}

@inproceedings{lee_coauthor_2022,
	title        = {{CoAuthor}: {Designing} a {Human}-{AI} {Collaborative} {Writing} {Dataset} for {Exploring} {Language} {Model} {Capabilities}},
	shorttitle   = {{CoAuthor}},
	author       = {Lee, Mina and Liang, Percy and Yang, Qian},
	year         = 2022,
	month        = apr,
	booktitle    = {Proceedings of the 2022 {CHI} {Conference} on {Human} {Factors} in {Computing} {Systems}},
	publisher    = {Association for Computing Machinery},
	address      = {New York, NY, USA},
	series       = {{CHI} '22},
	pages        = {1--19},
	doi          = {10.1145/3491102.3502030},
	isbn         = {978-1-4503-9157-3},
	url          = {https://dl.acm.org/doi/10.1145/3491102.3502030},
	urldate      = {2025-08-20},
}

@article{sawyer_iterative_2021,
	title        = {The iterative and improvisational nature of the creative process},
	author       = {Sawyer, R. Keith},
	year         = 2021,
	month        = dec,
	journal      = {Journal of Creativity},
	volume       = 31,
	pages        = 100002,
	doi          = {10.1016/j.yjoc.2021.100002},
	issn         = {2713-3745},
	url          = {https://www.sciencedirect.com/science/article/pii/S2713374521000029},
	urldate      = {2025-08-20},
}

@inproceedings{suh_sensecape_2023,
	title        = {Sensecape: {Enabling} {Multilevel} {Exploration} and {Sensemaking} with {Large} {Language} {Models}},
	shorttitle   = {Sensecape},
	author       = {Suh, Sangho and Min, Bryan and Palani, Srishti and Xia, Haijun},
	year         = 2023,
	month        = oct,
	booktitle    = {Proceedings of the 36th {Annual} {ACM} {Symposium} on {User} {Interface} {Software} and {Technology}},
	publisher    = {Association for Computing Machinery},
	address      = {New York, NY, USA},
	series       = {{UIST} '23},
	pages        = {1--18},
	doi          = {10.1145/3586183.3606756},
	isbn         = {979-8-4007-0132-0},
	url          = {https://dl.acm.org/doi/10.1145/3586183.3606756},
	urldate      = {2025-08-20},
}

@inproceedings{piper_illuminating_2002,
	title        = {Illuminating clay: a 3-D tangible interface for landscape analysis},
	author       = {Piper, Ben and Ratti, Carlo and Ishii, Hiroshi},
	year         = 2002,
	booktitle    = {Proceedings of the SIGCHI Conference on Human Factors in Computing Systems},
	location     = {Minneapolis, Minnesota, USA},
	publisher    = {Association for Computing Machinery},
	address      = {New York, NY, USA},
	series       = {CHI '02},
	pages        = {355–362},
	doi          = {10.1145/503376.503439},
	isbn         = 1581134533,
	url          = {https://doi.org/10.1145/503376.503439},
	numpages     = 8,
}

@inproceedings{ishii_tangible_1997,
	title        = {Tangible bits: towards seamless interfaces between people, bits and atoms},
	shorttitle   = {Tangible bits},
	author       = {Ishii, Hiroshi and Ullmer, Brygg},
	year         = 1997,
	month        = mar,
	booktitle    = {Proceedings of the {ACM} {SIGCHI} {Conference} on {Human} factors in computing systems},
	publisher    = {Association for Computing Machinery},
	address      = {New York, NY, USA},
	series       = {{CHI} '97},
	pages        = {234--241},
	doi          = {10.1145/258549.258715},
	isbn         = {978-0-89791-802-2},
	url          = {https://dl.acm.org/doi/10.1145/258549.258715},
	urldate      = {2025-08-20},
}

@incollection{gessler1998skeuomorphs,
	title        = {Skeuomorphs and Cultural Algorithms},
	author       = {Gessler, Nicholas},
	year         = 1998,
	booktitle    = {Evolutionary Programming VII},
	publisher    = {Springer, Berlin Heidelberg},
	address      = {Heidelberg, Germany},
	series       = {Lecture Notes in Computer Science},
	volume       = 1447,
	pages        = {229--238},
	doi          = {10.1007/BFb0040776},
	editor       = {Porto, V. W. and Saravanan, N. and Waagen, D. and Eiben, A. E.},
}

@incollection{marcus_principles_1995,
	title        = {Principles of {Effective} {Visual} {Communication} for {Graphical} {User} {Interface} {Design}},
	author       = {Marcus, Aaron},
	year         = 1995,
	month        = jan,
	booktitle    = {Readings in {Human}–{Computer} {Interaction}},
	publisher    = {Morgan Kaufmann},
	address      = {San Francisco, CA, USA},
	series       = {Interactive {Technologies}},
	pages        = {425--441},
	doi          = {10.1016/B978-0-08-051574-8.50044-3},
	isbn         = {978-0-08-051574-8},
	url          = {https://www.sciencedirect.com/science/article/pii/B9780080515748500443},
	urldate      = {2025-08-20},
	editor       = {Baecker, RONALD M. and Grudin, JONATHAN and Buxton, WILLIAM A. S. and Greenberg, SAUL},
}

@inproceedings{gross_skeu_nodate,
	title        = {Skeu the evolution: skeuomorphs, style, and the material of tangible interactions},
	author       = {Gross, Shad and Bardzell, Jeffrey and Bardzell, Shaowen},
	year         = 2014,
	booktitle    = {Proceedings of the 8th International Conference on Tangible, Embedded and Embodied Interaction},
	location     = {Munich, Germany},
	publisher    = {Association for Computing Machinery},
	address      = {New York, NY, USA},
	series       = {TEI '14},
	pages        = {53–60},
	doi          = {10.1145/2540930.2540969},
	isbn         = 9781450326353,
	url          = {https://doi.org/10.1145/2540930.2540969},
	numpages     = 8,
}

@article{urbano2022skeuomorphism,
	title        = {From skeuomorphism to flat design: age-related differences in performance and aesthetic perceptions},
	author       = {Urbano, In{\^e}s Cunha Vaz Pereira and Guerreiro, Jo{\~a}o Pedro Vieira and Nicolau, Hugo Miguel Aleixo Albuquerque},
	year         = 2022,
	journal      = {Behaviour \& Information Technology},
	publisher    = {Taylor \& Francis},
	volume       = 41,
	number       = 3,
	pages        = {452--467},
}

@article{norman1999affordance,
	title        = {Affordance, conventions, and design},
	author       = {Norman, Donald A},
	year         = 1999,
	journal      = {interactions},
	publisher    = {ACM New York, NY, USA},
	volume       = 6,
	number       = 3,
	pages        = {38--43},
}

@article{jung_metaphors_2017,
	title        = {Metaphors, materialities, and affordances: {Hybrid} morphologies in the design of interactive artifacts},
	shorttitle   = {Metaphors, materialities, and affordances},
	author       = {Jung, Heekyoung and Wiltse, Heather and Wiberg, Mikael and Stolterman, Erik},
	year         = 2017,
	month        = nov,
	journal      = {Design Studies},
	volume       = 53,
	pages        = {24--46},
	doi          = {10.1016/j.destud.2017.06.004},
	issn         = {0142-694X},
	url          = {https://www.sciencedirect.com/science/article/pii/S0142694X17300467},
	urldate      = {2025-08-20},
}

@inproceedings{masson_visual_2025,
	title        = {Visual Story-Writing: Writing by Manipulating Visual Representations of Stories},
	author       = {Masson, Damien and Zhao, Zixin and Chevalier, Fanny},
	year         = 2025,
	booktitle    = {Proceedings of the 38th Annual ACM Symposium on User Interface Software and Technology (UIST ’25)},
	location     = {Busan, Republic of Korea},
	publisher    = {Association for Computing Machinery},
	address      = {New York, NY, USA},
	doi          = {10.1145/3746059.3747758},
	isbn         = {979-8-4007-2037-6},
	url          = {https://doi.org/10.1145/3746059.3747758},
	numpages     = 15,
}

@book{braunThematicAnalysisPractical2022,
	title        = {Thematic {{Analysis}}: {{A Practical Guide}}},
	shorttitle   = {Thematic {{Analysis}}},
	author       = {Braun, Virginia and Clarke, Victoria},
	year         = 2022,
	publisher    = {SAGE Publications Ltd},
	address      = {London},
	isbn         = {978-1-5264-1729-9},
}

@article{KirshIntelligentUseOfSpace1995,
	title        = {The intelligent use of space},
	author       = {Kirsh, David},
	year         = 1995,
	month        = {feb},
	journal      = {Artif. Intell.},
	publisher    = {Elsevier Science Publishers Ltd.},
	address      = {GBR},
	volume       = 73,
	number       = {1–2},
	pages        = {31–68},
	doi          = {10.1016/0004-3702(94)00017-U},
	issn         = {0004-3702},
	url          = {https://doi.org/10.1016/0004-3702(94)00017-U},
	issue_date   = {Feb. 1995},
	numpages     = 38,
}

@book{HutchinsCognitionWild1996,
	title        = {Cognition in the Wild},
	author       = {Edwin Hutchins},
	year         = 1996,
	publisher    = {Bradford Books; Revised ed. edition (September 1, 1996)},
	address      = {St Bradford, PA, United States},
	isbn         = {978-0262581462},
	numpages     = 402,
}

@inproceedings{AdarBenevolentDeception2013,
	title        = {Benevolent deception in human computer interaction},
	author       = {Adar, Eytan and Tan, Desney S. and Teevan, Jaime},
	year         = 2013,
	booktitle    = {Proceedings of the SIGCHI Conference on Human Factors in Computing Systems},
	location     = {Paris, France},
	publisher    = {Association for Computing Machinery},
	address      = {New York, NY, USA},
	series       = {CHI '13},
	pages        = {1863–1872},
	doi          = {10.1145/2470654.2466246},
	isbn         = 9781450318990,
	url          = {https://doi.org/10.1145/2470654.2466246},
	numpages     = 10,
}

@article{hartson2003cognitive,
  title={Cognitive, physical, sensory, and functional affordances in interaction design},
  author={Hartson, Rex},
  journal={Behaviour \& information technology},
  volume={22},
  number={5},
  pages={315--338},
  year={2003},
  publisher={Taylor \& Francis}
}

@inproceedings{zimmerman2007research,
author = {Zimmerman, John and Forlizzi, Jodi and Evenson, Shelley},
title = {Research through design as a method for interaction design research in HCI},
year = {2007},
isbn = {9781595935939},
publisher = {Association for Computing Machinery},
address = {New York, NY, USA},
url = {https://doi.org/10.1145/1240624.1240704},
doi = {10.1145/1240624.1240704},
booktitle = {Proceedings of the SIGCHI Conference on Human Factors in Computing Systems},
pages = {493–502},
numpages = {10},
location = {San Jose, California, USA},
series = {CHI '07}
}

\clearpage
\newpage
\appendix
\onecolumn

\section{Case Study Walkthrough}

\begin{figure*}[ht!]
\vspace{12pt}
  \includegraphics[width=\textwidth]{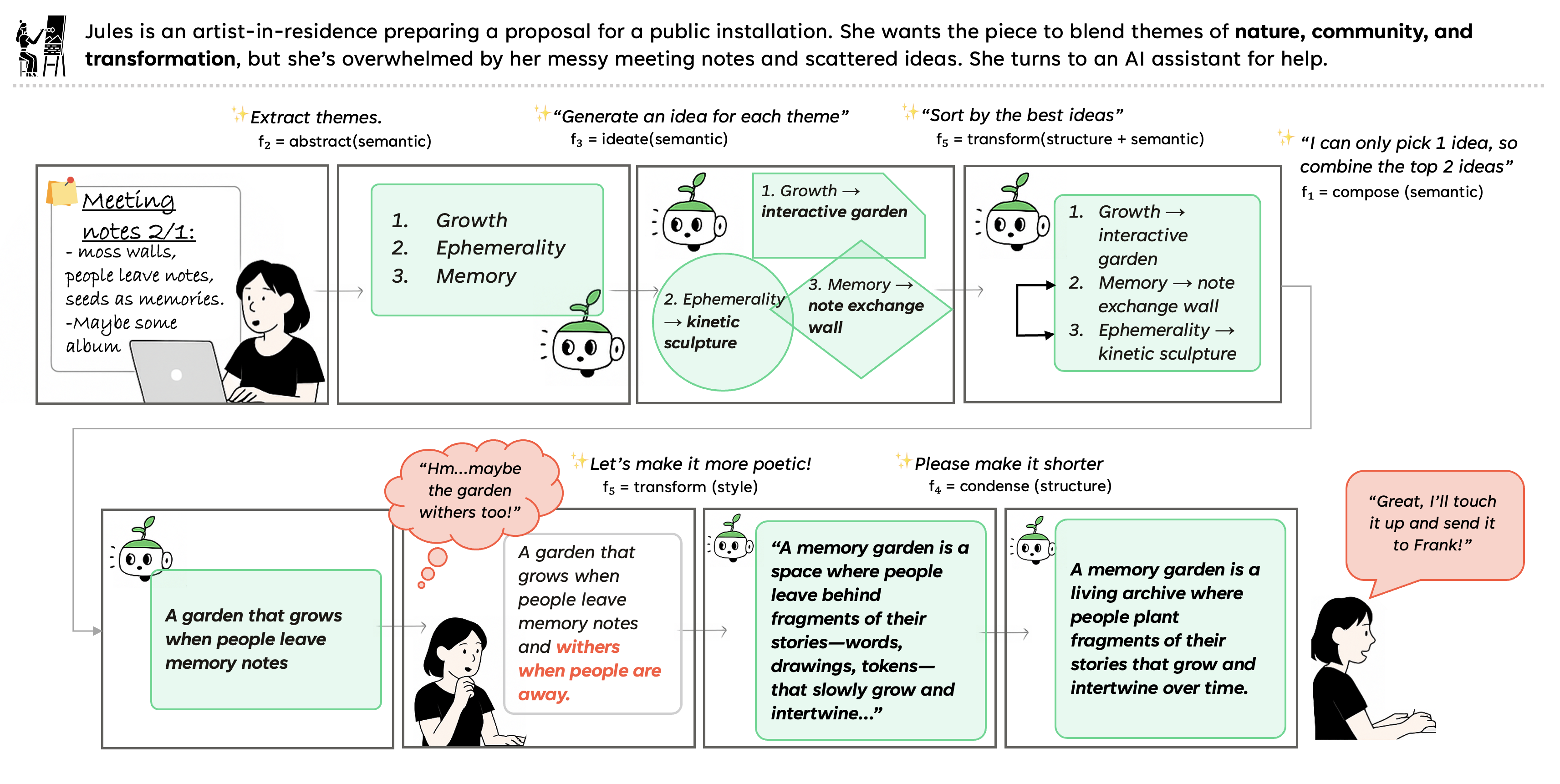}
  \caption{An example walkthrough of our conceptual framework using a case study of an artist-in-residence preparing a written proposal for an installation. This storyboard illustrates steps users might take in a typical writing process, and how these align with components of our framework.}
  \Description{DESCRIPTION}
  \label{fig:storyboard}
  \vspace{12pt}
\end{figure*}

In Figure \ref{fig:storyboard}, we illustrate how an artist-in-residence, Jules, applies our framework in a simple writing workflow with AI assistance. She begins by asking the system to \textsc{ideate} for each theme extracted from her notes—an \textsc{abstraction} operation in the \textit{semantic} space. Next, she has the agent sort the ideas by quality, which involves \textsc{transforming} the \textit{structure} of the text to rank alternatives. She then selects the top two ideas and asks the agent to \textsc{compose} them into a single proposal, while also adding her own manual edits. To tailor the draft for the exhibit context, she prompts the agent to make the idea more poetic, a \textit{stylistic} \textsc{transformation} that alters tone without changing the underlying meaning. Finally, because the draft is overly long, she asks the agent to \textsc{condense} the \textit{structure} so the proposal is succinct and focused.

\revision{
\section{Formative Study Details}
All materials, instructions, and cards used for our formative study are included in our Supplementary Files. To prompt participants, we chose examples spanning approachable skillsets (pottery, photography, gardening), and  advanced crafts (embroidery, beekeeping), and balancing inanimate materials (clay, threads), animate ones (bees, plants), and hybrid (scenes to photograph). For each set of cards, we provided examples of three characteristics of the materials (e.g. sticky golden honey, a working bee, a communicating bee), and three techniques or tools typical of the handcraft (e.g. placing hives strategically, smoking the hive, harvesting the honey) illustrated only with one image.

While providing concrete illustrations to work from can support idea elicitation from participants, we note that there is also a risk that the chosen set of cards could prime participants in certain directions. We deliberately chose diverse concepts, as described above, with imagery that conveyed ideas without prescribing one particular concept, and also encouraged participants to come up with their own material-inspired metaphors (which several participants did).

}

\newpage
\section{Initial Concepts}
\begin{figure}[ht!]
    \centering
    \includegraphics[width=\linewidth]{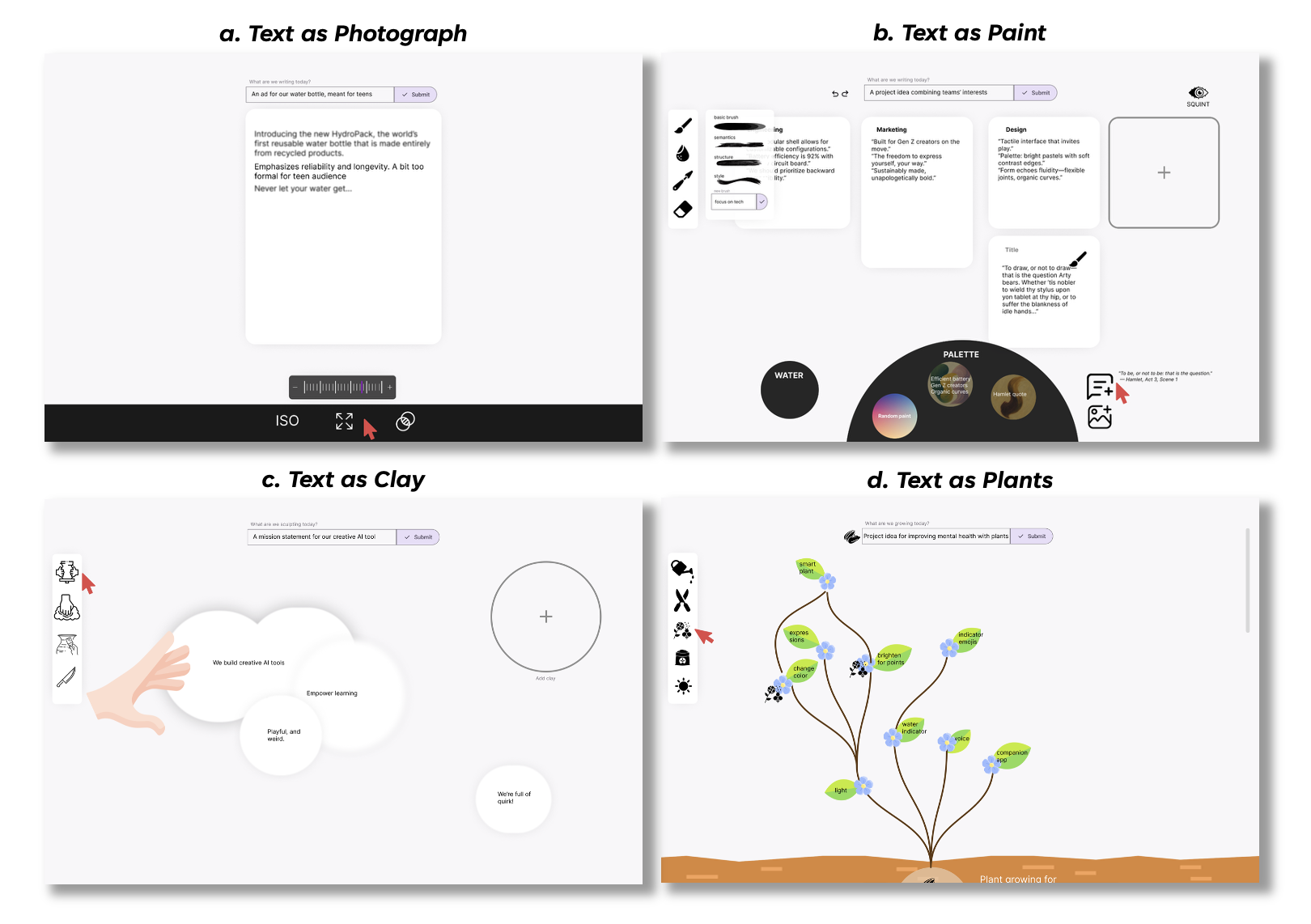}
    \caption{Low-fidelity prototypes of various material-inspired interaction techniques.}
    \label{fig:earlyprototypes}
    \vspace{12pt}
\end{figure}

Our framework and formative study structured our early explorations of concepts and guided the design of initial prototypes.
For each material metaphor, we asked three guiding questions: (1) Which components of text does it operate on? (semantic, structural, stylistic), (2) Which operations does it suggest, and how do material metaphors map to these operations (e.g. composing ideas = blending paint or combining clay)?, (3) What core interaction technique maps LLM operations to material operations?

In this sense, each metaphor served as an \textit{instantiation of the framework}, where core interaction techniques are inspired by the material’s affordances. Figure \ref{fig:earlyprototypes} shows four examples of our initial, low-fidelity prototypes that cover diverse materials: (a) \textit{Text as Photograph}, (b) \textit{Text as Paint}, (c) \textit{Text as Clay} and (d) \textit{Text as Plants}. 
For example, photography suggests operations such as zooming and transforming, while clay suggests direct sculpting and refinement.

These prototype concepts are again by no means comprehensive but instead serve as provocations of how the material metaphor can inspire various interaction techniques with text.

\clearpage

\section{Interaction Techniques User Manual}
In the following section, we show how an example of interaction designers can leverage our \textsc{Texterial} framework to implement material-inspired interaction techniques, as well as our own implementations for \textit{Text as Clay} and \textit{Text as Plants}. The following section provides all prompts used to implement such techniques.
\vspace{10pt}

\begin{table*}[h]
\vspace{12pt}
\centering
\small
\caption{\textbf{Single-metaphor example of a concrete technique.}—\emph{merge/combine blocks} (drag \& overlap)}

\setlength{\tabcolsep}{5pt}
\renewcommand{\arraystretch}{1.25}

\begin{tabularx}{\textwidth}{@{}p{2cm}YYYYYYY@{}}
\toprule
\textbf{Conceptual lens} &
\textbf{Material affordances \& constraints} &
\textbf{Embodied reasoning} &
\textbf{Direct manipulation} &
\textbf{Non-linear workflow} &
\textbf{Gestural-driven input} &
\textbf{Expressive control} &
\textbf{Immediate sensory feedback} \\
\midrule

\textbf{Merge / combine blocks} \newline \textit{(drag \& overlap)} &
``I can \emph{join/knead} clay. Smearing two lumps together blends the seam.
More overlap yields a more unified form; minimal overlap keeps pieces
distinguishable.'' \par &
Approach direction sets ordering. A slow, deliberate push suggests careful integration; a quick shove suggests
rough stitching. Where and how much I overlap determines what becomes the seam versus what remains intact. \par &
User drags one text block onto another. Contact initiates merging, and overlap defines the blending region (the seam), rather than issuing a verbal instruction. \par &
User can merge, rip to re-separate, then merge again
with a different direction or overlap. \par &
Drag vector (source$\rightarrow$target), drop position, overlap ratio, and anchor points corresponding to the intersection region
(seam). \par &
Drag direction determines ordering. Overlap magnitude controls degree of integration. Seam location biases where merge is strongest. \par &
Blocks visually react on contact with spreading/bending at the contact edges and color becomes denser.
After release, the merged text animates locally at the seam, highlights the
change. \par \\
\midrule
\multicolumn{8}{@{}p{\textwidth}@{}}{%
\textbf{\textsc{Texterial} operation.}
\textsc{COMPOSE} (f\textsubscript{1}) combines content, and under high overlap
} \\
\midrule
\multicolumn{8}{@{}p{\textwidth}@{}}{%
\textbf{Implementation hook (prompt routing + control variables).}\par
\textbf{Detect / tag overlap.} During drag, compute collision \emph{intensity} $i\in[0,1]$ from overlap extent (e.g., line/region intersection). Mark the visually overlapping lines in each block with \texttt{<overlap>}...\texttt{</overlap>} tags.\par
\textbf{Route by collision type + intensity.}
\begin{itemize}\setlength{\itemsep}{2pt}\setlength{\topsep}{2pt}
  \item \textbf{Vertical merge (source above target).} Call \texttt{getVerticalCollisionPrompt(topTextWithTags, bottomTextWithTags, i, context)}. Enforce ordering: TOP $\rightarrow$ BOTTOM in the final text. Edits should concentrate \emph{within/near} the tagged overlap region; avoid changes outside unless necessary for coherence. Output length $\le$ (top+bottom).
  \item \textbf{Full blend (extreme overlap).} For near-complete overlap, call \texttt{getFullBlendCollisionPrompt(firstText, secondText, i, context)} to generate a single unified passage that blends both sources (not simple insertion), with length $\le$ (first+second).
  \item \textbf{Horizontal insertion (source into target).} When the dragged block is inserted into a target block, call \texttt{getHorizontalCollisionPrompt(mainTextWithTags, insertTextWithTags, pos, i, context, insertLineText?)}. Use \texttt{pos} (or an anchor line) to place the insertion; apply blending primarily within/near \texttt{<overlap>} lines.
\end{itemize}
\textbf{Encode intensity guidance.} Convert $i$ to \texttt{intensityPercent} and select a blending regime:
light ($i<0.6$) $\rightarrow$ minimal edits for a smooth transition;
moderate ($0.6\le i<0.9$) $\rightarrow$ natural flow with moderate editing;
heavy ($i\ge 0.9$) $\rightarrow$ extensive merging/weaving.\par
\textbf{Constraints / outputs.} Return \emph{only} the combined text (no markup/tags/markdown). Keep length the same or shorter than the two inputs. If \texttt{userWritingContext} is provided, prepend it as a conditioning line and keep edits relevant to that project.\par
} \\

\bottomrule
\end{tabularx}

\label{tab:single-metaphor-merge-horizontal}
\end{table*}

\newpage
\subsection{\textit{Text as Clay}}

\begin{figure*}[h!]
  \includegraphics[width=\textwidth]{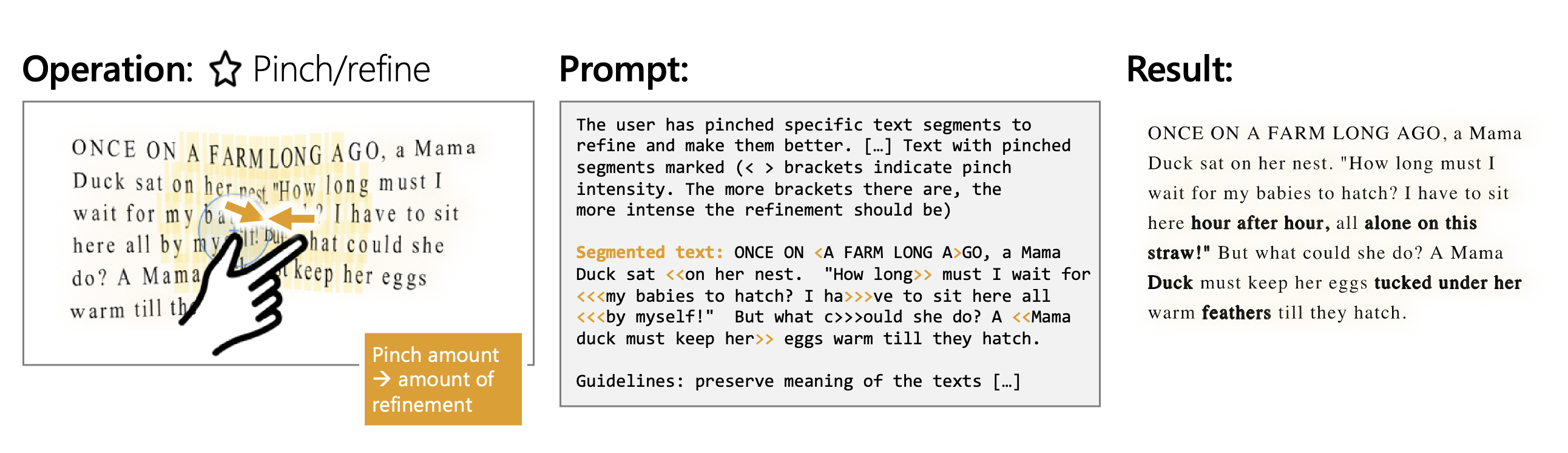}
  \caption{Pinching to refine ($f_2^{-1} = $ \textsc{concretize})}
  \Description{DESCRIPTION}
  \label{fig:pinching}
  \vspace{12pt}
\end{figure*}

\begin{figure*}[h!]
  \includegraphics[width=\textwidth]{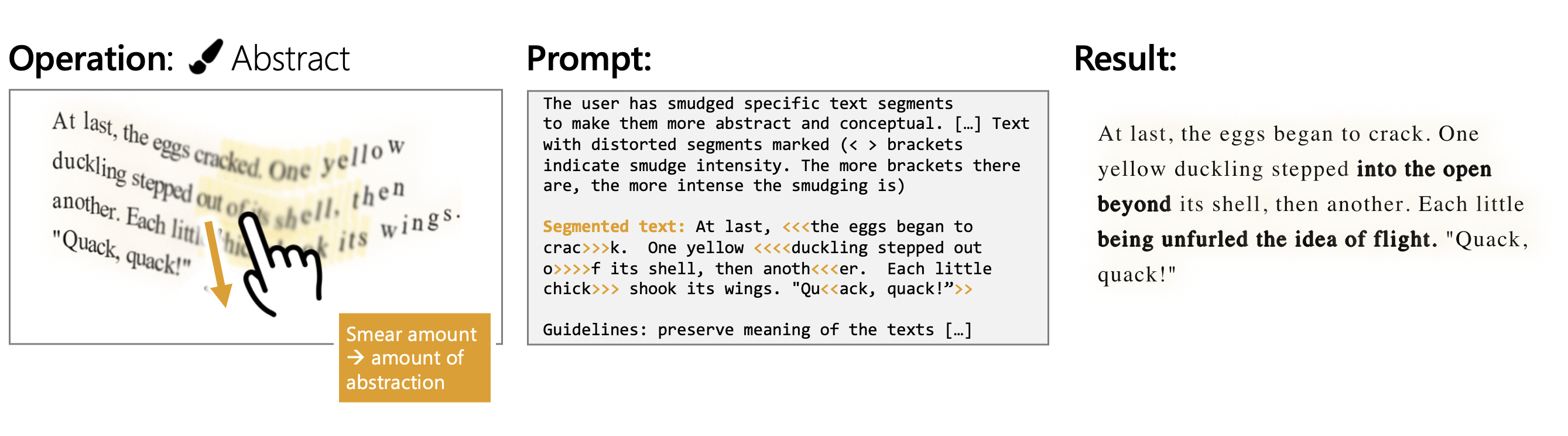}
  \caption{Smear to abstract  ($f_2 = $ \textsc{abstract})}
  \Description{DESCRIPTION}
  \label{fig:smear}
  \vspace{12pt}
\end{figure*}

\begin{figure*}[h!]
  \includegraphics[width=\textwidth]{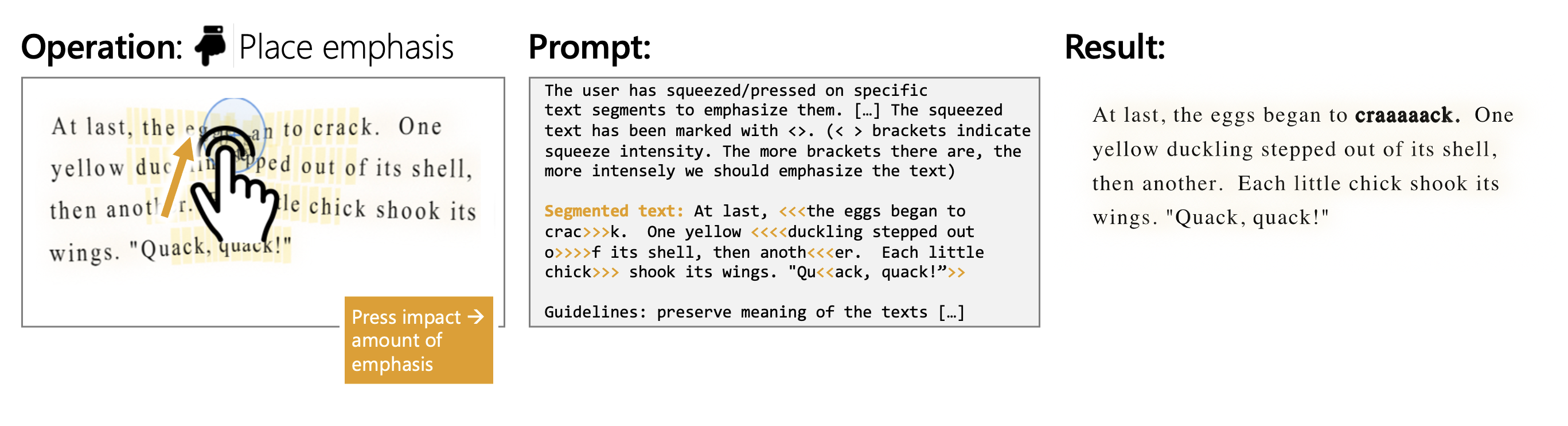}
  \caption{Press to emphasize}
  \Description{DESCRIPTION}
  \label{fig:emphasize}
  \vspace{12pt}
\end{figure*}

\begin{figure*}[h!]
  \includegraphics[width=\textwidth]{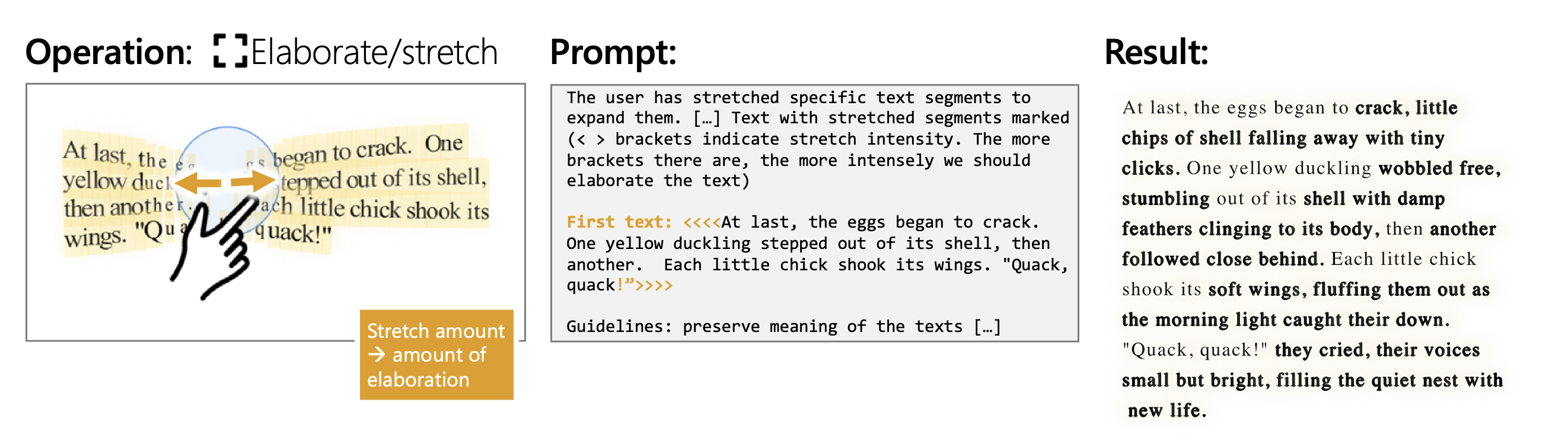}
  \caption{Stretch to elaborate ($f_4^{-1} = $ \textsc{elaborate})}
  \Description{DESCRIPTION}
  \label{fig:elaborate}
  \vspace{12pt}
\end{figure*}

\begin{figure*}[h!]
  \includegraphics[width=\textwidth]{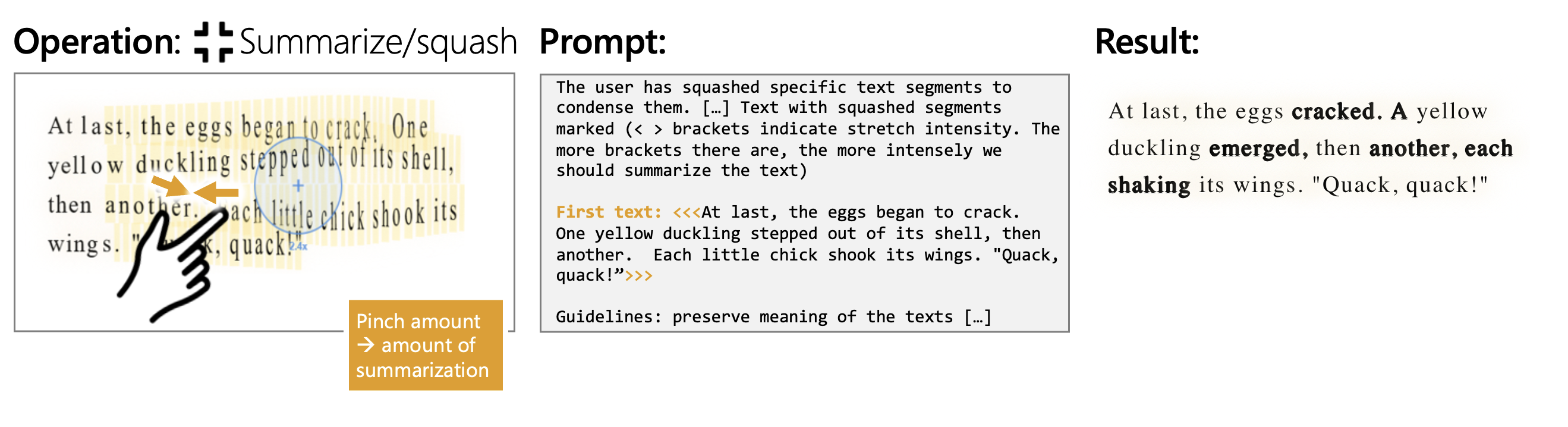}
  \caption{Squash to summarize ($f_2 = $ \textsc{condense})}
  \Description{DESCRIPTION}
  \label{fig:summarize}
  \vspace{12pt}
\end{figure*}

\begin{figure*}[h!]
\begin{flushleft}
  \includegraphics[width=.7\textwidth]{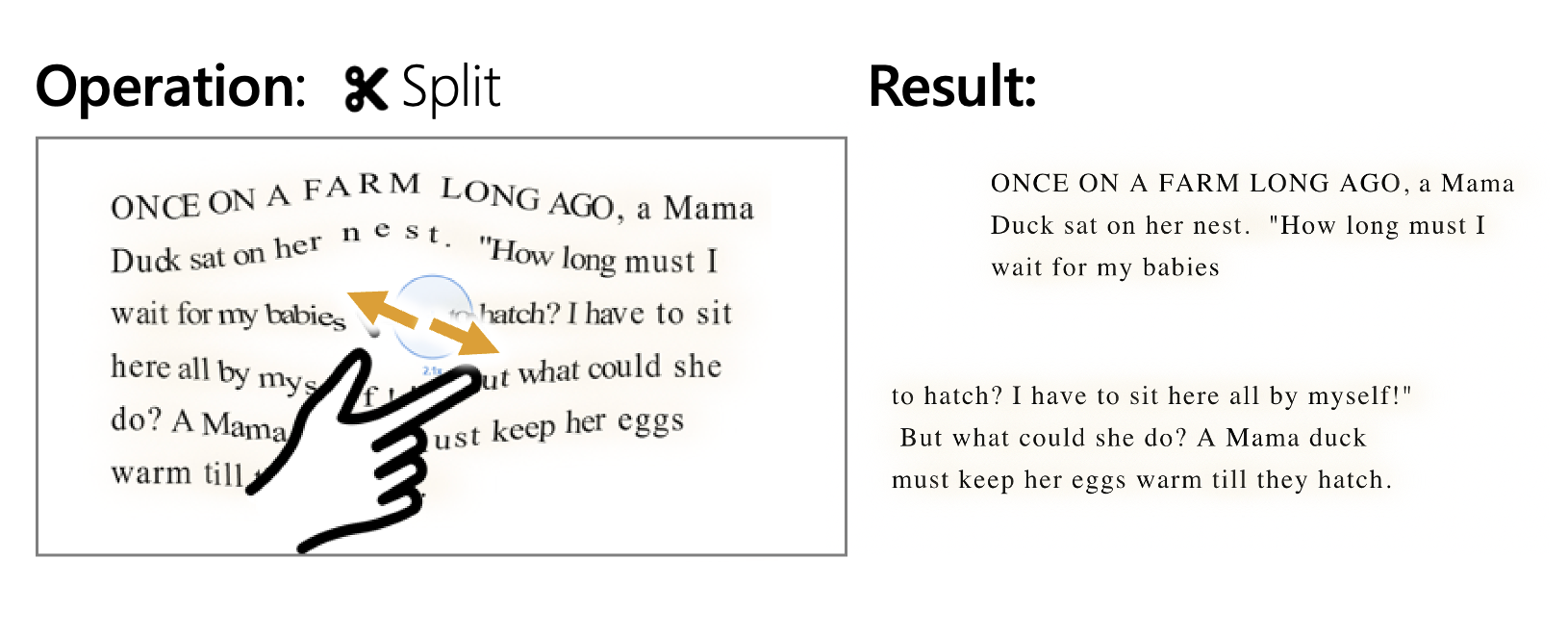}
  \caption{Split into chunks}
  \Description{DESCRIPTION}
  \label{fig:split}
  \end{flushleft}
  \vspace{12pt}
\end{figure*}

\clearpage
\newpage
\subsection{\textit{Text as Plants}}

\begin{figure*}[h!]
\begin{flushleft}
   \vspace{12pt}
  \includegraphics[width=\textwidth]{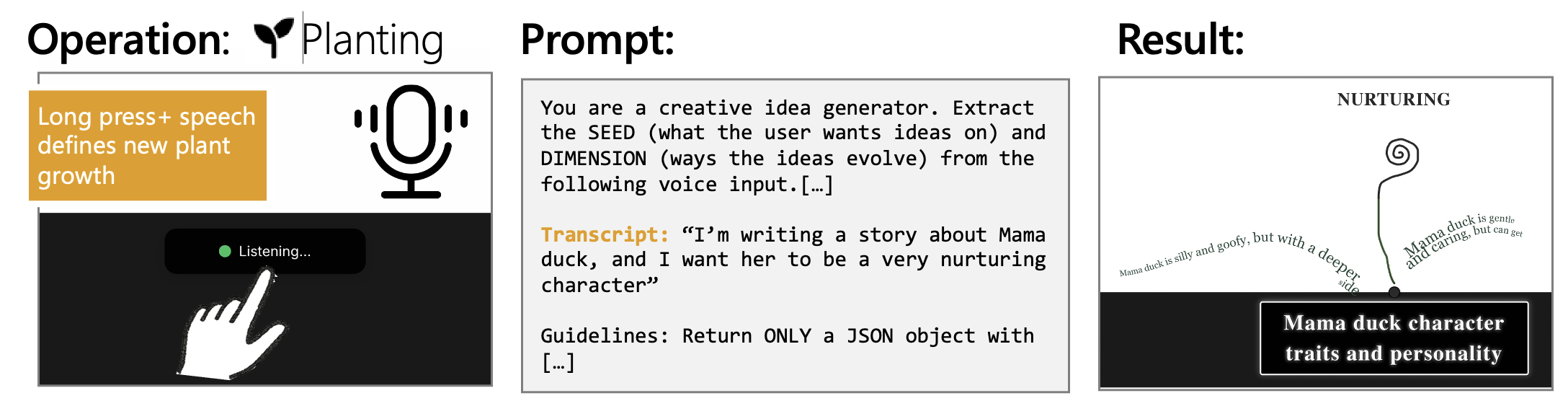}

  \caption{Planting a new idea fern via voice}
  \Description{DESCRIPTION}
  \label{fig:planting}
    \end{flushleft}
    \vspace{24pt}
\end{figure*}

\begin{figure*}[h!]
\begin{flushleft}

  \includegraphics[width=\textwidth]{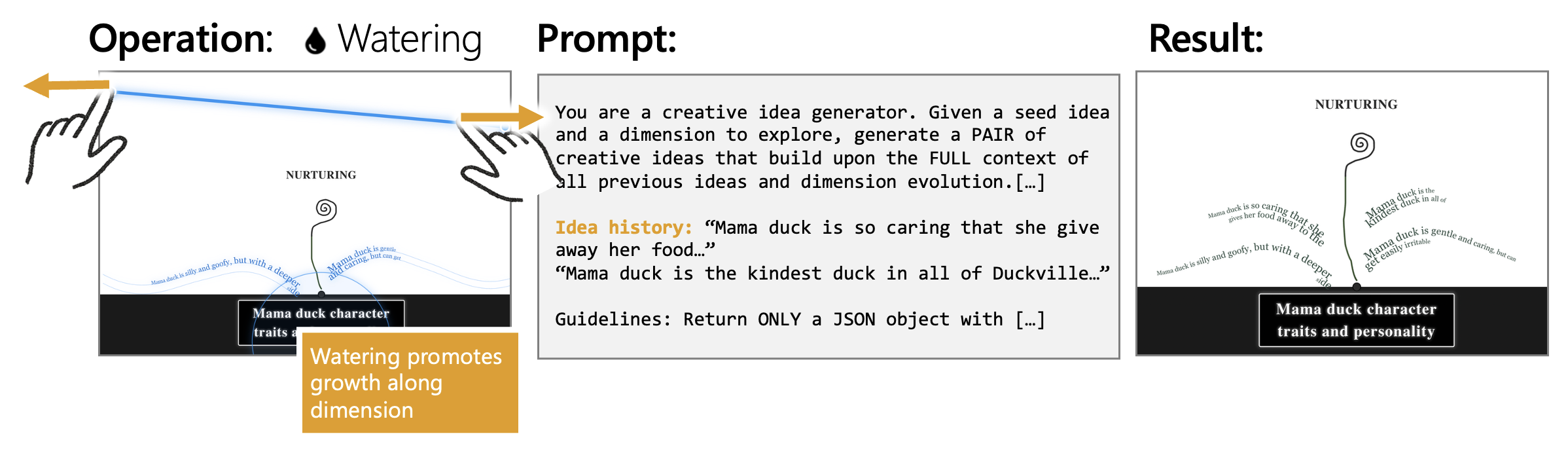}
  \caption{Watering to promote growth of new ideas ($f_3 = $ \textsc{ideate})}
  \Description{DESCRIPTION}
  \label{fig:watering}
    \end{flushleft}
    \vspace{24pt}
\end{figure*}
\begin{figure*}[h!]
  \includegraphics[width=\textwidth]{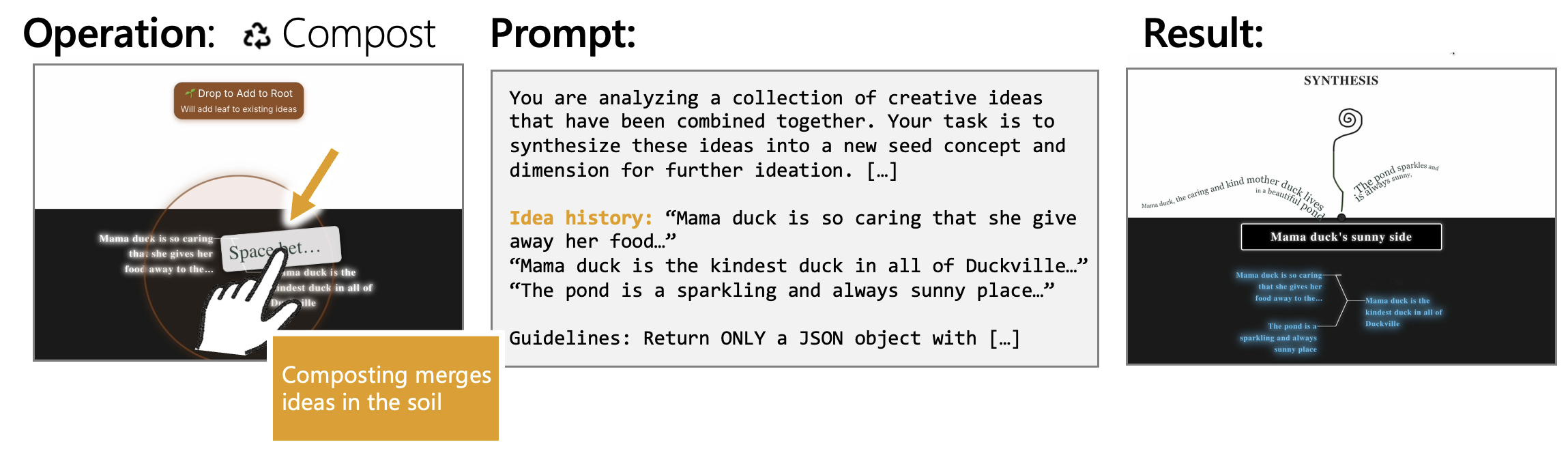}
  \caption{Composting to merge ideas together ($f_1 = $ \textsc{compose})}
  \Description{DESCRIPTION}
  \label{fig:composting}
  \vspace{12pt}
\end{figure*}

\begin{figure*}[h!]

    \vspace{20pt}
  \includegraphics[width=.7\textwidth]{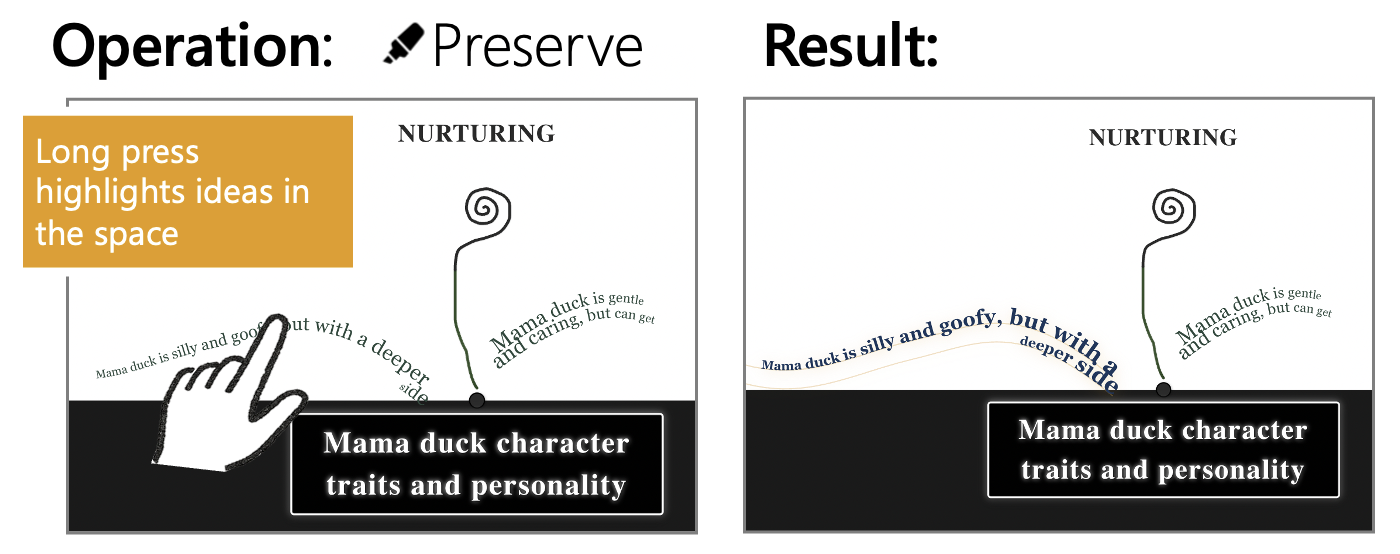}
  \caption{Preserving or highlighting favorite ideas}
  \Description{DESCRIPTION}
  \label{fig:preserving}
    \vspace{20pt}
\end{figure*}

\begin{figure*}[h!]

  \includegraphics[width=.7\textwidth]{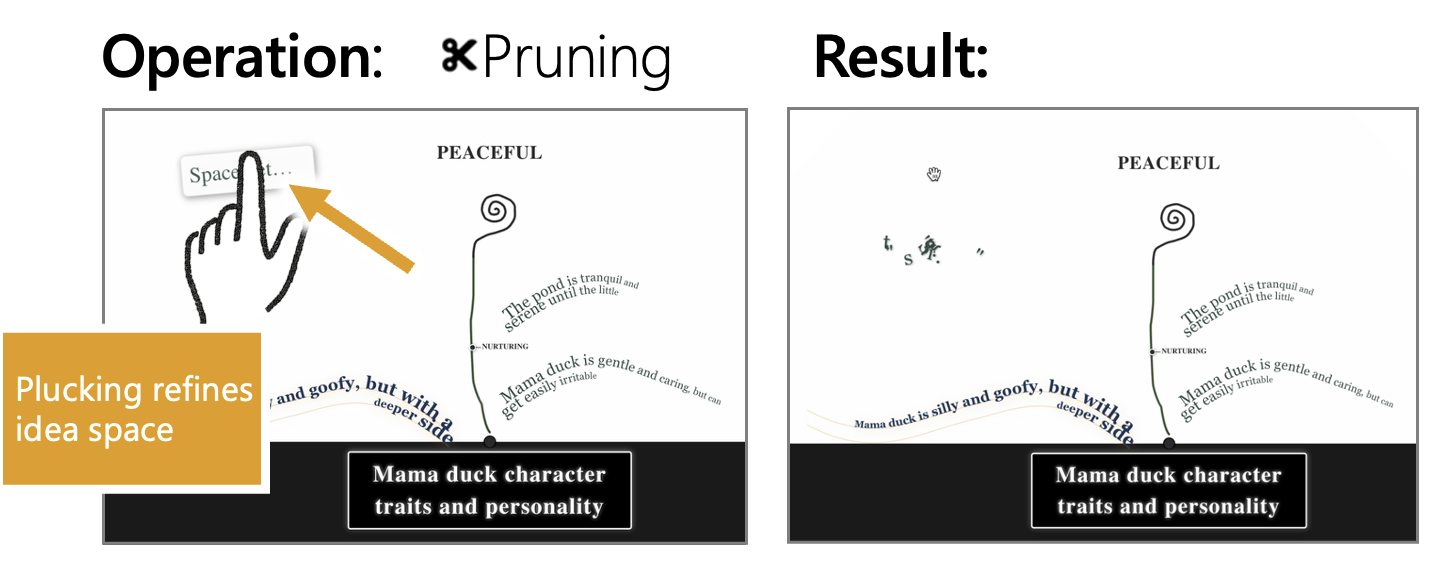}
  \caption{Pruning to refine idea space}
  \Description{DESCRIPTION}
  \label{fig:pruning}
    \vspace{20pt}
\end{figure*}

\begin{figure*}[h!]

  \includegraphics[width=.7\textwidth]{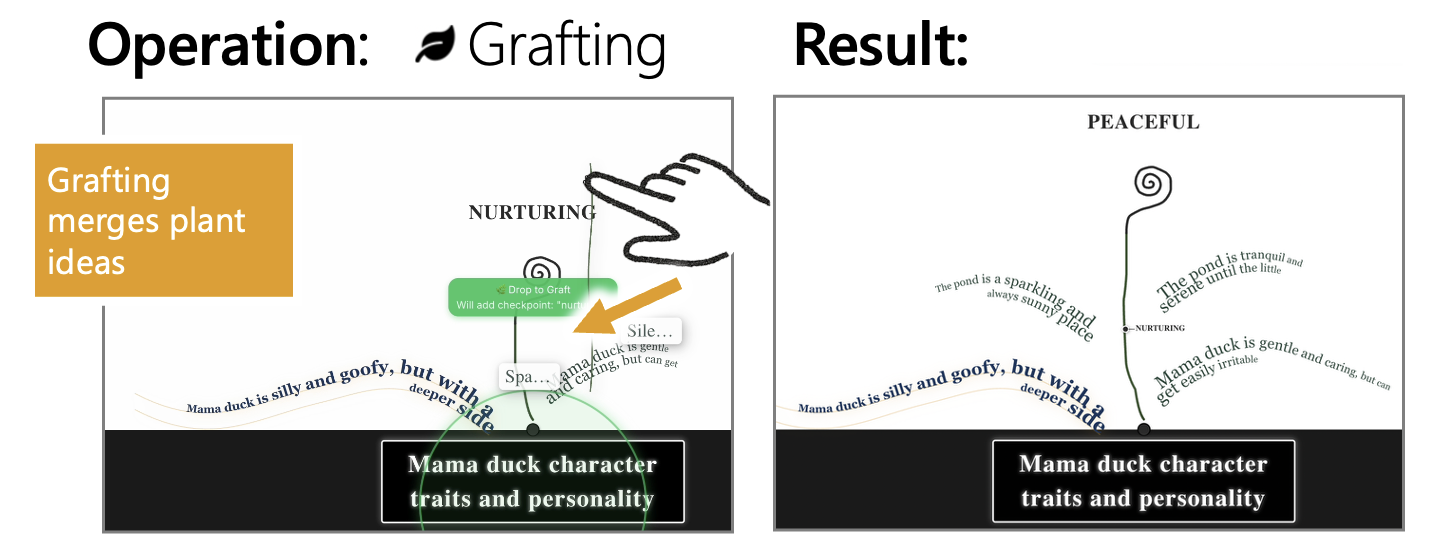}
  \caption{Grafting to organize idea chunks together ($f_1 = $ \textsc{compose})}
  \Description{DESCRIPTION}
  \label{fig:grafting}
    \vspace{12pt}
\end{figure*}

\clearpage
\newpage
\section{Prompts}

\subsection{\textit{Text as Clay}}
\begin{MyVerbatim}
export function getSqueezeEditPrompt(
  originalText: string,
  squeezedSegments: string[],
  squeezeIntensity: number,
  userWritingContext?: string
): string {
  const segmentsText = squeezedSegments.join(' ');
  const contextLine = userWritingContext ? `The user is working on: "${userWritingContext}". ` : "";
  
  return `${contextLine}You are a creative writing assistant. The user has squeezed/pressed on specific text segments to emphasize them.

Text with squeezed segments marked (< > brackets indicate squeeze intensity. The more brackets there are, the more intensely we should emphasize the text):
\${segmentsText}

Your task is to edit the text to EMPHASIZE and enhance the squeezed segments. You can:
- Add emphasis to the tagged parts
- Modify the language used, tone, or style to make those segments more prominent

CRITICAL GUIDELINES:
- ONLY EDIT TEXT WITHIN/NEAR the tagged segments (<>), and DO NOT EDIT outside of the tagged segments. Use your BEST JUDGMENT to determine which TAGGED segments to edit.
- Keep the same general structure and length of the text (DO NOT ADD TOO MUCH TEXT)
- Do NOT add markdown formatting or tags (like <>) to the text.
- IMPORTANT! If a [CONTEXT] is provided, make sure the edits are relevant to the user's current writing project.

Return only the edited text.`;
}
\end{MyVerbatim}

\begin{MyVerbatim}
export function getStretchEditPrompt(
  originalText: string,
  stretchedSegments: string[],
  stretchIntensity: number,
  userWritingContext?: string
): string {
  const segmentsText = stretchedSegments.join(' ');
  const contextLine = userWritingContext ? `The user is working on: "\${userWritingContext}". ` : "";
  
  // Check if segments are tagged (contain < >) or if it's the full text as fallback
  const hasTaggedSegments = segmentsText.includes('<') && segmentsText.includes('>');
  
  if (hasTaggedSegments) {
    return `\${contextLine}You are a creative writing assistant. The user has stretched specific text segments to expand them.

Text with stretched segments marked (< > brackets indicate stretch intensity. The more brackets there are, the more intensely we should elaborate the text):
\${segmentsText}

Your task is to expand and elaborate on the stretched segments. You can do any of the following:
- Add A BIT more elaboration to the stretched parts
- Expand descriptions and explanations as needed
- Add examples, metaphors, or additional context as needed
- Develop ideas more fully in the stretched areas as needed

CRITICAL GUIDELINES:
- Preserve the overall meaning and flow of the text
- Focus expansion ONLY WITHIN/NEAR the tagged segments and DO NOT EDIT outside of the tagged segments. Use your BEST JUDGMENT to determine which TAGGED segments to edit.
- Make sure the length of the text is ONLY SOMEWHAT LONGER THAN THE ORIGINAL TEXT (only add as many words as are in the TAGGED text)
- Do NOT add markdown formatting or tags (like <>) to the text.
- IMPORTANT! If a [CONTEXT] is provided, make sure the edits are relevant to the user's current writing project.

Return only the edited text. `;
  } else {
    // Fallback when no specific segments were detected - treat the entire text as needing expansion
    return `\${contextLine}You are a creative writing assistant. The user has performed a stretch gesture on this text to expand and elaborate it.

Original text:
\${originalText}

Your task is to expand and elaborate on the text. You can do any of the following:
- Add more elaboration and detail
- Expand descriptions and explanations
- Add examples, metaphors, or additional context
- Develop ideas more fully

CRITICAL GUIDELINES:
- Preserve the overall meaning and flow of the text
- Make the text SOMEWHAT LONGER but not excessively long (aim for about 20-40%
- Do NOT add markdown formatting or tags
- IMPORTANT! If a [CONTEXT] is provided, make sure the edits are relevant to the user's current writing project.

Return only the edited text.`;
  }
}
\end{MyVerbatim}

\begin{MyVerbatim}
export function getSquashEditPrompt(
  originalText: string,
  squashedSegments: string[],
  squashIntensity: number,
  userWritingContext?: string
): string {
  const segmentsText = squashedSegments.join(' ');
  const contextLine = userWritingContext ? `The user is working on: "\${userWritingContext}". ` : "";
  
  // Check if segments are tagged (contain < >) or if it's the full text as fallback
  const hasTaggedSegments = segmentsText.includes('<') && segmentsText.includes('>');
  
  if (hasTaggedSegments) {
    return `\${contextLine}You are a creative writing assistant. The user has squashed specific text segments to condense them.

Text with squashed segments marked (< > brackets indicate squash intensity. The more brackets there are, the more intensely we should summarize):
\${segmentsText}

Your task is to condense and summarize the squashed segments. You can do any of the following:
- Make the squashed parts more concise
- Remove redundant words or phrases
- Simplify complex descriptions
- Distill ideas to their essential elements

CRITICAL GUIDELINES:
- Preserve the overall meaning and flow of the text
- Focus condensing ONLY WITHIN/NEAR the tagged segments (<>) and DO NOT EDIT outside of the tagged segments. Use your BEST JUDGMENT to determine which TAGGED segments to edit.
- Do NOT add markdown formatting or tags (like <>) to the text.
- Make sure the new text is SHORTER than the original text
- IMPORTANT! If a [CONTEXT] is provided, make sure the edits are relevant to the user's current writing project.

Return only the edited text.`;
  } else {
    // Fallback when no specific segments were detected - treat the entire text as needing condensation
    return `\${contextLine}You are a creative writing assistant. The user has performed a squash gesture on this text to condense and summarize it.

Original text:
\${originalText}

Your task is to condense and summarize the text. You can do any of the following:
- Make the text more concise
- Remove redundant words or phrases
- Simplify complex descriptions
- Distill ideas to their essential elements

CRITICAL GUIDELINES:
- Preserve the overall meaning and flow of the text
- Make the text SHORTER than the original (aim for about 20-40%
- Do NOT add markdown formatting or tags
- IMPORTANT! If a [CONTEXT] is provided, make sure the edits are relevant to the user's current writing project.

Return only the edited text.`;
  }
}
\end{MyVerbatim}

\begin{MyVerbatim}
export function getPinchEditPrompt(
  originalText: string,
  pinchedSegments: string[],
  pinchIntensity: number,
  userWritingContext?: string
): string {
  const segmentsText = pinchedSegments.join(' ');
  const contextLine = userWritingContext ? `The user is working on: "\${userWritingContext}". ` : "";
  
  return `\${contextLine}You are a creative writing assistant. The user has pinched specific text segments to refine and make them better.

Text with pinched segments marked (< > brackets indicate pinch intensity. The more brackets there are, the more intense the refinement should be):
\${segmentsText}

Your task is to refine and make the pinched segments more concrete and specific. You can do any of the following:
- Replace abstract or vague language with concrete, specific details
- Add precise examples instead of general concepts  
- Use more vivid, sensory language
- Replace broad statements with specific, tangible descriptions
- Make metaphors more literal and grounded

CRITICAL GUIDELINES:
- Preserve the overall meaning and flow of the text
- Focus refinement ONLY WITHIN/NEAR the tagged segments (<>) and DO NOT EDIT outside of the tagged segments. Use your BEST JUDGMENT to determine which TAGGEDsegments to edit.
- Keep the same general structure and length of the text (DO NOT ADD TOO MUCH TEXT)
- Do NOT add markdown formatting or tags (like <>) to the text.
- IMPORTANT! If a [CONTEXT] is provided, make sure the edits are relevant to the user's current writing project.

Return only the edited text.`;
}
\end{MyVerbatim}

\begin{MyVerbatim}
export function getDistortEditPrompt(
  originalText: string,
  distortedSegments: string[],
  distortIntensity: number,
  userWritingContext?: string
): string {
  const segmentsText = distortedSegments.join(' ');
  const contextLine = userWritingContext ? `The user is working on: "\${userWritingContext}". ` : "";
  
  return `\${contextLine}You are a creative writing assistant. The user has smudged specific text segments to make them more abstract and conceptual.

Text with distorted segments marked (< > brackets indicate smudge intensity. The more brackets there are, the more intense the smudging is):
\${segmentsText}

Your task is to make the smudged segments more abstract. You can do any of the following:
- Replace specific details with broader, more conceptual language
- Transform literal descriptions into metaphorical or symbolic language
- Elevate practical details to theoretical or abstract concepts

CRITICAL GUIDELINES:
- Preserve the overall meaning and flow of the text
- Focus abstraction ONLY WITHIN/NEAR the tagged segments (<>) and DO NOT edit outside of the tagged segments. Use your BEST JUDGMENT to determine which TAGGED segments to edit.
- Keep the same general structure and length of the text (DO NOT ADD TOO MUCH TEXT)
- Do NOT add markdown formatting or tags (like <>) to the text.
- IMPORTANT! If a [CONTEXT] is provided, make sure the edits are relevant to the user's current writing project.

Return only the edited text.`;
}
\end{MyVerbatim}

\begin{MyVerbatim}
export function getVerticalCollisionPrompt(
  topTextWithTags: string,
  bottomTextWithTags: string,
  intensity: number,
  userWritingContext?: string
): string {
  const intensityPercent = Math.round(intensity * 100);
  
  let blendingGuidance = "";
  if (intensity < 0.6) {
    blendingGuidance = "Light blending: Make minimal changes to create a smooth transition while preserving most original content.";
  } else if (intensity < 0.9) {
    blendingGuidance = "Moderate blending: Create a natural flow by moderately editing the overlapping areas to merge the concepts.";
  } else {
    blendingGuidance = "Heavy blending: Extensively merge and blend the overlapping content to create a mixed/blended version of the two texts";
  }

  const contextLine = userWritingContext ? `The user is working on: "${userWritingContext}". ` : "";
  
  return `${contextLine}You are a creative writing assistant. The user has dragged one text block vertically into another to combine them. The overlapping visual lines have been marked with <overlap> tags.

COLLISION INTENSITY: \${intensityPercent}\% - \${blendingGuidance}

Top text block (with overlapping lines tagged):
\${topTextWithTags}

Bottom text block (with overlapping lines tagged):
\${bottomTextWithTags}

Your task is to combine these two text blocks vertically, creating a cohesive COMBINED TEXT that flows naturally from the top text into the bottom text.

CRITICAL GUIDELINES:
- Preserve the overall meaning and flow of the texts
- The TOP text block MUST ALWAYS come before the BOTTOM text block in the final combined text
- Focus your editing ONLY WITHIN/NEAR the content in <overlap> tags and DO NOT EDIT outside of the tagged areas unless absolutely necessary for coherence
- The tagged lines represent the visual overlap where the collision happened - merge these areas thoughtfully
- Apply the blending intensity guidance above to determine how extensively to merge the content
- KEEP THE TEXT THE SAME LENGTH OR SHORTER THAN THE ORIGINAL TWO TEXT BLOCKS
- DO NOT ADD MARKDOWN FORMATTING OR TAGS (like <>) TO THE TEXT.
- IMPORTANT! If a [CONTEXT] is provided, make sure the edits are relevant to the user's current writing project.

Return only the combined text.`;
}

export function getFullBlendCollisionPrompt(
  firstText: string,
  secondText: string,
  intensity: number,
  userWritingContext?: string
): string {
  const intensityPercent = Math.round(intensity * 100);
  
  const contextLine = userWritingContext ? `The user is working on: "\${userWritingContext}". ` : "";
  
  return `\${contextLine}You are a creative writing assistant. The user has created an EXTREMELY HIGH INTENSITY collision (\${intensityPercent}\%) with complete overlap between two text blocks. This requires a FULL BLEND approach where both texts are completely merged into a unified narrative.

First text:
\${firstText}

Second text:
\${secondText}

Your task is to create a completely new, unified text that seamlessly blends ALL content from both texts into a single, cohesive piece of text. This is not about insertion or simple combination - it's about creating something new that blends both sources.

CRITICAL GUIDELINES:
- Preserve the overall meaning and flow of the texts
- KEEP THE TEXT THE SAME LENGTH OR SHORTER THAN THE ORIGINAL TWO TEXT BLOCKS
- DO NOT ADD MARKDOWN FORMATTING OR TAGS (like <>) TO THE TEXT. Do NOT include any tags, markers, or explanations.
- IMPORTANT! If a [CONTEXT] is provided, make sure the edits are relevant to the user's current writing project.

Return only the fully blended text. `;
\}

export function getHorizontalCollisionPrompt(
  mainTextWithTags: string,
  insertTextWithTags: string,
  insertPosition: number,
  intensity: number,
  userWritingContext?: string,
  insertLineText?: string
): string {
  const positionDescription = insertPosition < 0.3 ? "beginning" : 
                             insertPosition > 0.7 ? "end" : "middle";
  
  let insertionInstruction;
  if (insertLineText) {
    insertionInstruction = `The second text MUST be inserted around the line: "\${insertLineText}", creating a cohesive transition between them.`;
  } else {
    insertionInstruction = `The second text MUST be inserted \${positionDescription} of the main text, creating a cohesive transition between them.`;
  }
  
  const intensityPercent = Math.round(intensity * 100);
  
  let blendingGuidance = "";
  if (intensity < 0.6) {
    blendingGuidance = "Light blending: Insert with minimal changes, preserving the original structure and making small adjustments for flow.";
  } else if (intensity < 0.9) {
    blendingGuidance = "Moderate blending: Blend the insertion more naturally by moderately editing both texts to create smoother integration.";
  } else {
    blendingGuidance = "Heavy blending: Extensively merge and weave the content together, creating a deeply integrated unified combined text.";
  }
  
  const contextLine = userWritingContext ? `The user is working on: "\${userWritingContext}". ` : "";
  
  return `\${contextLine}You are a creative writing assistant. The user has dragged one text block horizontally into another to insert and blend the content. The overlapping visual lines have been marked with <overlap> tags.

COLLISION INTENSITY: \${intensityPercent}%

Main text block (with overlapping lines tagged):
\${mainTextWithTags}

Text to insert (with overlapping lines tagged):
\${insertTextWithTags}

\${insertionInstruction}

CRITICAL GUIDELINES:
- Preserve the overall meaning and flow of the texts
- Focus your editing ONLY on the content WITHIN/NEAR <overlap> tags and DO NOT EDIT outside of the tagged areas unless absolutely necessary for coherence
- The tagged lines represent the visual overlap where the collision happened - merge these areas thoughtfully
- Apply the blending intensity guidance above to determine how extensively to merge the content
- KEEP THE TEXT THE SAME LENGTH OR SHORTER THAN THE ORIGINAL TWO TEXT BLOCKS
- DO NOT ADD MARKDOWN FORMATTING OR TAGS (like <>) TO THE TEXT.
- IMPORTANT! If a [CONTEXT] is provided, make sure the edits are relevant to the user's current writing project.

Return only the combined text.`;
}
\end{MyVerbatim}

\subsection{\textit{Text as Plants}}
\begin{MyVerbatim}
GENERATE_IDEA_PAIR = `You are a creative idea generator. Given a seed idea and a dimension to explore, generate a PAIR of creative ideas that build upon the FULL context of all previous ideas and dimension evolution.

The first idea should be a NEW idea of {seedText} representing the NEXT LEVEL of progression in the specified dimension, and combines prior ideas, making sure it prioritizes {seedText}. It does NOT need to combine ALL ideas, just the most relevant ones.
The second idea should be a VARIANT of that synthesis—a different approach to the combined concept.

What we are generating (the seed): "{seedText}"
How the ideas evolve over time (the dimension to progress in): "{dimension}"

PRIOR IDEAS:
{existingIdeas}

{rootContextLine}

CRITICAL FORMATTING REQUIREMENTS:
- Return ONLY a JSON object with two ideas, each containing "gist" and "full" fields
- GIST should be 1 phrase with keywords around 10 words long (doesn't have to be perfectly formatted sentence, and avoid filler words please) that captures the essence of the idea. Make sure idea gists are immediately distinct form each other.
- FULL should be the complete idea (at most 100 words)
- Example: {"ideas": [{"gist": "Flying car automotive + aviation for urban transport.", "full": "A vehicle that combines traditional car functionality with flight capabilities for urban transport, featuring vertical takeoff and landing capabilities, autonomous navigation systems, and hybrid propulsion technology."}, {"gist": "Underwater sustainable city design + marine ecosystem integration.", "full": "An architectural concept for sustainable underwater communities with transparent domes, renewable energy systems, and integrated marine life habitats."}]}
- Do NOT use numbering, bullets, or other prefixes

Guidelines:
- IMPORTANT! If a [CONTEXT] is provided, make sure the idea is relevant to the user's current writing project. However, the idea should still be a {seedText} (we are generating ideas to support the [CONTEXT], not directly generating ideas for the [CONTEXT] directly)
- IMPORTANT! If IDEAS FROM ROOT NETWORK is provided, make sure to include those in the ideas generated
- SPECIAL CASE: If the dimension is "creative", "new", "novel", "original", "fresh", "innovative", or any similar generic term for generating new ideas, then Idea 1 should be a completely NEW and DIFFERENT seed concept that is unrelated to both the current {seedText} and the PRIOR IDEAS. This should be a totally fresh starting point. Idea 2 should still be a variant of this new Idea 1.
- Idea 1 (New Progression): Should be MORE {dimension} than previous ideas, considering the FULL evolution of the ideas. However, do not only be constrained to the history, and be creative.
- Idea 2 (Variant): Must be a creative alternative to Idea 1.
- GIST FORMAT: Each gist should be one phrase with keywords (around 10 words long) that captures the core concept of the idea.

Generate the PAIR of ideas now.`,

\end{MyVerbatim}

\begin{MyVerbatim}
INITIAL_IDEA_PAIR = `You are a creative idea generator. Generate a PAIR of creative ideas based on the seed and dimension.
The first idea should be a NEW idea formed from the seed text ({seedText}), based on the dimension ({dimension}).
The second idea should be a VARIANT of the first idea (e.g. "A dog with wings" and "A cat with claws").

What we are generating (the seed): "{seedText}"
How the ideas evolve over time (the dimension to progress in): "{dimension}"

{rootContextLine}

CRITICAL FORMATTING REQUIREMENTS:
- Return ONLY a JSON object with two ideas, each containing "gist" and "full" fields
- GIST should be 1 phrase with keywords around 10 words long (doesn't have to be perfectly formatted sentence, and avoid filler words please) that captures the essence of the idea. Make sure idea gists are immediately distinct form each other.
- FULL should be the complete idea (at most 100 words)
- Example: {"ideas": [{"gist": "Flying car automotive + aviation for urban transport.", "full": "A vehicle that combines traditional car functionality with flight capabilities for urban transport, featuring vertical takeoff and landing capabilities, autonomous navigation systems, and hybrid propulsion technology."}, {"gist": "Underwater sustainable city design + marine ecosystem integration.", "full": "An architectural concept for sustainable underwater communities with transparent domes, renewable energy systems, and integrated marine life habitats."}]}
- Do NOT use numbering, bullets, or other prefixes

Guidelines:
- IMPORTANT! If a [CONTEXT] is provided, make sure the idea is relevant to the user's current writing project. However, the idea should still be a {seedText} (we are generating ideas to support the [CONTEXT], not directly generating ideas for the [CONTEXT] directly)
- IMPORTANT! If IDEAS FROM ROOT NETWORK is provided, make sure to include those in the ideas generated
- GIST FORMAT: Each gist should be one phrase with keywords (around 10 words long) that captures the core concept of the idea.

Generate the PAIR of ideas now.`,
\end{MyVerbatim}

\begin{MyVerbatim}
VOICE_PLANT_CREATION_PROMPT = `
Extract the SEED (what the user wants ideas on) and DIMENSION (ways the ideas evolve) from the following voice input.
- The SEED should be the core concept the user wants to ideate on.
- The DIMENSION should be a phrase that explains how ideas can vary/be generated along a certain axis. For example, more believable, more creative, etc.
IMPORTANT! If a [CONTEXT] is provided, make sure the seed text and dimension are relevant to the user's current writing project. However, the seed should still be what the user specified (they are generating ideas to support the [CONTEXT], not directly generating ideas for the [CONTEXT] directly)

Example inputs and outputs:
Input: "I want to explore how technology affects human relationships"
Output: { seed: "technology's impact on human relationships", dimension: "digital impact" }

Input: "Let's think about sustainable energy solutions"
Output: { seed: "sustainable energy soltuions", dimension: "sustainability" }

Input: "Consider the role of art in society"
Output: { seed: "art in society", dimension: "creativity" }

If you can't figure out the dimension, just default the dimension to "creativity".

Voice input: "{transcript}"

CRITICAL FORMATTING REQUIREMENTS:
- Keep the SEED short and concise. AVOID BEING VERBOSE. (at most 100 words)
- Keep the DIMENSION short and concise. AVOID BEING VERBOSE. (at most 10 words)
- Return ONLY a JSON object with seed and dimension fields, nothing else.`;

\end{MyVerbatim}

\begin{MyVerbatim}
export const ROOT_COMBINE_PROMPT = `
You are analyzing a collection of creative ideas that have been combined together. Your task is to synthesize these ideas into a new seed concept and dimension for further ideation.
IMPORTANT! If a [CONTEXT] is provided, make sure the seed text and dimension are relevant to the user's current writing project. However, the seed should still be what the user specified (they are generating ideas to support the [CONTEXT], not directly generating ideas for the [CONTEXT] directly)

The ideas to combine:
{ideas}

Analyze the patterns, themes, and connections between these ideas. Then create:
1. A SEED: The synthesized concept that captures the essence of the combined ideas, and what we want to generate more of
2. A DIMENSION: A phrase that represents the key axis along which this concept can be explored further (e.g. "creativity", "sustainability", "accessibility" etc)

CRITICAL FORMATTING REQUIREMENTS:
- Keep the SEED short and concise. AVOID BEING VERBOSE. (at most 100 words)
- Keep the DIMENSION short and concise. AVOID BEING VERBOSE. (at most 10 words)
- Return ONLY a JSON object with seed and dimension fields, nothing else. Example: { "seed": "sustainable urban mobility", "dimension": "accessibility" }`;
\end{MyVerbatim}

\end{document}